\documentclass[sigconf]{acmart}

\AtBeginDocument{%
  }

\copyrightyear{2026}
\acmYear{2026}
\setcopyright{cc}
\setcctype{by-nc-nd}
\acmConference[CHI '26]{Proceedings of the 2026 CHI Conference on Human Factors in Computing Systems}{April 13--17, 2026}{Barcelona, Spain}
\acmBooktitle{Proceedings of the 2026 CHI Conference on Human Factors in Computing Systems (CHI '26), April 13--17, 2026, Barcelona, Spain}
\acmPrice{}
\acmDOI{10.1145/3772318.3790824}
\acmISBN{979-8-4007-2278-3/2026/04}

\acmISBN{978-1-4503-XXXX-X/18/06}

\usepackage{listings}
\usepackage{makecell}
\usepackage{url}
\usepackage{longtable}
\usepackage{color}
\usepackage{tabu} 
\usepackage{tabularx}
\usepackage{multirow}
\usepackage{enumitem}
\usepackage{cleveref}

\definecolor{codegreen}{rgb}{0,0.6,0}
\definecolor{codegray}{rgb}{0.5,0.5,0.5}
\definecolor{codepurple}{rgb}{0.58,0,0.82}
\definecolor{backcolour}{rgb}{0.95,0.95,0.92}
\lstdefinestyle{mystyle}{
  backgroundcolor=\color{backcolour}, commentstyle=\color{codegreen},
  keywordstyle=\color{magenta},
  numberstyle=\tiny\color{codegray},
  stringstyle=\color{codepurple},
  basicstyle=\ttfamily\footnotesize,
  breakatwhitespace=false,         
  breaklines=true,                 
  captionpos=b,                    
  keepspaces=true,                 
  numbers=left,                    
  numbersep=5pt,                  
  showspaces=false,                
  showstringspaces=false,
  showtabs=false,                  
  tabsize=2
}
\definecolor{PurpleColor}{RGB}{0,0,0}
\newcommand{\RR}[1]{{\color{PurpleColor}#1}}
\definecolor{Purple}{RGB}{0,0,0}
\newcommand{\RX}[1]{{\color{Purple}#1}}
\begin{document}

\title{ClassAid: A Real-time Instructor-AI-Student Orchestration System for Classroom Programming Activities}

\author{Gefei Zhang}
\affiliation{%
  \institution{Zhejiang University of Technology}
  \city{Hangzhou}
  \state{Zhejiang}
  \country{China}}
\email{gefei@zjut.edu.cn}
\orcid{1234-5678-9012}

\author{Guodao Sun}
\authornote{Corresponding author.}
\affiliation{%
  \institution{Zhejiang Key Laboratory of Visual Information Intelligent Processing\\Zhejiang University of Technology}
   \city{Hangzhou}
   \state{Zhejiang}
   \country{China}}
% \affiliation{%
%   \institution{Zhejiang Key Laboratory of Visual Information Intelligent Processing}
%   \city{Hangzhou}
%   \state{Zhejiang}
\email{guodao@zjut.edu.cn}

\author{Meng Xia}
\authornote{Corresponding author.}
\affiliation{%
  \institution{Texas A\&M University}
  \city{College Station}
  \state{Texas}
  \country{USA}}
\email{mengxia@tamu.edu}

\author{Ronghua Liang}
\affiliation{%
  \institution{Zhejiang University of Technology}
  \city{Hangzhou}
  \state{Zhejiang}
  \country{China}}
\email{rhliang@zjut.edu.cn}

\renewcommand{\shortauthors}{Trovato et al.}

\begin{abstract}
Generative AI is reshaping education, but it also raises concerns about instability and overreliance. In programming classrooms, we aim to leverage its feedback capabilities while reinforcing the educator's role in guiding student–AI interactions.
We developed \textit{ClassAid}, a real-time orchestration system that integrates TA Agents to provide personalized support and an AI-driven dashboard that visualizes student–AI interactions, enabling instructors to dynamically adjust TA Agent modes. Instructors can configure the Agent to provide technical feedback (direct coding solutions), heuristic feedback (hint-based guidance), automatic feedback (autonomously selecting technical or heuristic support), or silent operation (no AI support).
We evaluated \textit{ClassAid} through three aspects: (1) the TA Agents' performance, (2) feedback from 54 students and one instructor during a classroom deployment, and (3) \RR{interviews with eight educators}. 
Results demonstrate that dynamic instructor control over AI supports effective real-time personalized feedback and provides design implications for integrating AI into authentic educational settings.

\end{abstract}

\begin{CCSXML}
<ccs2012>
   <concept>
       <concept_id>10003120.10003145.10003151</concept_id>
       <concept_desc>Human-centered computing~Visualization systems and tools</concept_desc>
       <concept_significance>500</concept_significance>
       </concept>
   <concept>
       <concept_id>10010405.10010489.10010492</concept_id>
       <concept_desc>Applied computing~Interactive learning environments</concept_desc>
       <concept_significance>500</concept_significance>
       </concept>
 </ccs2012>
\end{CCSXML}

\ccsdesc[500]{Human-centered computing~Visualization systems and tools}
\ccsdesc[500]{Applied computing~Interactive learning environments}

\ccsdesc[500]{Human-centered computing~Visualization systems and tools}
\begin{comment}

\end{comment}
\keywords{Programming education, AI assistants, class deployment, orchestration system}

\begin{teaserfigure}
 \includegraphics[width=\textwidth]{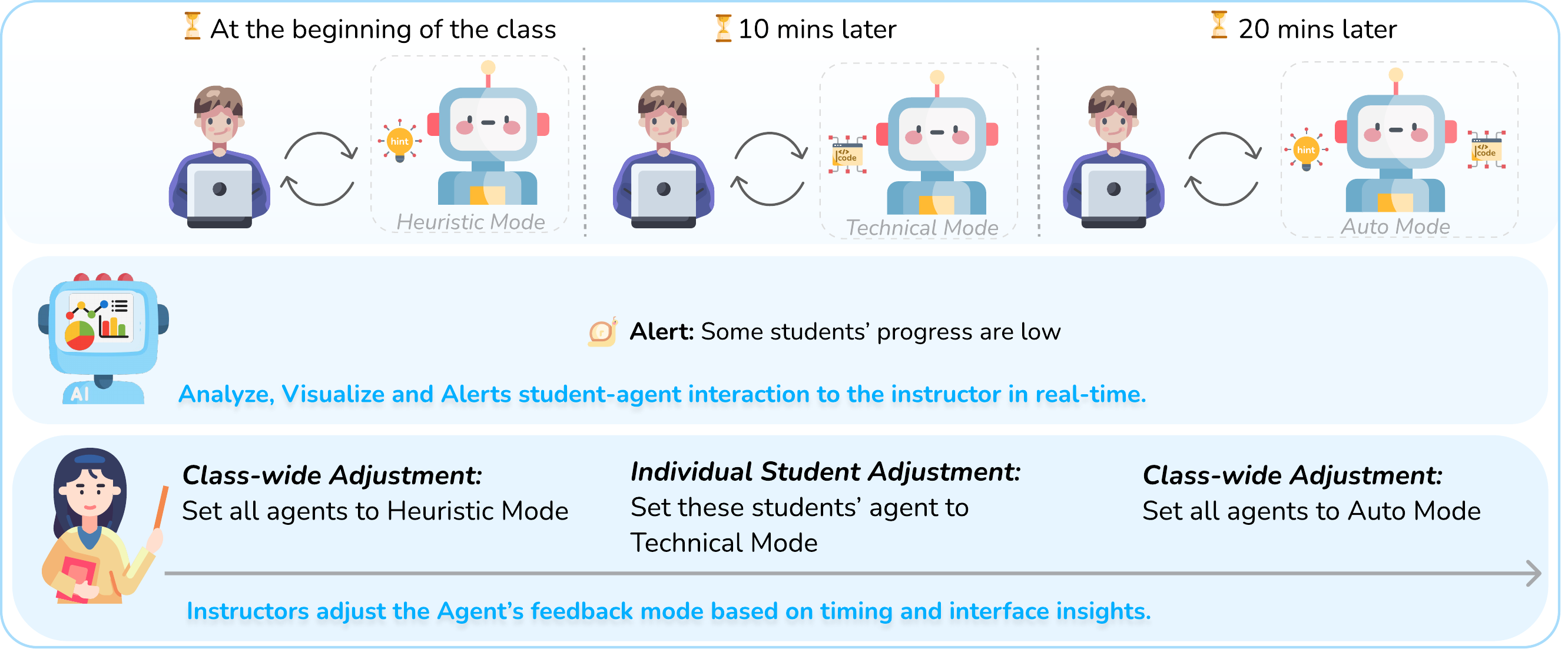}
  \caption{A real-world classroom example of using the \textit{ClassAid} system is shown here. At the beginning of the class, the instructor set all TA agents to Heuristic Mode, which provides high-level hints to encourage independent thinking. After 10 minutes, noticing that some students were progressing slowly, the instructor switched their agents to Technical Mode, which offers code examples. At the 20-minute mark, all agents were changed to Auto Mode, allowing the system to adaptively support students based on their real-time performance.}
  \Description{A real-world classroom example of using the \textit{ClassAid} system is shown here. At the beginning of the class, the instructor set all TA agents to Heuristic Mode, which provides high-level hints to encourage independent thinking. After 10 minutes, noticing that some students were progressing slowly, the instructor switched their agents to Technical Mode, which offers code examples. At the 20-minute mark, all agents were changed to Auto Mode, allowing the system to adaptively support students based on their real-time performance.}
  \label{fig:teaser}
\end{teaserfigure}


\maketitle
% \newtcbox{\inlinecode}{on line, colback=gray!10, colframe=gray!10, boxrule=0.1mm, 
% 	rounded corners, fontupper=\bfseries, left=0.1mm, right=0.1mm, top=0.1mm, bottom=0.1mm}

\section{Introduction} 
\label{sec:introduction}
Programming is increasingly recognized as a foundational literacy of the digital era, swelling beginner enrollments and straining the capacity of large courses to provide timely, individualized feedback~\cite{kumar2023bridging,liffiton2023codehelp}.
Conventional supports can be insufficient for novices, as limited instructor time, help-seeking hesitancy, and repeated requests from a small subset of students often result in uneven feedback~\cite{zhang2025cpvis,zheng2022exploratory,kazemitabaar2024codeaid,liu2024teaching}.
Generative AI, particularly large language models (LLMs) such as ChatGPT, has opened significant opportunities for programming education by improving information retrieval, serving as capable programming assistants, and reducing instructor workload~\cite{moore2024teaching,neyem2024towards,woodrow2024ai}.
LLMs achieve high accuracy on beginner tasks and readily complete small programming exercises~\cite{jury2024evaluating,jacobs2024evaluating}.
% \RR{However, AI instability and student overreliance remain major concerns in classroom settings~\cite{liu2024teaching,hou2024codetailor}.
% CodeAid provides accurate, solution-free feedback but lacks real-time monitoring for instructors, while SPHERE supports large-scale personalized feedback without real-time interactivity or impact assessment~\cite{kazemitabaar2024codeaid,tang2024sphere}. }
% Although such systems improve feedback quality and reduce copying risks, they remain insufficiently dynamic, failing to adjust feedback in real time to students' unfolding performance~\cite{kuramitsu2023kogi,liao2024scaffolding,lan2024teachers} and overlook the instructor's role as an active and autonomous participant in the classroom~\cite{lan2024teachers}.
\RR{
Despite this promise, classroom use remains challenging due to persistent AI instability and student overreliance~\cite{liu2024teaching,hou2024codetailor}, highlighting the need for more reliable, pedagogically aligned support systems.
Systems such as CodeAid leverage LLMs to provide accurate, solution-free feedback for post-class learning, yet they lack mechanisms for real-time instructor monitoring and adaptive adjustment to students' evolving progress~\cite{kazemitabaar2024codeaid}. Other systems, such as SPHERE, utilize LLMs to help instructors generate large-scale, high-quality personalized feedback. However, they did not analyze how students interact with LLM-driven agents and remain insufficiently dynamic to adjust responses in real time.
Nonetheless, none of these systems is designed to continuously interpret ongoing student–AI interactions or to support instructors in real-time orchestration of the AI's behavior~\cite{lan2024teachers}, which is essential for preventing overreliance and enabling more targeted, personalized support~\cite{kuramitsu2023kogi,liao2024scaffolding,lan2024teachers}.

% CodeAid provides accurate, solution-free feedback for post-class learning, yet it lacks mechanisms for real-time instructor monitoring and adaptive adjustment to students' ongoing progress~\cite{kazemitabaar2024codeaid}.
% In contrast, SPHERE supports instructors in generating large-scale, high-quality personalized feedback, centering on using LLMs to assist instructors in reviewing and validating feedback rather than facilitating in-class adaptation~\cite{tang2024sphere}.
% Although such systems effectively reduce risks of code copying and enhance feedback quality, they remain insufficiently dynamic to adjust responses in real time to students' evolving needs~\cite{kuramitsu2023kogi,liao2024scaffolding,lan2024teachers} and tend to overlook the instructor's role as an active and autonomous participant in the classroom~\cite{lan2024teachers}.
}

Moreover, most LLM-based programming assistants remain passive, responding only to explicit prompts~\cite{liu2024proactive,liu2024teaching}. In contrast, human instructors in classrooms actively circulate, detect when students are stuck or disengaged, and intervene proactively~\cite{yan2024practical}. 
To mitigate AI passivity and uncertainty while ensuring responsible use, AI systems should incorporate instructor-like reasoning while remaining under instructor control, allowing it to adapt to students' evolving learning states~\cite{lyu2024evaluating}.
Accordingly, there is a need for classroom-oriented AI that operates under instructor oversight to provide real-time, personalized feedback, reduce instructors' workload, and foster student engagement~\cite{chen2024learning}.

Based on formative and dynamic assessment theories~\cite{black2009developing,minick1987implications}, we developed an intelligent TA Agent within the \textit{ClassAid} student interface to deliver personalized and adaptive support.
The agent continuously \textit{monitors} \RR{students' interactions with AI} to \textit{identify} metacognitive levels and potential obstacles, \textit{reviews} prior work and current performance to diagnose issues and assess progress, \textit{considers} alternative feedback responses, \textit{selects} the one most aligned with the student's current needs, and implements targeted \textit{interventions} to help students adjust strategies and strengthen metacognitive abilities.
In parallel, \textit{ClassAid} provides an instructor dashboard \RR{that offers real-time visibility into student–AI interaction patterns and the TA Agent's response.}
% for real-time supervision of TA Agents and students' learning performance. 
% Instructors can dynamically adjust feedback by switching among four modes (\textit{technical, heuristic, automatic, silent}) to match classroom context, supporting cognitive development while discouraging overreliance on AI.
Instructors can dynamically regulate the Agent's behavior by switching among four feedback modes (\textit{technical, heuristic, automatic, silent}), ensuring pedagogically aligned guidance while preventing student overreliance on AI.
% We deployed \textit{ClassAid} in a real classroom setting with 54 students. Course instructors actively used the instructor dashboard to monitor and manage the TA Agents, validating the system's usefulness and usability in authentic teaching scenarios.
We deployed \textit{ClassAid} in a class activity with 54 students, where the instructor actively monitored and adjusted Agent behaviors through the dashboard, demonstrating the system's practicality and effectiveness in authentic teaching scenarios.
\RR{Follow-up interviews with eight programming educators further validated its instructional value and highlighted its potential for broader application.}
% In summary, this paper has three main contributions. 
\RR{The contributions of this study are summarized as follows:}
\begin{itemize}

\item We propose a six-stage intelligent TA Agent framework based on formative and dynamic assessment theories~\cite{black2009developing,minick1987implications}, 
which operationalizes instructors' diagnostic reasoning into dynamic analysis and personalized feedback.
% \item \RR{We develop \textit{ClassAid}, an instructor-AI-student orchestration system that integrates student interaction and instructor monitoring, allowing real-time adjustment of Agent feedback modes to enable human-AI collaborative personalized teaching while preventing student overreliance on AI.}
\item \RR{We develop \textit{ClassAid}, an instructor-AI-student orchestration system that analyzes students' interactions with AI in real time and enables instructors to dynamically adjust the Agent's behavior for personalized support while preventing overreliance.}
\item Through classroom deployment (n = 54), TA Agents' feedback-quality evaluation, and educator interviews (n = 8), we demonstrate that \textit{ClassAid}'s instructor–AI collaboration provides effective real-time personalized support for classroom programming.

\end{itemize}

% First, we propose a six-stage intelligent TA Agent framework based on formative and dynamic assessment theories~\cite{black2009developing,minick1987implications}, 
% which operationalizes instructors' diagnostic reasoning into dynamic analysis and personalized feedback.

% Second, we develop \textit{ClassAid}, a system with two interfaces that integrates student interaction and instructor monitoring, allowing real-time adjustment of Agent feedback modes to enable human-AI collaborative personalized teaching while preventing student overreliance on AI.

% Third, we conduct classroom deployment, TA Agents' feedback quality, and educator interviews, demonstrating that \textit{ClassAid}'s dynamic instructor-AI collaboration effectively provides real-time, personalized support in classroom programming.

\section{Related Work}
\label{sec:related}
% In this section, we discuss the relevant research, including LLM-based Conversational Agent in Programming Education and Human-AI Co-Orchestration in the Classroom.
\subsection{LLM-based Conversational Agent in Programming Education}

LLMs show promise in programming education by automating content generation~\cite{denny2023promptly}, improving error explanations and debugging~\cite{pu2025assistance}, and enabling strategies such as AI-guided learning and code refactoring~\cite{liffiton2023codehelp}. 
But instructors worry that, especially for beginners~\cite{nguyen2024beginning,kasneci2023chatgpt}, AI's excessive helpfulness can foster overreliance and weaken critical thinking~\cite{harvey2025don}. 
For example, novices using tools like GitHub Copilot embedded in the IDE may quickly accept automatic suggestions without understanding the underlying logic, leading to passive engagement~\cite{dakhel2023github}.

To address this issue, researchers design ``guardrails'' to ensure academic integrity and promote meaningful learning~\cite{liu2024teaching}. 
For instance, Liu et al. developed CS50.ai, which not only provides fast and accurate AI-generated answers but also incorporates ``teaching guardrails'' to encourage students to think critically rather than simply providing answers~\cite{estevez2024evaluation}. 
\RR{Similarly, CodeAid designed various query functions to prevent AI from directly offering solutions~\cite{kazemitabaar2024codeaid}.}
% Although these ``coding assistants'' effectively avoid directly providing answers~\cite{kazemitabaar2024codeaid}, existing systems often lack the kind of interactive engagement seen with human instructors and are still not sufficiently intelligent. 
% Research shows that students tend to prefer interacting with human instructors rather than relying on automated systems~\cite{echeverria2023designing}. 
% This is mainly because current LLM outputs are typically constrained by fixed prompts and struggle to adapt flexibly to the ever-changing needs of students~\cite{lyu2024evaluating}. 
% Additionally, most current programming assistance systems treat AI as a passive assistant, responding only when students explicitly ask. 
Although such coding assistants avoid giving direct answers \cite{estevez2024evaluation,kazemitabaar2024codeaid}, existing systems still lack human-like \RR{teaching interaction capabilities and exhibit insufficient intelligence.}
% interactive engagement and remain insufficiently intelligent. 
% Students often prefer interacting with human instructors rather than automated systems \cite{echeverria2023designing}. 
\RR{A key limitation is that outputs are tied to fixed prompts and adapt poorly to evolving student needs in real class\cite{lyu2024evaluating}. }
Most systems also keep AI reactive, responding only to explicit requests.
% Recent work~\cite{liu2024proactive, chen2024need} has explored the idea of proactively detecting code and offering proactive help. 
% However, their feedback tends to be task completion-oriented and lacks relevance and consistency with specific learning objectives~\cite{beck2023backtalk}.
Recent studies explore proactive code detection and help \cite{liu2024proactive,chen2024need}, but their feedback often remains oriented toward task completion and weakly aligned with specific learning objectives \cite{beck2023backtalk}.

% In reality, the support that instructors provide to students during classroom programming activities is dynamic and complex~\cite{tan2024more}. 
Instructors' support during classroom programming is dynamic and context dependent \cite{tan2024more}.
It must account for students' cognitive level, learning objectives, progress, and overall class performance, which calls for personalized and adaptive guidance \cite{mogavi2024chatgpt}.
% It involves considering multiple factors, such as the student's cognitive level, specific learning objectives, progress, and the overall performance of the class, requiring dynamic and personalized support~\cite{mogavi2024chatgpt}.
% Several studies have investigated the effectiveness of providing personalized learning assistance to students. 
For example, 
% Code Tailor uses LLMs to provide Parsons puzzles based on personalized scenarios to help students struggling with learning~\cite{hou2024codetailor}, and 
% SPHERE supports instructors in providing personalized feedback based on ``evidence'' from the student learning processes~\cite{tang2024sphere}. 
% \RR{SPHERE leverages students' process-based evidence to support instructors in generating high-quality personalized feedback~\cite{tang2024sphere}. However, it primarily focuses on instructors' feedback creation and verification, without providing real-time interactivity or adaptive adjustments for students.}
\RR{SPHERE leverages students' process-based evidence to help instructors craft personalized feedback~\cite{tang2024sphere}, but its instructor-driven, asynchronous workflow requires manual review, which prevents it from delivering real-time or adaptively responsive support to students' evolving needs or their interactions with AI.
% However, these approaches still rely entirely on AI judgment, with limited oversight and few avenues for timely adjustment.
% feedback mechanisms often rely entirely on AI's judgment, lacking oversight of AI's performance and the ability to make timely adjustments.
% This paper examines how to design more intelligent AI that participates proactively in students' work and how to let instructors dynamically adjust and supervise its assistance in classroom programming. 
This paper examines how to design more intelligent AI that can proactively participate in students' work in real-time while remaining dynamically adjustable and supervisable by instructors in classroom programming.
}

% It also investigates how AI can participate proactively in students' work through multidimensional performance analysis to deliver more personalized and flexible support.
% In this paper, we focus on how to design more intelligent AI systems to support instructors in dynamically adjusting and supervising AI based on the complex needs of students in classroom programming activities. We also explore how AI can proactively participate in students' programming processes based on multidimensional student performance analysis, providing more personalized and flexible support.

\RR{
\subsection{Classroom Orchestration in Education}

\subsubsection{Real-Time Classroom Orchestration}
Classroom orchestration encompasses individual assignments~\cite{head2017writing}, group tasks, and whole-class interactions~\cite{lui2023facilitated}. 
Instructors need to plan~\cite{glassman2015overcode}, monitor, and adjust activities to achieve instructional goals~\cite{martinez2014mtfeedback}. 
Real-time orchestration and monitoring are essential for providing timely feedback~\cite{hmelo2004problem}. 
For example, FACT monitors learners' behaviors, such as handwriting and typing, to help instructors stay aware of classroom dynamics~\cite{vanlehn2018can}.
Codeopticon displays each learner's actions on a dashboard with real-time collage views of editing and debugging processes, supported by chat for one-to-many tutoring~\cite{guo2015codeopticon}. 
VizGroup offers group-level anomaly detection through alerts and notifications~\cite{tang2024vizgroup}. 
Tools such as RIMES~\cite{kim2015rimes} and VizProg~\cite{zhang2023vizprog} provide dashboards for real-time observation, enabling instructors to identify key learning behaviors. 
Building on these systems, SPARK introduces a checkpoint-based progress monitoring framework that dynamically visualizes students' progress across stages~\cite{yangspark}.
These systems improve instructors' monitoring and analytical efficiency but remain focused primarily on real-time learning analytics, with limited support for in-situ classroom adjustments. \textit{ClassAid} extends this line of work by enabling instructors to not only view learning analytics but also dynamically adjust classroom strategies and feedback in response to emerging instructional needs.

}

% Classroom orchestration refers to the management of classroom workflows, including individual activities, group tasks, and whole-class interactions. 
% Instructors need to plan, monitor, and dynamically adjust classroom activities to ensure the achievement of teaching objectives~\cite{martinez2014mtfeedback,lui2023facilitated}.

% Classroom orchestration manages individual work, group tasks, and whole-class interactions~\cite{lui2023facilitated}.
% Instructors need to plan, monitor, and adapt activities to meet objectives~\cite{martinez2014mtfeedback}.

% Codeopticon supported one-to-many tutoring by showing tiled views of students' editing and debugging with chat, but the interface became overwhelming at scale~\cite{guo2015codeopticon}.

% Mixed initiative tools let instructors write reusable comments and propagate them across similar submissions~\cite{head2017writing}. 

% OverCode clustered student solutions to reveal high-level patterns and enable batch feedback~\cite{glassman2015overcode}. 
% These tools improved review and feedback efficiency, but provided limited real-time visibility into classroom dynamics.

% FACT in mathematics learning identifies students who need help in real time and recommends interventions such as targeted or whole-class broadcasts, combining prompting, questioning, and feedback with rule-based or data-driven strategies to reduce decision errors \cite{vanlehn2018can}. 

\subsubsection{Human–AI Co-Orchestration}

% Holstein et al. developed a mixed-reality orchestration tool called Lumilo, which can monitor students' status and behaviors (such as task deviations or learning difficulties) in real-time within intelligent tutoring systems (ITS), helping instructors quickly identify and address students who need additional support~\cite{holstein2019co}.
% With the progress of AI, coordination tools have evolved into a new human-AI collaborative orchestration model, designed to assist instructors in managing complex classroom tasks~\cite{van2019orchestration,holstein2019co}. 
With advances in AI, coordination tools now support a human and AI collaborative orchestration model that helps manage complex classrooms \cite{van2019orchestration} \cite{holstein2019co}.
For example, 
% VizGroup in programming classrooms provides instructors with group anomaly analysis through alerts and notifications, enhancing classroom insights~\cite{tang2024vizgroup}.
Pair Up in mathematics education automates peer matching and assigns roles such as mentor and problem solver, easing group management \cite{yang2023pair}.
Despite demonstrated benefits, AI-based orchestration tools remain inflexible.
% One key limitation is that they tend to focus on helping instructors adjust lesson plans while overlooking the evolving cognitive needs of novice students in classroom programming environments~\cite{holstein2019designing,wei2022emergent}, and they fail to provide personalized support and adaptive guidance~\cite{wei2022chain}.
They center on lesson-plan adjustments rather than novices' evolving cognitive needs, offering little personalized or adaptive support \cite{holstein2019designing,wei2022emergent,wei2022chain}.
Research also seldom examines how students interact with generative AI or how instructors can monitor and manage these interactions in real time \cite{chen2024stugptviz,williamson2024time}.
% More importantly, existing systems and research rarely examine how students interact with generative AI during learning activities, or how instructors can effectively monitor and manage these interactions in real time~\cite{chen2024stugptviz,williamson2024time}. 
% As generative AI becomes increasingly prevalent in classrooms, this oversight not only risks missed opportunities for timely pedagogical intervention, but also raises concerns about the quality, reliability, and safety of student-AI interactions—such as the potential for misinformation, cognitive overload, or overreliance on AI tools~\cite{wu2022ai,park2024generative}.
As generative AI becomes increasingly common in classrooms, this gap risks missed timely interventions and raises concerns about the quality, reliability, and safety of interactions between students and AI, including the potential for misinformation, cognitive overload, and overreliance \cite{wu2022ai} \cite{park2024generative}.
Instructors, therefore, need real-time insights into individual progress, whole-class dynamics, and how students engage with AI systems \cite{yang2021surveying}.
% Therefore, instructors need real-time insights not only into individual learning progress and whole-class dynamics~\cite{yang2021surveying}, but also into how students engage with AI systems. 
Such visibility enables more purposeful interventions and helps ensure the appropriate, productive, and safe use of AI \cite{khosravi2022explainable}.
% This kind of oversight supports more purposeful interventions and helps ensure that students are using AI in appropriate, productive, and safe ways~\cite{khosravi2022explainable}.
To address these challenges, we study students' cognitive processes during classroom programming and introduce a multi-level TA Agent framework. 
\textit{ClassAid} provides interactive AI guidance for students and a transparent dashboard for instructors to observe and understand interactions between students and AI.

% To address these challenges, our study investigates students' dynamic cognitive during classroom programming activities, and introduces a multi-level TA Agent framework. 

% \textit{ClassAid} not only supports students directly through interactive AI guidance, but also serves as a transparent dashboard for instructors to observe and understand student-AI interactions. 
% By supervising how students engage with the TA Agent, instructors are better equipped to manage both individual learning and class-wide instruction—while maintaining visibility and control over the role of generative AI in the learning process, thereby safeguarding instructional quality and student well-being.

\section{Formative Study}
% We conducted a formative study to explore the challenges instructors face during classroom programming activities, as well as their concerns and expectations regarding the integration of AI tools into real-world teaching scenarios.
% A total of seven participants with experience in teaching programming were recruited for this study (four female, age: 35.42 ± 6.50). The group consisted of seven university instructors (T1–T7, experience teaching programming: 8.14 ± 6.67). Participants were recruited through snowball sampling from the research team's professional network and were compensated \$20 each for their participation~\cite{given2008sage}.
% Details on the teaching backgrounds of the participants and demographic information are provided in the supplementary materials. This study was approved by the Institutional Review Board of the host university and adhered to established ethical research guidelines.

We conducted a formative study to investigate challenges in classroom programming and to understand instructors' concerns and expectations about integrating AI into teaching.
% We recruited seven university programming instructors (T1–T7) via snowball sampling from our professional network (4 female; mean age: 35.42 years, SD = 6.50 years; teaching and programming experience: mean = 8.14 years, SD = 6.67 years) and each received \$20 compensation. 
Seven university programming instructors (T1–T7) were recruited through snowball sampling (4 female; mean age = 35.42, SD = 6.50; teaching and programming experience: mean = 8.14 years, SD = 6.67), each receiving \$20 compensation.
% Teaching background and demographic details are provided in the Appendix.
Additional demographic details are provided in the Appendix.
% The study was approved by the host university's IRB and adhered to established ethical guidelines.
The study was approved by the host university's IRB.

\subsection{Procedure}
% To better understand the challenges instructors face during real classroom programming activities, we conducted one-on-one semi-structured interviews with each participant. These interviews were held online via Zoom and aimed to gather in-depth insights into instructors' authentic needs and feedback during the teaching process.
% The interviews covered a variety of topics, including participants' experiences with classroom programming activities, instructional design goals, student feedback and assessment methods, and the specific difficulties encountered during teaching. In particular, we focused on participants' attitudes toward the use of AI tools in the classroom and their potential concerns. Follow-up questions were asked during the interviews to probe deeper and obtain more detailed responses as needed.
% Each interview lasted approximately 40 to 60 minutes and was recorded in text, audio, and video formats to ensure data completeness and accuracy.

We conducted semi-structured Zoom interviews to understand challenges in classroom programming and to elicit instructors' needs and feedback. 
The protocol covered experiences with in-class programming, instructional design goals, assessment practices, and common difficulties. 
We also explored attitudes toward using AI in class and related concerns, posing follow-up questions as needed to obtain details \RR{(shown in the Appendix~\ref{Instructor})}. 
% Each interview lasted 40–60 minutes and was recorded in text, audio.
Each session lasted 40–60 minutes and was documented through typed notes and audio recordings.

\subsection{Findings}
% The following presents instructors' perspectives on challenges in programming activities and the use of AI tools to support teaching.
The following summarizes instructors' perspectives on programming challenges and on using AI tools to support teaching.

\subsubsection{Challenges in In-Class Programming Activities}

\textbf{C1: Limited Feedback Capacity – Challenges in Providing Timely and Personalized Support.}
% All participants noted that limited class time makes it difficult to provide timely and effective feedback to every student. T1 mentioned that he is usually only able to respond to students who actively ask questions, while those who are less expressive or too shy to speak up are often overlooked—raising concerns about the equity of feedback.
% Moreover, students' willingness to ask questions is often hindered by social pressures, such as the fear of appearing ``not smart,'' leading to a reluctance to speak up. To reduce this psychological burden, T7 introduced an anonymous question submission mechanism, which helped encourage student expression to some extent but still fell short of enabling real-time responses. T5 attempted to dedicate specific Q\&A sessions to address students' questions collectively, but the lack of continuity made it difficult to offer sustained support.
% When students do not receive timely feedback, they are more likely to fall into a cycle of misunderstanding, and instructors typically become aware of the issue only when the problem has already escalated, compromising teaching effectiveness.
Consistent with prior work, participants reported that limited class time hindered timely and personalized feedback.
Peer assessment offers some support but does not meet students' personalized needs~\cite{wang2021puzzleme}.
\textit{
% T1 said he can respond only to students who proactively ask, which leaves quieter or shy students overlooked and raises equity concerns. 
% Participants linked low help-seeking to social pressures such as fear of seeming ``not smart.'' 
T1 noted that only students who proactively seek help receive responses, leaving quieter students overlooked. Participants linked low help-seeking to social pressures, such as fear of seeming ``not smart.''
% T7 tried anonymous question submission, which increased willingness to ask but did not enable real-time responses. 
% T5 used dedicated Q\&A segments to batch questions, but the lack of continuity limited sustained support. 
T7 found that anonymous question submission increased willingness to ask but did not allow real-time responses. T5 used dedicated Q\&A segments to batch questions, but the lack of continuity limited ongoing support.
}
% Without timely feedback, students can enter cycles of misunderstanding, and instructors often notice only after issues escalate, reducing instructional effectiveness.
Without timely feedback, students may fall into cycles of misunderstanding, and instructors often detect issues only after they escalate.

% ,VizProg 在教师带宽不足时提供学生编程进度可视化,以增强教师对课堂的监控,Vizgroup通过通知和警报

\textbf{C2: Instructional Blind Spots – Difficulty Monitoring Student and Class Progress.}
% Consistent with prior work, instructors reported difficulty in monitoring progress and task completion in real-time, which limited immediate pedagogical adjustments~\cite{zhang2023vizprog,tang2024vizgroup}.
Consistent with prior work, instructors reported difficulty monitoring progress and task completion in real time, limiting immediate pedagogical adjustments~\cite{zhang2023vizprog,tang2024vizgroup}.
\textit{
% T2 explained that in large-class settings, instructors and students often become ``disconnected,'' making it difficult to adapt the pace of instruction in real time based on students' actual needs.
T2 noted that large-class settings often create a ``disconnect'' that prevents instructors from adjusting instruction to students' needs.
}
% Participants also expressed concerns about not being able to identify common issues across the class. 
Participants also struggled to identify common classwide issues.
These challenges highlight the need for real-time support that captures individual and class progress and provides efficient, personalized feedback.

\subsubsection{Challenges of Using AI Tools in Real-World Classrooms}

\textbf{C3: Lack of Trust – Concerns About the Accuracy and Appropriateness of AI-Generated Content.}
% All participants had prior experience using AI tools. While they generally acknowledged the potential of AI in providing feedback on programming-related issues—and saw it as a possible aid for students tackling technical problems—they also unanimously expressed concerns about the reliability and educational appropriateness of AI-generated content. 
% On one hand, they were aware of the possibility that AI might produce biased or incorrect information (i.e., ``hallucinations'')~\cite{ji2023survey}. 
All participants had prior experience with AI and acknowledged its potential for programming feedback, but they questioned the reliability and pedagogical appropriateness of AI outputs. 
% They noted risks of biased or incorrect content (``hallucinations'')~\cite{ji2023survey} and the possibility of academic dishonesty or overreliance. 
They cited risks of biased or incorrect content, academic dishonesty, and overreliance~\cite{ji2023survey}.
Emphasizing that the process matters more than the answer \cite{harvey2025don}, \textit{T7 allowed AI use but required students to submit conversation logs to monitor motivation and ability.}
Participants also worried that AI can weaken higher-order thinking. 
% \textit{T2 observed that some students, once accustomed to AI, shift to ``mechanical questioning and mechanical receiving,'' which undermines independent and critical reasoning.}
\textit{T2 noted that some students become accustomed to ``mechanical questioning and mechanical receiving,'' which undermines independent and critical reasoning.}

% On the other hand, and more critically, they were concerned that the use of AI could lead to academic dishonesty or overreliance on students.
% For students, the process of arriving at an answer is often more important than the answer itself \cite{harvey2025don}. T7 noted that although she could not entirely prohibit students from using AI tools, she required them to submit conversation logs with the AI to monitor their usage and assess their learning motivation and actual abilities.
% In addition, all participants pointed out that AI tools may undermine students' higher-order thinking skills—particularly critical thinking. T2 observed that once students become accustomed to relying on AI to complete tasks, they tend to lose the ability to think independently, falling into a pattern of ``mechanical questioning and mechanical receiving,'' which makes it difficult for them to grasp the underlying logic and methodology of the content.

\textbf{C4: Limited Intelligence – Inability of AI to Offer Dynamic and Contextualized Feedback.}
% Participants commonly believed that, compared to human instructors, current AI tools lack flexibility in feedback and the ability to understand classroom context.
Participants noted that current AI tools lack the feedback flexibility and contextual awareness that human instructors provide.
% Human instructors typically integrate multiple factors, such as student background, task progress, and course pacing, when providing adaptive feedback. In contrast, existing AI tools often generate fixed responses based on preset prompts, lacking the capacity for real-time adjustment.
Whereas instructors integrate student background, task progress, and course pacing when adapting support, AI tools often generate fixed responses based on preset prompts and cannot adjust in real time. 
% For example, at the beginning of an activity, when students are just being introduced to new material, instructors tend to offer encouraging and generalized guidance. Later in the activity, they shift to more detailed and directive support to ensure task completion. 
Instructors also vary their guidance throughout an activity, starting with broad encouragement and shifting to detailed, directive support.
% T4 added, \textit{``AI is usually passive, but classroom interactions are active. Instructors observe students' states and intervene proactively, AI still struggles with that.''}
T4 commented, \textit{``AI is usually passive, but classroom interactions are active. Instructors intervene proactively; AI still struggles with that.''}

\textbf{C5: Role Conflict – Risk of Undermining Instructor Authority.}
% When the research team proposed the idea of introducing a TA Agent to provide real-time feedback to students, several participants voiced concerns that it might diminish their authority in the classroom. With the rise of intelligent tools, some instructors were concerned that their role in classroom teaching might become marginalized. 
% When we proposed a TA Agent for real-time feedback, several participants worried it could weaken their classroom authority and marginalize the instructor's role.
When we introduced the idea of a TA Agent for real-time feedback, several participants worried it could undermine their classroom authority and marginalize their role.
% As T5 candidly put it, \textit{``If students can get timely and accurate feedback from AI, then what do they still need us for?''}
As T5 said, \textit{``If students can get timely and accurate feedback from AI, then what do they still need us for?''}
% Moreover, participants noted that because AI interaction is free of social pressures, students might prefer engaging with AI over instructors, which could gradually erode student–instructor communication. 
% They also noted that AI interactions lack social pressure, so students may prefer AI over instructors, which can erode communication.
They also noted that AI lacks social pressure, causing some students to favor AI over instructors.
% T6 remarked, \textit{``After asking the AI, students stop coming to me. No matter how detailed I explain things, they don't listen—they just go straight to the AI.''}
T6 remarked, \textit{``After asking the AI, students stop coming to me. They just go straight to the AI.''}
% Such shifts reduce instructors' control over course pacing and depth and hinder adaptation to class needs.
These shifts reduce instructors' control over pacing and depth and limit their ability to adjust to class needs.

% This shift could reduce instructors' control over the pacing and depth of course content, making it harder to adapt instruction based on the actual needs of the class.

% These challenges suggest that, when introducing AI tools into real-world classrooms, it is essential to safeguard instructors' authority and oversight. AI should serve as a supportive assistant—not a replacement. Moreover, systems should be designed to allow for dynamic behavior adjustment and contextual adaptation in AI responses, in order to provide feedback that is both targeted and educationally meaningful.

% These challenges suggest the classroom AI that preserves instructors' authority and oversight, functions as a supportive assistant rather than a substitute, and enables dynamically adjustable, context-aware responses that deliver targeted, educationally meaningful feedback.

These challenges highlight the need for classroom AI that preserves instructors' authority, supports rather than replaces them, and offers context-aware, adaptive feedback that is targeted and instructionally meaningful.

\subsection{Design Goals}

% Based on the challenges identified by the participants and their expectations for the \textit{ClassAid} system. With insights from related research, we propose the following four design goals.
Drawing on participant feedback and related research, we propose the following four design goals.

\RR{
\textbf{DG1: Provide Real-Time Personalized Feedback to Support Student Learning.}
Instructors struggle to provide timely, individualized feedback during class due to limited time and large class sizes (C1). \textit{ClassAid} should deliver a TA Agent that offers real-time, context-aware feedback tailored to each student's questions, cognitive level, and learning progress. To address concerns about AI quality and pedagogical appropriateness (C3, C4), the system must ensure feedback is trustworthy, pedagogically sound, and aligned with instructional goals.

\textbf{DG2: Enable Real-Time Monitoring of Individual and Class-Wide Student-AI Learning Dynamics.} 
% Instructors lack effective tools to track students' learning progress, task completion, and common difficulties in real time (C2). \textit{ClassAid} should provide instructors with comprehensive visibility into both individual and student-AI instraction states, enabling them to identify learning challenges early, adjust instructional strategies dynamically, and implement targeted interventions based on data-driven insights.
Instructors currently lack effective tools to monitor how students engage with AI during problem solving, including their progress, misconceptions, and reliance patterns (C2). \textit{ClassAid} should offer real-time visibility into both individual student-AI interactions and aggregated class-wide learning dynamics, allowing instructors to detect emerging challenges early, adjust instructional strategies in real-time, and deliver targeted, data-informed interventions.

\textbf{DG3: Empower Instructors with Control and Oversight of AI-Generated Feedback.} Instructors express concerns about losing instructional authority and control when AI agents interact directly with students (C5). \textit{ClassAid} should preserve instructor leadership by providing flexible mechanisms for supervising and adjusting AI behavior in line with pedagogical needs and classroom context. This ensures that AI augments rather than replaces, the instructor's role, maintaining accountability and alignment with learning objectives.

\textbf{DG4: Facilitate Student Engagement and Feedback to Build Trust.}
Students' acceptance and trust in AI-generated feedback directly impact learning outcomes. 
To address concerns about AI intelligence and appropriateness (C3, C4) while strengthening the instructor-student feedback loop (C5), \textit{ClassAid} should enable students to actively evaluate and respond to AI feedback. 
This bidirectional feedback mechanism enhances student agency, provides instructors with signals to assess AI reliability, and fosters a harmonious learning environment.

}
\section{ClassAid Design and Implementation}
\begin{figure*}
    \centering
    \includegraphics[width=1\linewidth]{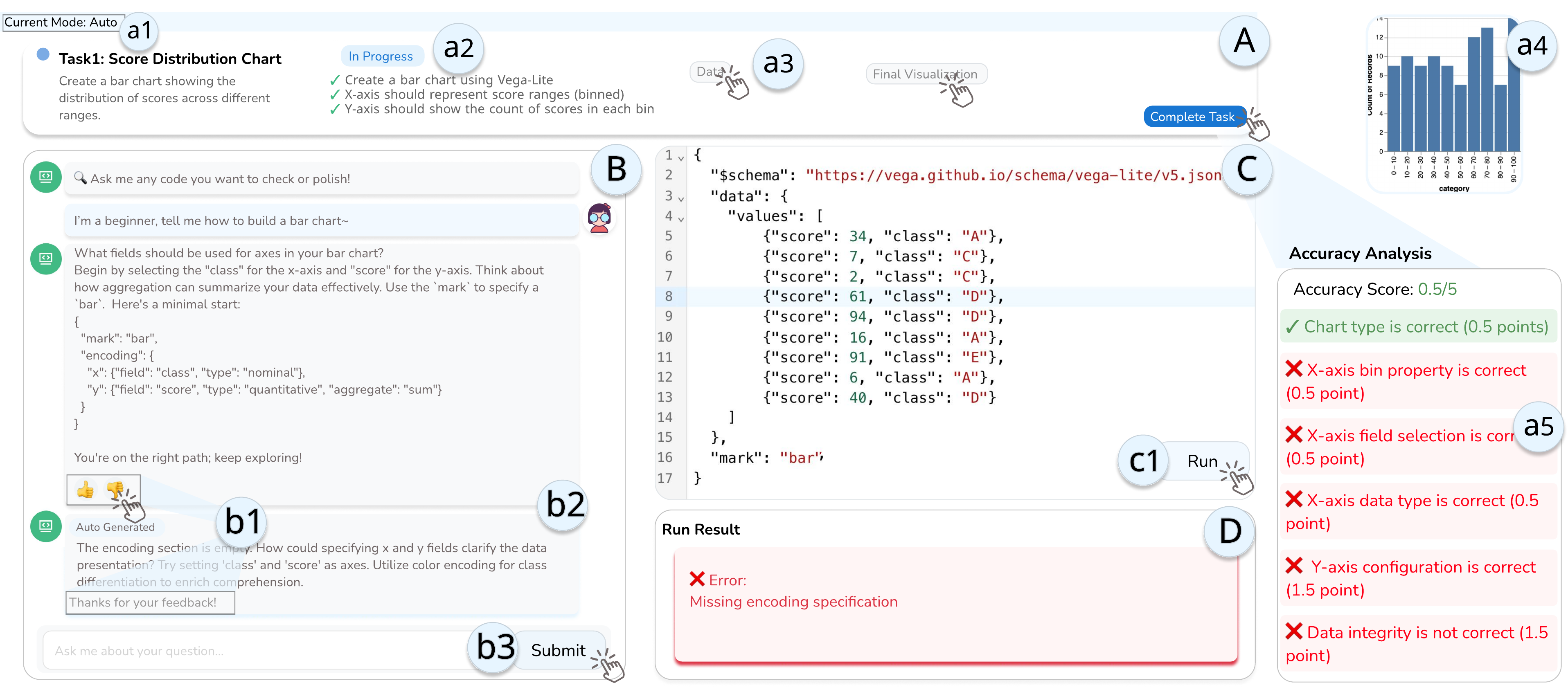}
    \caption{Student interface during in-class programming activities.
(A) The task panel shows the current feedback mode (a1), task description (a2), data (a3), and expected output (a4).
(B) The chat panel supports student questions and TA Agent responses, with options to rate messages and receive proactive feedback (b1).
(C) The code panel allows students to write and run code.
(D) The output panel displays execution results and error messages.
}
\Description{
Student interface during in-class programming activities.
(A) The task panel shows the current feedback mode (a1), task description (a2), data (a3), and expected output (a4).
(B) The chat panel supports student questions and TA Agent responses, with options to rate messages and receive proactive feedback (b1).
(C) The code panel allows students to write and run code.
(D) The output panel displays execution results and error messages.
}
    \label{fig:student}
\end{figure*}

% This section focuses on \textit{ClassAid}, which underwent several updates following a pilot study. Details of these updates are provided in Section~\ref{pilot}.

% \subsection{System Usage Introduction}

We designed \textit{ClassAid}, a real-time orchestration system that integrates a student interface (Fig.~\ref{fig:student}) and an instructor dashboard (Fig.~\ref{fig:teacher}) to support programming instruction in live classroom settings.

\RR{\subsection{Student Interface Design}}
Once the programming activity begins, students access an interactive interface (Fig.~\ref{fig:student}). 
The top of the interface displays the current TA Agent feedback mode (Fig.~\ref{fig:student}-a1), helping students understand how the agent will respond. 
\RR{Before writing any code, students can view the task description in the Task Panel (Fig.~\ref{fig:student}-a2), open datasets through the ``Data'' button (Fig.~\ref{fig:student}-a3), and preview the expected output via the ``Final Visualization'' button (Fig.~\ref{fig:student}-a4).
The interface also provides partial starter specifications that allow students to edit an existing JSON structure (Fig.~\ref{fig:student}-c1), helping them begin the task more effectively.}
Together, these features help students clearly understand the task objectives and available resources.

% During the problem-solving process, if students encounter difficulties, they can type questions to the TA Agent using the Chat Panel (Fig.~\ref{fig:student}-B) (Fig.~\ref{fig:student}-b3). 

During the problem-solving process, if students encounter difficulties, they can ask questions to the TA Agent through the Chat Panel (Fig.~\ref{fig:student}-B, b3).
% 受形成性评估理论和动态评估理论的启发，TA Agent反馈的设计模拟了讲师的自适应策略，如脚手架搭建、反馈调整和逐步消退，使反馈能够更好地与课堂动态相适应。TA Agent根据每个学习者的问题、认知水平和学习轨迹调整反应，缩小反馈差距并支持自主学习以促进高阶思维而不是替代学生推理。具体而言，TA Agent融合了四种响应模式：启发式模式通过开放式提示鼓励反思，技术模式提供直接的任务特定指导，自动模式根据实时上下文在两者之间动态平衡，以及静默模式不提供反应并有意保护学生自主性。
\RR{
% Drawing on formative and dynamic assessment theories~\cite{black2009developing,minick1987implications}, the TA Agent's feedback mechanism integrates instructor adaptive teaching strategies.
% % including scaffolding, feedback adjustment, and fading to achieve deep alignment with classroom dynamics. 
% By analyzing each learner's cognitive level, error type, and learning progression, the TA Agent dynamically adjusts feedback content to effectively close feedback gaps, support autonomous learning, and promote higher-order thinking while preserving student independent reasoning \textbf{(DG1)}. 
Drawing on formative and dynamic assessment theories~\cite{black2009developing,minick1987implications}, the TA Agent incorporates instructors' adaptive teaching strategies. 
By analyzing each learner's cognitive level, error type, and learning progression, it dynamically adjusts feedback to close feedback gaps, support autonomous learning, and promote higher-order thinking while preserving independent reasoning \textbf{(DG1)}.
% The Agent implements four response modes: Heuristic Mode stimulates reflection through open-ended prompts, Technical Mode delivers precise task-specific guidance, Auto Mode intelligently balances both approaches based on real-time context, and Silent Mode intentionally preserves space to maintain student autonomy (details in Section~\ref{stage4}).
The Agent provides four response modes: Heuristic Mode encourages reflection through open-ended prompts; Technical Mode provides concrete code solutions guidance; Auto Mode balances both approaches based on real-time context; and Silent Mode intentionally withholds responses to maintain student autonomy (details in Section~\ref{stage4}). The TA Agent responds according to the active mode.
}
% The TA Agent will respond according to the current feedback mode.
\RR{
% \textit{ClassAid} enables students to actively evaluate TA Agent feedback and provide their own responses, strengthening the student-instructor feedback loop on AI-generated content.
% Specifically, Students can choose to ``like'' or ``dislike'' the feedback (Fig.~\ref{fig:student}-b1), which helps monitor the TA Agent's performance \textbf{(DG4)}. }
% This feedback is optional. If a student provides a rating, a ``Thanks for your feedback!'' message will pop up to confirm the submission.
\textit{ClassAid} also allows students to evaluate TA Agent feedback, strengthening the student–instructor feedback loop on AI-generated content. Students can ``like'' or ``dislike'' the response (Fig.~\ref{fig:student}-b1), helping instructors monitor the TA Agent's performance \textbf{(DG4)}. }
Providing feedback is optional, if a student submits a rating, a confirmation message (``Thanks for your feedback!'') appears.

% Code editing takes place in the Code Panel (Fig.~\ref{fig:student}-C), where students write their solutions and execute them using the ``Run'' button (Fig.~\ref{fig:student}-c1). 
% The execution results are presented in the Output Panel (Fig.\ref{fig:student}-D).

Code editing occurs in the Code Panel (Fig.~\ref{fig:student}-C), where students write their solutions and execute them using the ``Run'' button (Fig.~\ref{fig:student}-c1). The execution results are shown in the Output Panel (Fig.~\ref{fig:student}-D).
% If an error occurs, the system provides a specific message; if the code executes successfully, the corresponding result is displayed. 
% Unlike typical console reports that return only generic information, our system classifies errors into five categories: schema, data, mark, encoding, and JSON syntax, and it provides targeted recommendations accordingly. 
If an error occurs, the system returns a specific message; if the code runs successfully, the result appears.
Unlike typical console reports that provide only generic feedback, our system classifies errors into five categories (schema, data, mark, encoding, and JSON syntax) and offers targeted recommendations.
% For example, the code in Fig.\ref{fig:student}-B would yield only ``\texttt{Error: Invalid Vega-Lite specification}'' in a standard console, which provides limited guidance for beginners. 
% In our interface (Fig.~\ref{fig:student}-D), the system further reports ``\texttt{Error: Missing encoding specification},'' explicitly identifying the missing encoding declaration and helping students correct the issue efficiently.
For example, the code in Fig.~\ref{fig:student}-B would yield only ``\texttt{Error: Invalid Vega-Lite specification}'' in a standard console, which provides limited guidance for beginners. 
In our interface (Fig.~\ref{fig:student}-D), the system instead reports ``\texttt{Error: Missing encoding specification},'' explicitly identifying the missing encoding declaration and helping students fix the issue more efficiently.

% The results appear in the Output Panel (Fig.~\ref{fig:student}-D) below: if there are errors, specific error messages are shown; if the code runs correctly, the corresponding result is displayed. 
Except in Silent mode, the TA Agent can also generate proactive feedback.
For example, if a student submits incorrect code multiple times within a short period, the system automatically provides suggestions (Fig.~\ref{fig:student}-b2). 
This proactive feedback aligns with the active mode: in Heuristic mode it is heuristic, in Technical mode it is technical, and in Auto mode the system selects the appropriate type based on context. 
These messages are labeled as ``Auto Generated'' and visually distinguished with a blue background, helping students recognize that the feedback is system-initiated.

% During the activity, instructors can dynamically adjust the TA Agent's feedback mode. The mode indicator (Fig.~\ref{fig:student}-a1) on the student interface will update in real time to reflect these changes, enabling immediate pedagogical adjustments. After completing a task, students can click the ``Complete Task'' button to move on to the next phase (Fig.~\ref{fig:student}-a5). 
% Upon task completion, the system automatically archives the current conversation and code and presents a task summary that includes the core concepts and the detailed score (see Fig.~\ref{fig:student}-a5), helping students deepen understanding and consolidate mastery. 
% The system then resets and refreshes all interface panels to provide a clean workspace for the next task, maintaining a structured and coherent learning process.

Instructors can adjust the TA Agent's feedback mode in real time, and the mode indicator on the student interface (Fig.~\ref{fig:student}-a1) updates immediately. 
After finishing a task, students click ``Complete Task'' (Fig.~\ref{fig:student}-a5) to move to the next phase. The system then archives the current conversation and code, generates a task summary with core concepts and scoring, and refreshes the interface to create a clean workspace for the next task.

% the system automatically archives the current conversation and code, and refreshes all interface panels to provide a clean workspace for the next task—ensuring a structured and coherent learning experience.

\begin{figure*}[t]
    \centering
    \includegraphics[width=1\linewidth]{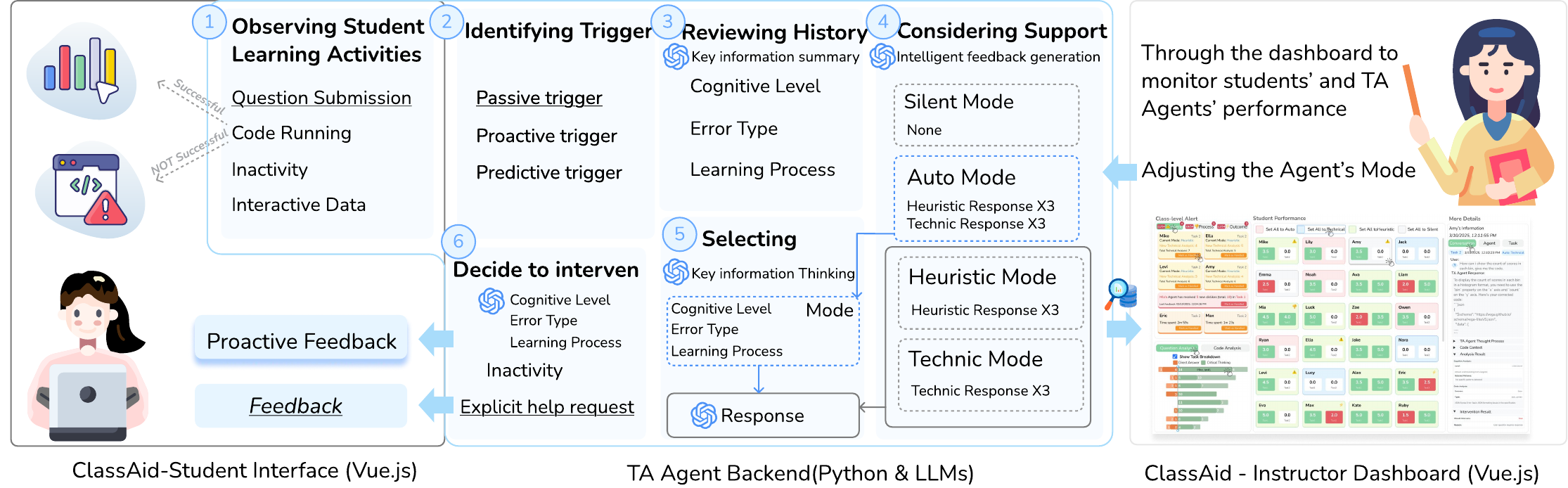}
    \caption{
Overview of the TA Agent's six-stage orchestration pipeline for student learning support within the \textit{ClassAid} system.
Student activity data, such as question submissions, code execution, and interaction traces, are first collected through the \textit{ClassAid} Student Interface and passed to the TA Agent backend. The agent then enters a six-stage pipeline that observes, analyzes, and responds to student learning behaviors.
Meanwhile, instructors monitor student progress and the TA Agent's performance through the \textit{ClassAid} Instructor Dashboard and can adjust feedback modes in real time to align with pedagogical goals.}
\Description{
Overview of the TA Agent's six-stage orchestration pipeline for student learning support within the \textit{ClassAid} system.
Student activity data, such as question submissions, code execution, and interaction traces, are first collected through the \textit{ClassAid} Student Interface and passed to the TA Agent backend. The agent then enters a six-stage pipeline that observes, analyzes, and responds to student learning behaviors.
Meanwhile, instructors monitor student progress and the TA Agent's performance through the \textit{ClassAid} Instructor Dashboard and can adjust feedback modes in real time to align with pedagogical goals.
}
    \label{fig:s-agent}
\end{figure*}

% AI 触发与反馈
% 初始系统将不活动触发阈值设为 30 秒。但当参与者查阅 Vega-Lite 文档时，往往会超过该时长，导致频繁且打断工作的 AI 提示。
% 我们将不活动检测阈值延长至 4 分钟，并引入时间窗口限制：在 5 分钟内最多触发两次与暂停相关的触发器。此外，我们为非被动触发增加了“冷却期”。若在冷却期内再次触发，系统将返回空触发列表且不继续后续分析，从而尽量减少对学生的无谓打断并降低提醒频率。
% 增加学习理论
% 不同出发的例子response..//有一个不同模式下回答的表格
% 修改图片~~~~!!!

\subsection{TA Agent Framework}

% To build an intelligent teaching assistant (TA Agent) capable of providing students with real-time, personalized feedback (DG1, DG2), we designed and implemented a six-stage framework grounded in theories of formative and dynamic assessment~\cite{black2009developing,minick1987implications}.
% under the guidance of an educational technology expert and powered by LLMs. 

To build an intelligent TA Agent capable of providing students with real-time, personalized feedback (DG1, DG2), we designed and implemented a six-stage framework grounded in formative and dynamic assessment theories~\cite{black2009developing,minick1987implications}.
% This framework was developed under the guidance of an educational technology expert with over a decade of experience in learning analytics and powered by LLMs, aiming to emulate the diagnostic reasoning loop of human instructors in classroom settings (show in Table~\ref{tab:stage}). 
The framework was developed with guidance from an educational technology expert with over a decade of experience in learning analytics and is powered by LLMs to emulate the diagnostic reasoning loop of human instructors in classroom settings (shown in Table~\ref{tab:stage}).
% Specifically, formative assessment informed the evidence collection and feedback–adjustment cycle, while dynamic assessment inspired the use of adaptive triggers and scaffolding interventions to model learners' zones of proximal development (ZPD). 
The expert also contributed to defining and validating the weighting of key indicators to ensure their pedagogical soundness. 
% By structuring the TA Agent around these six stages, we extend traditional teacher practices into a scalable, human–AI collaborative model capable of supporting fine-grained, real-time decision making.
The full prompt templates are provided in the Appendix.

\renewcommand{\arraystretch}{1.25}
\begin{table*}[ht]
\centering
\small
\caption{Alignment between the Six-Stage TA Agent Framework and Formative / Dynamic Assessment Theories.}
\begin{tabular}{p{0.18\linewidth} p{0.28\linewidth} p{0.48\linewidth}}
\toprule
\textbf{Stage} & \textbf{Theoretical Alignment} & \textbf{Implementation} \\
\midrule

Observing Student Learning Activities & 
Formative Assessment – Evidence Collection &
\begin{itemize}[leftmargin=*]
  \item Monitor editing, running, questioning, and inactivity
  \item Classify questions with Bloom's taxonomy
  \item Analyze syntax/semantic errors
  \item Inactivity timer flags stagnation
\end{itemize} \\

Identifying Learning Obstacles & 
Dynamic Assessment – Exploring ZPD Boundaries &
\begin{itemize}[leftmargin=*]
  \item Passive triggers: student requests
  \item Proactive triggers: inactivity or extended editing
  \item Predictive triggers: historical patterns, repeated errors
  \item Simulates teacher probing of ZPD
\end{itemize} \\

Reviewing and Assessing Student History & 
Formative Assessment – Accumulating Evidence \newline Dynamic Assessment – Simulating ZPD &
\begin{itemize}[leftmargin=*]
  \item Aggregate logs of past interactions
  \item Track Bloom's taxonomy level shifts (stagnation/regression)
  \item Compile error statistics and distributions
  \item Distinguish mastered vs. problematic concepts
  \item Model ZPD dynamically using historical patterns
\end{itemize} \\

Considering Appropriate Forms of Learning Support & 
Formative Assessment – Feedback and Adjustment &
\begin{itemize}[leftmargin=*]
  \item Generate feedback based on cognitive level, error type, and knowledge gaps
  \item Heuristic feedback: open-ended question, prompts, optional code, supportive tone
  \item Technical feedback: brief explanation, code fix (3–5 lines), respectful tone
  \item Four instructor modes: Auto, Technical, Heuristic, Silent
\end{itemize} \\

Selecting Adaptive Feedback Modes & 
Formative Assessment – Quantifying Diagnostic Judgments &
\begin{itemize}[leftmargin=*]
  \item Auto mode: applies cognitive psychology and instructional strategy framework
  \item Weighted scheme: cognitive level (50\%), error type (20\%), learning history (30\%)
  \item Score candidate feedback: relevance (40\%), complexity (20\%), consistency (20\%), clarity (15\%), urgency (5\%)
  \item Select the highest-quality response
\end{itemize} \\

Intervening to Support Learning Progress & 
Formative Assessment – Feedback Implementation \newline Dynamic Assessment – Scaffolding and Extending Potential &
\begin{itemize}[leftmargin=*]
  \item If inactivity/help request: immediate intervention
  \item Otherwise compute intervention score (error severity 40\%, cognitive level 30\%, history 30\%)
  \item Score > 0.5 $\rightarrow$ proactive intervention
  \item Score $\leq$ 0.5 $\rightarrow$ support autonomous exploration
  \item Scaffolding extends learning potential
\end{itemize} \\

\bottomrule
\end{tabular}
\label{tab:stage}
\end{table*}

\subsubsection{Stage 1: Observing Student Learning Activities}

Once a student begins a task, the TA Agent enters an observation stage and continuously tracks behaviors such as editing inactivity, question submissions, code modifications, and execution events to capture real-time learning dynamics.
An inactivity timer with a 240-second threshold flags periods without keyboard, mouse, or click interactions as inactivity.
This value was empirically chosen to reduce false alarms when students pause to read or consult materials, helping identify learning bottlenecks or motivational issues.
% helping identify learning bottlenecks or motivational issues.
% 一个设定为 30/4分钟 秒阈值的不活跃计时器，会将无键盘、鼠标或点击交互的时段标记为“不活跃”，以帮助识别学习瓶颈或动机问题。
When a question is submitted, the TA Agent classifies it using Bloom's taxonomy (Remember, Understand, Apply, Analyze, Evaluate, Create)~\cite{forehand2010bloom} and analyzes any accompanying code for syntax or semantic errors, labeling specific error types to clarify student difficulties.
% Code changes are tracked incrementally, and every time the student clicks ``Run,'' the TA Agent performs a multi-level check: verifying JSON format and required fields, conducting runtime analysis, and providing detailed error feedback or visual output, depending on the result. This integrated monitoring supports precise assessment and personalized feedback.
Code changes are recorded incrementally. Each time the student clicks ``Run,'' the TA Agent performs a multi-step check that verifies JSON format and required fields, conducts runtime analysis, and returns either error details or the visual output. This integrated monitoring enables precise assessment and personalized feedback.

\subsubsection{Stage 2: Identifying Learning Obstacles}
\label{trigger}

Building on Stage 1 behavioral monitoring, the TA Agent in Stage 2 identifies learning obstacles through three types of dynamically triggered interventions. 
% passive, proactive, and predictive.
\textit{Passive triggers} respond to explicit student actions, such as question submissions or code execution failures, and are treated as high-priority help requests. 
\textit{Proactive triggers} arise from the Agent's own judgment. For example, prolonged inactivity or extended editing without queries require autonomous intervention. 
\textit{Predictive triggers} rely on historical patterns, such as repeated errors or shifts in cognitive level (e.g., a two-level shift across five interactions), signaling potential comprehension gaps.
When multiple triggers occur, the TA Agent prioritizes them in the order of passive, proactive, and predictive, ensuring prompt responses to direct help-seeking while still addressing subtler issues.
%此外,为了减少对学生的无谓打断并降低提醒频率。
%我们在不活动检测的基础上，引入时间窗口限制.
% 在 5 分钟内最多触发两次与暂停相关的触发器。此外，我们为非被动触发增加了“冷却期”。若在冷却期内再次触发，系统将返回空触发列表且不继续后续分析，从而尽量减少对学生的无谓打断并降低提醒频率。
% To minimize unnecessary interruptions and reduce reminder frequency, we introduced a time window and a cooling mechanism based on inactivity detection. Within any five-minute period, no more than two pause-related triggers are allowed. For non-passive triggers, a two-minute cooling period is applied. If triggered again within this period, the system returns an empty trigger list and halts further analysis, thereby reducing redundant interventions.
To minimize unnecessary interruptions, we added a time window and cooling mechanism based on inactivity detection. Within any five-minute period, no more than two pause-related triggers are permitted, and non-passive triggers are subject to a two-minute cooling period. If triggered again within this interval, the system returns an empty trigger list and halts further analysis, reducing redundant interventions.

\subsubsection{Stage 3: Reviewing and Assessing Student History}

Once activated by a trigger, the TA Agent enters the review stage and analyzes the student's historical interactions to support targeted feedback.
It automatically retrieves and organizes behavioral data and Student-AI interaction logs, extracting and summarizing key information across multiple dimensions.
The Agent monitors the student's cognitive level using Bloom's taxonomy to detect stagnation or regression~\cite{forehand2010bloom}. It also compiles statistics on error types, frequencies, and distributions to identify recurring patterns and infer potential conceptual misunderstandings.
To evaluate task performance, the Agent considers both progress and code execution success to estimate the student's learning stage and mastery level. From a conceptual perspective, it distinguishes between mastered and problematic concepts, highlighting areas that require further study.

% This multi-dimensional analysis, which inclused cognition, behavior, and knowledge, enables the TA Agent to construct a comprehensive learning profile. It provides instructors with reliable diagnostic insights and equips the system to deliver personalized feedback and intelligent interventions.
% , completing the loop from observation to interven.

\subsubsection{Stage 4: Considering Appropriate Forms of Learning Support}
\label{stage4}

\begin{table*}[ht]
\centering
\caption[TA agent support modes]{\RR{TA agent support modes derived from formative classroom observations, aligned with theoretical foundations and design implementations.}}
\label{tab:ta-agent-modes}
\begin{tabular}{
  >{\color{PurpleColor}}p{0.14\textwidth}
  >{\color{PurpleColor}}p{0.16\textwidth}
  >{\color{PurpleColor}}p{0.28\textwidth}
  >{\color{PurpleColor}}p{0.28\textwidth}
}
\hline
\textbf{Mode Type} & \textbf{Theoretical Alignment} & \textbf{Formative Findings (C4)} & \textbf{Design Implementation} \\
\hline
Heuristic Mode 
& Socratic Method~\cite{benson2011socratic}; constructivist Learning theory~\cite{hein1991constructivist}. 
& At the beginning of the task, instructors often use open ended questions and \textit{Socratic dialogue} to stimulate students' thinking while deliberately avoiding giving direct answers. 
& The agent uses open ended prompts to trigger reflection, offering directions for thinking instead of direct solutions and guiding students to reason and revise on their own. \\
\hline
Technical Mode 
& Direct Instruction~\cite{carnine1997direct}. 
& When students encounter specific technical obstacles such as syntax errors or logic errors, instructors shift to precise and directive support to ensure that the task can be completed. 
& The agent provides focused and task specific guidance, including procedural steps, code examples and error corrections, which helps students debug quickly and move the task forward. \\
\hline
Auto Mode 
& Formative Assessment Theory~\cite{black2009developing}; Dynamic Assessment~\cite{minick1987implications}. 
& Instructors continuously assess students' states and flexibly adjust support strategies based on real time context such as learning stage, type of difficulty and current performance, switching between different levels and types of help. 
& The agent intelligently selects between heuristic and technical modes based on context, taking into account learning stage, problem type and students' past performance, in order to dynamically adjust the form and intensity of support. \\
\hline
Silent Mode 
& Fading of Scaffolding~\cite{noroozi2018promoting}; Metacognition and Self regulated Learning Theory~\cite{sperling2004metacognition}. 
& Classroom observations indicate that frequent immediate help can foster AI dependence and reduce students' independent attempts, so instructors sometimes delay or withhold responses to prompt autonomous exploration.
& For a period of time, the agent intentionally does not respond. This encourages students to search, debug and reflect on their own first, avoiding over reliance on AI. \\
\hline
\end{tabular}
\end{table*}

\RR{
Based on our formative study and related educational theories, we found that instructors consider students' cognitive level, error type, and learning progress when deciding how to provide support. 
In practice, we identified three recurring support patterns. 
At the beginning of the class activity, instructors often used open-ended and encouraging prompts that guided students to construct understanding independently, consistent with constructivist teaching and aligns with heuristic feedback~\cite{hein1991constructivist}. 
As the activity progressed and students encountered concrete syntax or logic problems, instructors shifted to more detailed and directive guidance to help them resolve problems efficiently, corresponding to technical feedback~\cite{carnine1997direct}. 
When instructors noticed signs of overreliance on AI support, they deliberately delayed or temporarily withheld direct answers to preserve space for autonomous exploration and productive struggle~\cite{noroozi2018promoting,sperling2004metacognition}. 
Building on these findings and on formative and dynamic assessment theories~\cite{black2009developing,minick1987implications}, we modeled instructional support as four feedback modes that integrate both feedback type and triggering mechanism: Auto, Heuristic, Technical, and Silent feedback (details in Table~\ref{tab:ta-agent-modes}).

}

In this stage, the TA Agent generates targeted feedback based on the student's cognitive level, error types, and knowledge gaps. 
To ensure pedagogically grounded and context-sensitive responses, we developed a structured prompt-based framework that considers two key dimensions: \textit{feedback type} (heuristic vs. technical) and \textit{triggering mechanism} (proactive vs. user-triggered).
Examples are shown in Table~\ref{tab:example}.

\begin{enumerate}
\item \textbf{Context-setting Prompt:} 
Defines the TA's instructional persona and pedagogical goals. For example:
\begin{quote}
You are a helpful and encouraging teaching assistant for a Vega-Lite data visualization course. Depending on the situation, you may proactively highlight issues or respond directly to students' questions.
\end{quote}

\item \textbf{Heuristic Feedback Prompts:} 
Encourage critical thinking and reflection. 
\begin{itemize}
    \item \textit{Proactive:} Short observations and a focused guiding question (under 50 words).
    \item \textit{User-triggered:} One open-ended question, 2--3 concise thinking prompts, an optional code snippet, and a supportive tone (under 100 words).
\end{itemize}

\item \textbf{Technical Feedback Prompts:} 
Provide immediate, task-specific support. 
\begin{itemize}
    \item \textit{Proactive:} A brief explanation and one direct code suggestion (3--5 lines).
    \item \textit{User-triggered:} A detailed explanation of the issue, a complete code correction, and reasoning for why it works.
\end{itemize}

\item \textbf{Response Constraints:} 
Each feedback instance produces three concise responses, written in clear, respectful, and stylistically consistent language.

\item \textbf{Feedback Mode Configuration:} 
The generation strategy is governed by four instructor-controlled modes:
\textit{Auto Mode:} Both heuristic and technical responses (6) are generated;
\textit{Technical Mode:} Only technical responses (3) are generated;
\textit{Heuristic Mode:} Only heuristic responses (3) are generated;
\textit{Silent Mode:} No feedback is provided.

\end{enumerate}

\begin{table*}[ht]
\centering
\caption{Examples of Technical and Heuristic Responses under Proactive vs. Normal Templates.}
\begin{tabular}{p{0.18\linewidth} p{0.38\linewidth} p{0.38\linewidth}}
\toprule
\textbf{Response Type} & \textbf{Proactive Template (auto-triggered)} & \textbf{Normal Template (user-triggered)} \\
\midrule
\textbf{Technical} &
I noticed your chart does not have a Y encoding, which prevents Vega-Lite from drawing the bars. You can try adding an aggregate count on the Y-axis: \newline
\quad \{ \newline
\quad\quad ``y'': \{``aggregate'': ``count'', ``type'': ``quantitative''\} \newline
\quad\} \newline
This small change will allow the bars to show the number of records in your dataset. &
The issue happens because Vega-Lite does not know what to plot on the Y-axis. To make the bar chart display properly, you should specify an aggregate count for the Y encoding. For example: \newline
\quad \{ \newline
\quad\quad ``x'': \{``field'': ``category'', ``type'': ``nominal''\}, \newline
\quad\quad ``y'': \{``aggregate'': ``count'', ``type'': ``quantitative''\} \newline
\quad\} \newline
Here, ``x'' shows categories, and ``y'' counts how many records fall into each. This way, the height of each bar corresponds to the frequency of that category. \\
\midrule
\textbf{Heuristic} &
I noticed the Y-axis is not defined. What would happen if you tried adding a count aggregation for Y? Could that make the bars appear as expected? &
What do you expect the Y-axis to represent in your bar chart? Should it show raw counts, averages, or something else? If you want counts, you might consider using an aggregate function. Which option best matches the story you want your chart to tell? You are making good progress. \\
\bottomrule
\end{tabular}
\label{tab:example}
\end{table*}
% \begin{figure}
%     \centering
%     \includegraphics[width=1\linewidth]{figs/table2.pdf}
%     % \caption{Enter Caption}
%     \label{tab:example}
% \end{figure}

% To enhance instructional flexibility, the TA Agent adapts its feedback strategy based on the feedback mode selected by the instructor. In Auto Mode, it generates both heuristic and technical feedback (three responses of each type). In Technical Mode, it provides only three technical responses, while Heuristic Mode outputs only three heuristic responses. In Silent Mode, the Agent withholds all feedback, allowing students to work independently. This instructor-controlled configuration enables responsive, goal-aligned support, bridging the gap between automated analysis and adaptive instructional strategies.

\subsubsection{Stage 5: Selecting Adaptive Feedback Modes}

In Stage 5, the TA Agent selects the most appropriate feedback from the candidate responses based on the student's current learning state. 
In Auto Mode, it assumes the role of a senior educational expert and uses a decision framework grounded in cognitive psychology~\cite{solso2005cognitive} and instructional strategy~\cite{moore2014effective} to determine whether heuristic or technical feedback is more suitable.
% \RR{We instantiate this decision rule using a simple heuristic weighting scheme, informed by our formative study and discussions with course instructors.}
% This decision is informed by a weighted scheme considering three factors: current cognitive level (50\%), error types (20\%), and learning history (30\%) .
\RR{We operationalize this decision rule using a heuristic weighting scheme derived from our formative study and discussions with course instructors. 
The scheme considers three factors: current cognitive level, error types, and learning history, with default weights of 50\%, 20\%, and 30\%, respectively, which can be adjusted for specific course contexts.}
% reflecting pedagogical considerations discussed with the expert that prioritize cognitive level as the most direct determinant of feedback needs, with learning history and error types providing complementary signals.
For example, if the student is operating at the Apply level or above, shows mainly design or logic errors, and is making steady progress, the Agent favors heuristic feedbac. 
In contrast, lower cognitive levels, frequent syntax errors, and fluctuating performance lead to technical feedback.
% After determining the mode, the TA Agent scores the three candidate responses across five dimensions: relevance (40\%), complexity (20\%), consistency with prior behavior (20\%), clarity (15\%), and urgency (5\%). It then selects the highest-quality response. This ensures feedback aligns with both instructional intent and the student's individual needs.

\RR{After determining the mode, the TA Agent evaluates the three candidate responses along five dimensions. 
These dimensions draw from prior work on effective feedback and instructional design: relevance and clarity reflect established principles of high-quality formative feedback~\cite{hattie2007power}; complexity relates to cognitive load~\cite{sweller1988cognitive}; consistency with prior behavior supports self-regulated learning~\cite{vanlehn2011relative}; and urgency highlights the importance of timely intervention~\cite{shute2008focus}. The default weighting configuration (40\%, 20\%, 20\%, 15\%, and 5\%) is informed by our formative study and can be adapted to instructional priorities.}

\subsubsection{Stage 6: Intervening to Support Learning Progress}

% In this Stage, the TA Agent completes the intelligent selection of both the feedback mode and content, and proceeds to the decision implementation phase—determining whether to formally intervene in the student's current learning process to provide effective support and foster improvements in learning strategies and metacognitive skills.

% The TA Agent first checks for the presence of special conditions, such as detecting that the student is inactive or has submitted a direct request for help. In these situations, the TA Agent determines that immediate intervention is necessary and delivers the selected feedback directly to the student, either in response to an explicit need or to break potential learning stagnation.

% If no explicit signals are present, the TA Agent activates a ``motivation-to-intervene'' evaluation mechanism to make a more cautious judgment about whether to intervene proactively. This mechanism takes into account the severity of the student's current errors (40\%), cognitive level (30\%), and historical learning performance (30\%). Each factor is quantified with a score between 0 and 1. If the total score exceeds 0.5, the TA Agent concludes that the student's current state may hinder task progress or conceptual understanding and proceeds with proactive intervention. If the score falls below the threshold, the TA Agent opts not to intervene, respecting the student's pace of autonomous exploration.

At this stage, the TA Agent finalizes the selection of feedback mode and content and then enters the decision implementation phase. 
During this phase, the system must determine whether to intervene in the student's learning process to provide timely support and promote metacognitive development.
The Agent first checks for special conditions, such as signs of inactivity or explicit help requests. 
If any of these conditions are detected, the system intervenes immediately to address learning stagnation or respond to the student's needs.

% If no such signals are present, the Agent activates a motivation-to-intervene mechanism, which evaluates error severity (40\%), cognitive level (30\%), and historical performance (30\%). Each factor is scored from 0 to 1, and a combined score above 0.5 triggers proactive intervention. Otherwise, the Agent refrains from intervening, supporting autonomous exploration.

\RR{
If none of these signals are present, the Agent activates a motivation-based intervention mechanism.
% This mechanism conducts a composite assessment based on three factors: cognitive level, error severity, and historical performance. 
% These factors mirror those used in selecting the feedback mode. However, instructor interviews indicate that their relative importance differs across the two stages. When determining the feedback mode, cognitive level is typically the dominant consideration. 
This mechanism conducts a composite assessment based on three factors: cognitive level, error type, and learning history. 
These factors mirror those used to select the feedback mode, although with different weightings.
Instructor interviews show that cognitive level carries the greatest weight when determining the feedback mode, while error type and learning history become more influential during assessment because reveal persistent misconceptions and recurring difficulties.

In contrast, for intervention decisions, instructors emphasized that error type is especially important. 
If a student is making only simple or low-level mistakes, immediate intervention is neither necessary nor desirable, as allowing more time for independent exploration may better support productive struggle and self-discovery.
The weights assigned to the three factors are informed by empirical observations from our formative study and can be adjusted to fit specific course needs. 
The default weighting configuration is 40\%, 30\% and 30\%. Each factor is normalized to a 0 to 1 range, and a weighted composite score is computed. 
When the score exceeds the intervention threshold (default = 0.5, configurable), the Agent initiates a proactive intervention; otherwise, it refrains from intervening to preserve autonomous exploration.

}

% \begin{figure}
%     \centering
%     \includegraphics[width=1\linewidth]{figs/1-s-t.pdf}
%     \caption{Enter Caption}
%     \label{fig:enter-label}
% \end{figure}

\begin{figure*}
    \centering
    \includegraphics[width=1\linewidth]{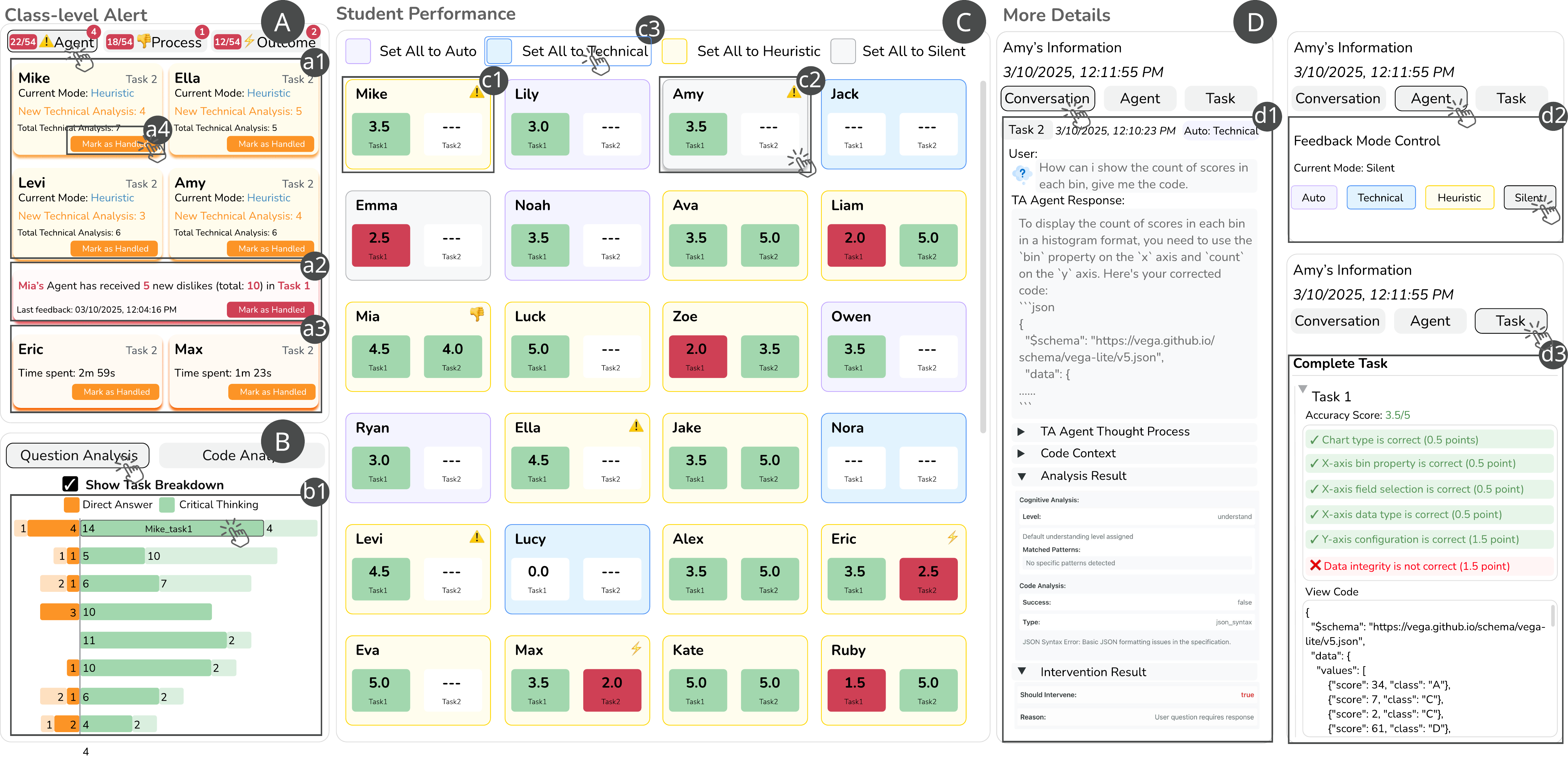}
    \caption{Instructor dashboard for real-time classroom orchestration.
(A) Class-Level Alerts highlight potential learning risks through Agent, Process, and Outcome alerts.
(B) Class-Level Analysis aggregates question (b1) and code (b2) issues to reveal class-wide bottlenecks.
(C) Student Performance Cards display each student's task score and current feedback mode, with global controls for mode switching (c3).
(D) More Details Panel provides drill-down views of individual students, including agent interactions (d1), mode control (d2), and task-level analysis (d3).
Together, these components enable timely intervention and data-informed teaching decisions.}
\Description{
Instructor dashboard for real-time classroom orchestration.
(A) Class-Level Alerts highlight potential learning risks through Agent, Process, and Outcome alerts.
(B) Class-Level Analysis aggregates question (b1) and code (b2) issues to reveal class-wide bottlenecks.
(C) Student Performance Cards display each student's task score and current feedback mode, with global controls for mode switching (c3).
(D) More Details Panel provides drill-down views of individual students, including agent interactions (d1), mode control (d2), and task-level analysis (d3).
Together, these components enable timely intervention and data-informed teaching decisions.
}
    \label{fig:teacher}
\end{figure*}

\subsection{Instructor Dashboard Design}
\textit{Real-Time Student Overview via Dynamic Performance Cards (DG2).}
When a student joins the interactive dashboard shown in Fig.~\ref{fig:teacher}, a corresponding student card (Fig.~\ref{fig:teacher}-C) is generated in the instructor's Student Performance panel. Each card displays the student's name and their current task state. 
\RR{If the student has not completed any tasks, the system shows ``---'' and the task background remains white.}
Once a task is completed, a score out of five appears, scores below three are shown in red and scores from three to five in green, providing a intuitive indication of student performance.
The card background color also reflects the TA Agent's current feedback mode:
\RR{purple for Auto Mode, blue for Technical Mode, yellow for Heuristic Mode, and gray for Silent Mode.}
This design helps instructors quickly understand both student progress and the support they are receiving.
Instructors can also manage AI behavior at the class level (Fig.~\ref{fig:teacher}-c3). 
By clicking the ``Set Mode to All Students'' button, they can apply the selected mode to the entire class in one action, improving operational efficiency and enabling timely adjustments.
% As shown in Fig.~\ref{fig:teaser}, at the beginning of the class, the instructor sets the AI mode for the entire class to Heuristic mode, providing high-level hints and encouraging independent thinking. 
% About ten minutes later, the instructor notices that some students are showing anomalous behavior and adjusts the AI mode for those students individually. 
% In the second half of the class, the instructor changes the AI mode for the entire class to Auto mode so that the system can provide adaptive support based on students' real-time performance.
As shown in Fig.~\ref{fig:teaser}, the instructor sets the class-wide AI mode to Heuristic Mode at the beginning of the activity to encourage independent thinking. About ten minutes later, the instructor notices unusual behavior in several students and adjusts their modes individually. In the second half of the class, the instructor switches the entire class to Auto Mode so the system can deliver adaptive support based on real-time performance.

\textit{Live Alerts to Surface Critical Teaching Moments (DG3, DG4).}
\label{alert}
As the activity progresses, the system continuously updates two panels, Class-Level Alerts (Fig.~\ref{fig:teacher}-A) and Class-Level Analysis (Fig.~\ref{fig:teacher}-B). These panels help instructors identify irregular student behaviors and potential anomalies in TA Agent feedback. The alert mechanism supports timely instructional intervention through three types of alerts.
Agent Alert (Fig.~\ref{fig:teacher}-a1) appears when the TA Agent in Auto Mode produces technical feedback three consecutive times, suggesting potential overreliance on direct answers.
Process Alert (Fig.~\ref{fig:teacher}-a2) notifies instructors when a student gives three ``dislikes'' on feedback within a single task, signaling potential mismatches between the AI's support and the student's needs; 
Outcome Alert (Fig.~\ref{fig:teacher}-a3) is shown when a student completes a task in under three minutes, raising concerns about shallow engagement. This threshold was determined in consultation with instructors during the formative study and can be adjusted to meet specific instructional needs.
All alert information is reflected on the corresponding student card (Fig.~\ref{fig:teacher}-c1), enabling instructors to detect issues quickly. Instructors can toggle across alert tabs to view class-wide summaries. 
Each tab shows how many students have ever triggered that alert type (e.g., 22 out of 54 in Agent Alert), and a red badge indicates the number of unresolved alerts requiring attention. Once an alert is addressed, instructors can click ``Mark as Handled'' (Fig.~\ref{fig:teacher}-a4) to remove it from view, ensuring only unresolved alerts remain visible.

% All alert information is synchronized on the corresponding student card (Fig.~\ref{fig:teacher}-c1), ensuring instructors can quickly detect and intervene. Instructors can toggle between different alert tabs to view class-wide summaries. 
% Each tab shows the number of students affected, and the number of unresolved alerts is shown at the top of the tab as a visual cue for instructor attention. 
% Each tab shows how many students have ever triggered this type of alert (e.g., 22 out of 54 students in Agent Alert), while the red badge indicates the number of currently unresolved alerts to guide instructor attention.
% Once an alert has been addressed, the instructor can click the ``Mark as Handled'' button to remove it from view (Fig.~\ref{fig:teacher}-a4), ensuring that only unresolved alerts remain~visible.

\begin{figure}
    \centering
    \includegraphics[width=1\linewidth]{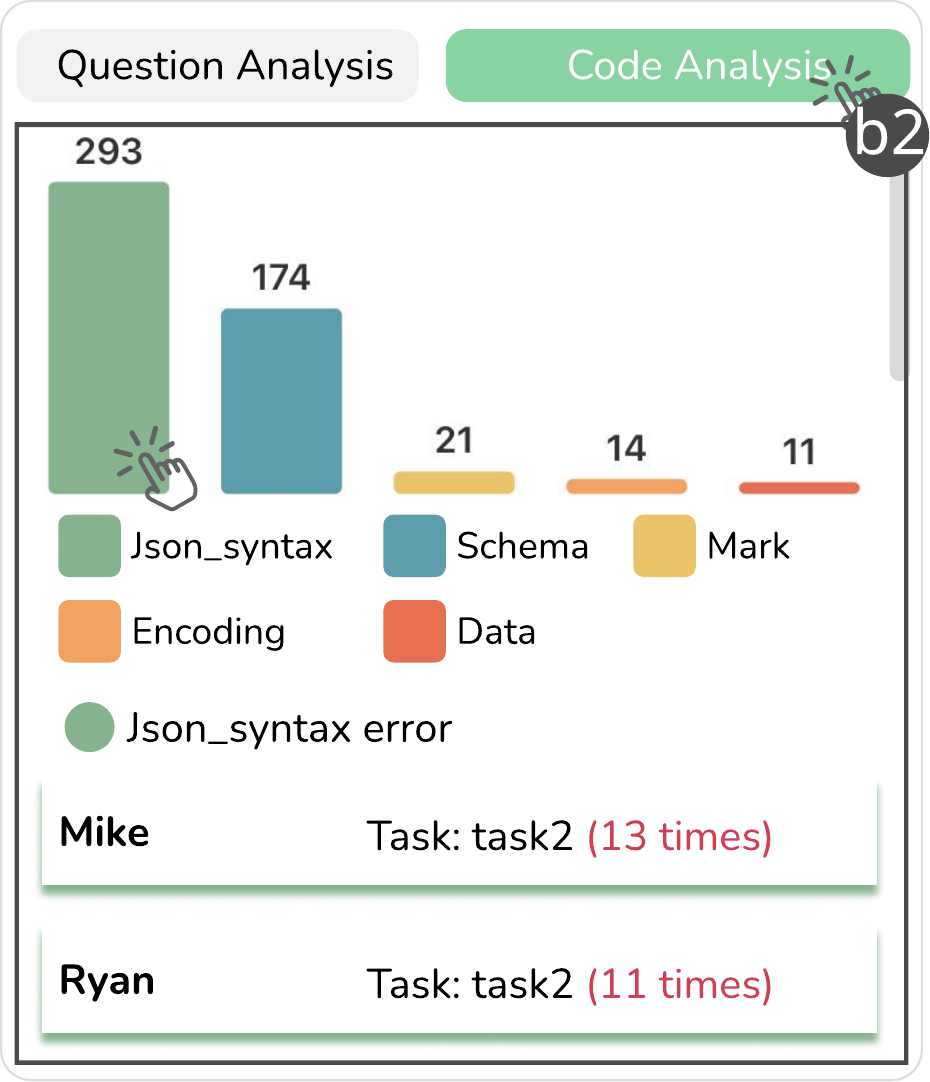}
    \caption{The Code Analysis view shows how often different code problems appear in student submissions. Instructors can click on each bar to see which students had that issue and and how many times it occurred.}
    \Description{
    The Code Analysis view shows how often different code problems appear in student submissions. Instructors can click on each bar to see which students had that issue and and how many times it occurred.
    }
    \label{fig:code}
\end{figure}

\textit{Class-Level Analysis to Identify Group-Wide Bottlenecks (DG2).}
To help instructors identify common issues across the class, the Class-Level Analysis panel provides two subviews: Question Analysis and Code Analysis (Fig.~\ref{fig:teacher}-B).
% In the Student Question Analysis view, the system uses a TF-IDF algorithm to extract high-frequency keywords from the questions students ask the TA Agent. These keywords are visualized in a word cloud along with their frequencies, allowing instructors to quickly identify cognitive bottlenecks and conceptual confusion points in the class (Fig.~\ref{fig:teacher}-b1). 
To monitor students' use of AI and detect abnormal behaviors, we designed a pyramid bar chart that visualizes question types.
As shown in Fig.~\ref{fig:teacher}-b1, this view uses LLM-based analysis to show whether students are primarily engaging in critical thinking or requesting direct answers.
Each bar represents a student: the orange section on the left indicates the number of answer-seeking questions, and the green section on the right represents critical thinking questions.
% This classification is based on content analysis of students' submissions. 
% When instructors hover over a bar, they can view the student's name and the total number of interactions. 
% The panel also provides a ``Show Task Breakdown'' toggle, which displays the distribution of question types across different tasks.
% The chart updates dynamically when data changes, and students are sorted by the total number of questions asked, with the most active students placed at the top to help instructors quickly identify and track engagement levels.
This classification is derived from content analysis of students' submissions. 
When instructors hover over a bar, they can see the student's name and total interactions. 
A ``Show Task Breakdown'' toggle reveals the distribution of question types across tasks. The chart updates dynamically, and students are sorted by total question count, with the most active students listed at the top.
In the Code Analysis view, student code errors are categorized and displayed in a bar chart (Fig.~\ref{fig:code}-b2).
Instructors can click an error category to view the list of affected students, sorted by error frequency so that those with the highest counts appear first and can be prioritized for intervention.

% Instructors can click on specific error categories to view the list of students affected and prioritize reviewing the most recent error examples shown at the top of the chart, enabling timely intervention.
% Instructors can click on specific error categories to view the list of affected students, ranked by error frequency, so that those with the highest number of errors appear first and can be prioritized for intervention.

\textit{Drill-Down View for Diagnosing Individual Student Needs (DG3).}
When an instructor identifies unusual behavior or performance issues through alerts or analysis panels, 
they can click the corresponding student card to open the More Details panel (Fig.~\ref{fig:teacher}-D), which provides an in-depth view of the student's learning trajectory and interaction history. This panel includes three subviews. 
The Conversation Information section logs the full history of the student's interactions with the TA Agent, including the student's questions, the Agent's responses, system-generated cognitive and code analyses, and the feedback mode or intervention applied.
This detailed record helps instructors understand the reasoning behind the TA Agent's feedback and supports subsequent instructional decisions (Fig.~\ref{fig:teacher}-d1).
The Agent Information section allows instructors to view the student's current feedback mode and adjust the TA Agent's support level based on learning progress (Fig.~\ref{fig:teacher}-d2) or class pacing (Fig.~\ref{fig:teacher}-c3). 
For example, if class time is nearly over and the student has not made progress, the instructor may switch to Technical Mode to provide more direct code guidance (Fig.~\ref{fig:teacher}-c3).
% Lastly, the Task Information panel presents detailed scores and final code submissions for each task. 
% Using the system's code quality evaluation model, it provides scores across four dimensions: correctness, efficiency, readability, and structure. 
% With this comprehensive information, instructors can trace back through the student's coding process when performance is low, identify learning obstacles and skill gaps, and offer targeted guidance and interventions to student (Fig~\ref{fig:teacher}-d3).
Finally, the Task Information panel presents detailed scores and final code submissions for each task. 
With this comprehensive information, instructors can trace the student's coding process when performance is low, identify learning obstacles and skill gaps, and offer targeted guidance and intervention to the student (Fig.~\ref{fig:teacher}-d3).

% 修改一下
% 增加更多细节
\subsection{ClassAid Implementation}
\textit{ClassAid} is a real-time programming support platform built on a client–server architecture, with a Vue.js frontend and a Python Flask backend. It supports core features including user authentication, interaction logging, and feedback collection. Client–server communication uses RESTful APIs, and Firebase manages user verification and data persistence. 
\RR{The system integrates OpenAI's GPT-4o to perform intelligent code analysis and generate feedback.}
\RX{For further details on the model selection process, refer to Appendix~\ref{comparison}.}
As shown in Fig.~\ref{fig:s-agent}, student activities such as question submissions, code executions, and interface interactions are captured and processed by the TA Agent through a six-stage orchestration pipeline, enabling fine-grained, personalized support.
The instructor dashboard provides real-time visibility into student progress and Agent activity. Instructors can adjust feedback modes (Silent, Auto, Heuristic, or Technical) individually or class-wide. These configurations are written to Firebase, which the TA Agent references in real time to align its responses with instructional intent.
To support adaptive feedback, \textit{ClassAid} maintains detailed contextual data for each student, including cognitive state changes, code quality metrics, error patterns, question types, learning history, and prior feedback. At each decision point, the system logs agent reasoning, student behavior, triggers, selected feedback, and intervention outcomes. All data is stored and visualized in real-time through the instructor dashboard, and rule-based evaluations prioritize critical events to ensure timely and targeted interventions.
This human-AI collaborative design provides instructors with strategic control and immediate insight into student learning, enabling alignment between pedagogical goals and AI-driven support. \textit{ClassAid} thus enhances orchestration capacity in hybrid programming classrooms.

\section{Evaluation Design}
To evaluate the effectiveness of \textit{ClassAid} in classroom programming and understand student experiences, we employed a mixed-method approach. First, a quantitative study assessed the performance of the LLM-based TA Agent. Second, a classroom study with 54 students, one instructor, and two TAs examined system's usefulness and usability. Finally, semi-structured interviews with eight experienced educators provided insights into its instructional value.
% All study protocols were approved by the university's IRB.

Our study is guided by the following research questions:

\textit{Student Interface: How do students perceive the effectiveness of the ClassAid student interface and TA Agent feedback during in-class programming activities?}

\textit{Instructor Dashboard: How do instructors perceive the accuracy and pedagogical utility of ClassAid's instructor dashboard in real-time classroom monitoring?}

\subsection{User Study}

We conducted a user study in a graduate-level data visualization course at a research university to evaluate \textit{ClassAid}'s usefulness and usability.
% , identify its strengths and limitations, and guide future improvements. 
During a 75-minute session, students completed two Vega-Lite visualization tasks using its declarative grammar for interactive graphics~\cite{vega-lite}.

\subsubsection{Participants}

% A total of 54 computer science graduate students participated (35 males, 19 females; M = 25.95, SD = 1.31). 
% The course was taught by a female assistant professor and supported by two TAs (one male, one female). Most students had some experience with AI tools but were largely unfamiliar with Vega-Lite.
% In particular, regarding prior experience with AI-assisted programming, 49\% reported frequent use, 40.8\% occasional use, 8.2\% had tried it once or twice, and 6.1\% had never used such tools. In contrast, 49\% had never heard of Vega-Lite, 38.8\% had heard of it but never used it, and 12.2\% had brief exposure.
Fifty-four computer science graduate students participated (35 males, 19 females; M = 25.95 years, SD = 1.31). 
The course was taught by a female assistant professor with support from two TAs. Most students had experience with AI tools but were unfamiliar with Vega-Lite.
Regarding AI-assisted programming, 49\% reported frequent use, 40.8\% occasional use, 8.2\% limited use, and 6.1\% no use. In contrast, 49\% had never heard of Vega-Lite, 38.8\% had heard of it but never used it, and 12.2\% had brief exposure.

\subsubsection{Study Design and Procedure}
% We did not include a baseline system in our study due to methodological and practical constraints. First, the study was embedded in an authentic 75-minute classroom session, where adding experimental conditions or a baseline system was infeasible due to time and resource limitations. Repeating similar tasks could also induce fatigue and learning effects, potentially biasing results.
% Moreover, conducting a comparable follow-up study with different participants would introduce significant variability in student background, programming proficiency, and instructional context, which would make meaningful comparisons unreliable and unfair.

We did not include a baseline system due to methodological and practical constraints.  
The 75-minute classroom session could not accommodate additional conditions, and repeating tasks risked fatigue and learning effects. A follow-up study with new participants would introduce variability in background and proficiency, making comparisons unreliable.

% \RR{The study began with a 20-minute instructor-led introduction to Vega-Lite and links to official documentation. One of the authors then demonstrated \textit{ClassAid} and introduced the two tasks (in Appendix~\ref{task}). }Students had 50 minutes to complete both tasks using the \textit{ClassAid} student dashboard; use of other AI tools was prohibited. They were asked to complete Task 1 before starting Task 2, and were informed that performance would not affect their final grades to encourage natural engagement.
% Throughout the study, the instructor and TAs monitored progress via the \textit{ClassAid} instructor dashboard and adjusted AI response modes as needed. Instructor interactions and discussions were recorded. After the study, students completed a post-study questionnaire using a 7-point Likert scale to assess feedback quality and overall experience (in Fig.~\ref{fig:result}). We also conducted follow-up interviews with 10 students (20 minutes each) and post-study interviews with the instructor and TAs to evaluate the instructor-facing dashboard.

\RR{Before the in-class activity, the instructor delivered a twenty-minute tutorial that introduced Vega-Lite's key concepts and worked examples based on its official documentation, ensuring that students entered the session with essential prior knowledge. The tutorial slides are included in the Supplementary Material.}
One of the authors demonstrated \textit{ClassAid} and introduced the two tasks (Appendix~\ref{task}).
Students were given 50 minutes to complete both tasks using the \textit{ClassAid} student interface, with other AI tools prohibited. They completed Task 1 before Task 2 and were informed that their performance would not affect their final grades to encourage natural engagement.
% Throughout the session, the instructor and TAs monitored progress through the instructor dashboard and adjusted AI modes as needed. Afterward, students completed a 7-point Likert questionnaire on feedback quality and overall experience, and we conducted follow-up interviews with 10 students and post-study interviews with the instructor and TAs.
Throughout the activity, the instructor and TAs monitored progress and adjusted AI response modes through the instructor dashboard. Students completed a 7-point Likert questionnaire afterward, followed by interviews with 10 students \RX{(S1-S10)} and post-study interviews with the instructor and TAs.

\subsection{TA Agents' Cognitive-Level Assessment and Feedback Quality}

\RR{
To mitigate potential trust issues associated with LLM-based feedback and better understand how students interact with the TA Agent~\cite{ji2023survey}, we conducted expert evaluations of four components: (1) the accuracy of the Agent's Bloom-based estimation of students' cognitive levels, (2) the correctness of its feedback across modes, (3) the appropriateness of its feedback-type selection in Auto mode, and (4) the accuracy of its question-type classification (critical thinking vs. answer seeking).
We drew a simple random sample comprising approximately 50\% of all student questions collected in the user study (n = 274).
% For each sampled question, we recorded the Bloom cognitive level label assigned by the TA Agent as well as its corresponding technical and heuristic feedback. 
% For each question, we extracted the Agent's Bloom label and its technical and heuristic feedback. 
% We also randomly sampled approximately 50\% of all student questions and asked I1 and I2 to verify the system's question-type classification.
Following StuGPTVis~\cite{chen2024stugptviz}, two instructors with data visualization teaching experience (I1–I2) independently evaluated the accuracy of the Agent's Bloom-level classification for each question and coded each piece of feedback and question type using a three-point scale (1 = mostly correct, 0.5 = partially correct, 0 = mostly incorrect). To assess the Agent's ability to emulate instructor decision-making in Auto mode, we drew an additional simple random sample comprising about 50\% of Auto-mode questions (n = 106), and I1 and I2 applied the same scale to rate the appropriateness of the selected feedback type.

% We also randomly sampled approximately 50\% of the student questions and asked I1 and I2 to verify the accuracy of the system's question-type classification.

}

% To address potential trust concerns with LLM-based feedback and better understand student–TA Agent interactions~\cite{ji2023survey}, we evaluated the response accuracy of the TA Agent. Following StuGPTVis~\cite{chen2024stugptviz}, feedback was categorized as mostly correct (1), partially correct (0.5), or mostly incorrect (0).
% \RR{From the full set of student questions collected across user study, we drew a simple random sample of 20\% of questions without replacement. 
% For each sampled question, we collected the corresponding technical and heuristic responses from the TA Agent.
% Two experienced data visualization instructors (I1–I2) independently assessed the accuracy of these responses.
% To evaluate the Agent's ability to emulate human instructors in choosing appropriate feedback modes, we similarly drew a simple random sample of 30\% of student queries handled under Auto mode. I1 and I2 assessed only the correctness of the feedback type selected by the Agent. Additionally, they evaluated a simple random sample of 20\% of student questions to verify the accuracy of the system's question-type classification.}

\begin{figure}
    \centering
    \includegraphics[width=1\linewidth]{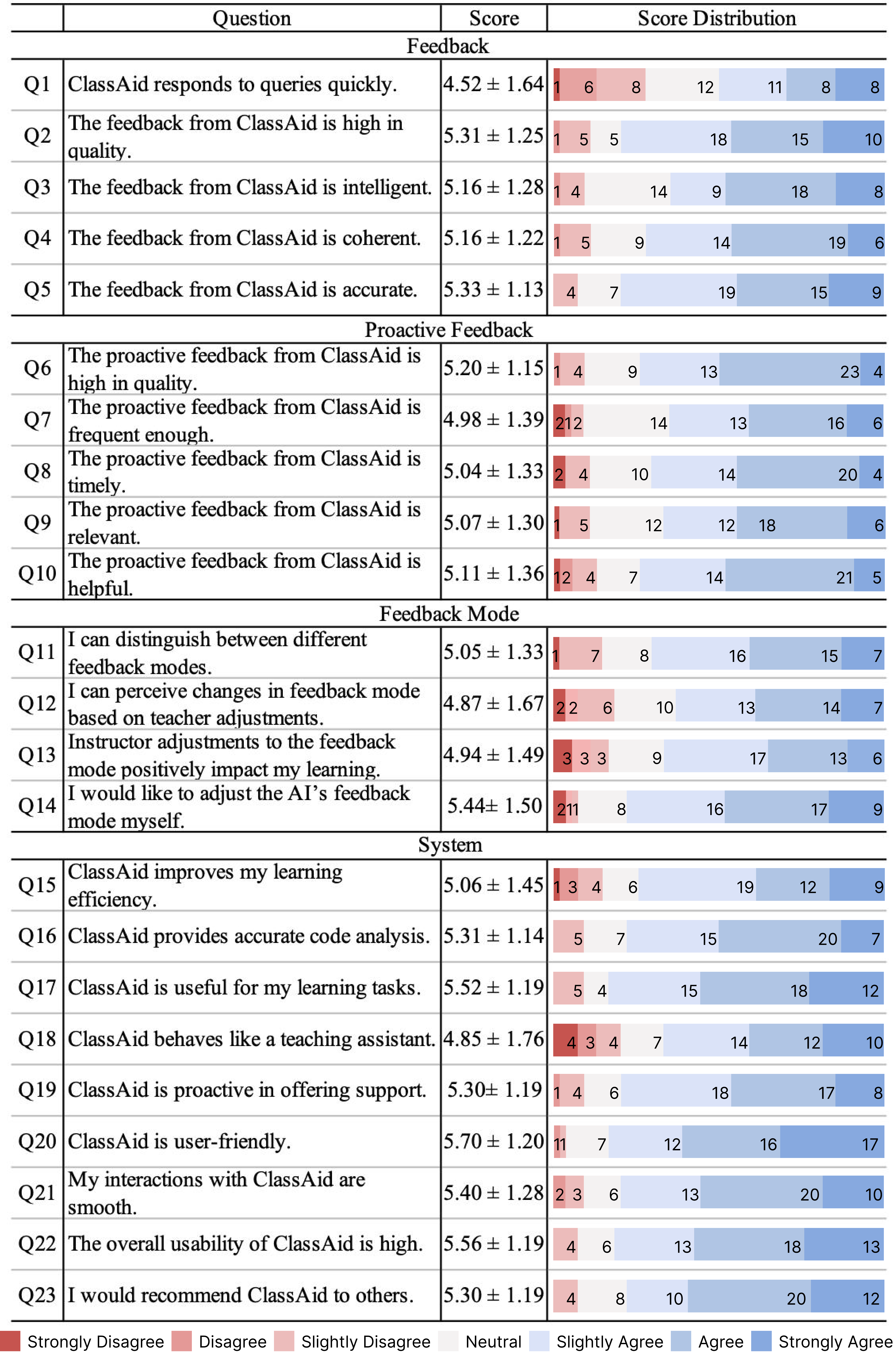}
    \caption{Student ratings on the \textit{ClassAid}.}
    \Description{
    Student ratings on the \textit{ClassAid}.
    }
    \label{fig:result}
\end{figure}

\subsection{Educator Interview}

To gather additional feedback and improve the generalizability of our findings, 
% we conducted semi-structured interviews with five university-level programming educators (E1–E5) after the user study.
\RR{we invited eight university-level programming educators (E1–E8) to participate in semi-structured interviews following the user study.}

\subsubsection{Participants}

% All were actively teaching programming-related courses: E1–E3、E6-E8 were assistant professors, E4-E5、E9-E10 were full professors, with gender distribution (three female and seven male). Teaching experience ranged from 1 to 23 years (mean = 7.6 years). E1, E2, and E4 had participated in the formative study; others were newly recruited.

Participants were recruited through snowball sampling within the authors' professional networks~\cite{given2008sage}. 
\RR{All were actively teaching programming-related courses. The sample included five assistant professors (E1–E2, E4–E6) and three full professors (E3, E7–E8), with a gender distribution of three female and five male.
Their teaching experience ranged from 1 to 23 years (M = 7.1). 
E1 and E3 had participated in the formative study, while the remaining participants were newly recruited.
Course topics included data visualization (4), Python programming (2), and Java programming (2).} Each educator received a \$50 honorarium.

\subsubsection{Procedure}

% Each interview (~90 minutes) began with an introduction to the study's goals, a system overview, and an explanation of how to use \textit{ClassAid}. 
% % Participants then had 10 minutes to explore the system freely. 
% \RR{We then invited participants to freely explore the system without any time constraints, and most concluded their exploration around ten minutes.}
% Next, we showed selected video excerpts from the user study, demonstrating how instructors adjusted TA Agent modes in real time, along with preliminary findings.
% Participants were invited to share feedback on system design goals (DG1–DG4), usefulness, usability, limitations, and suggestions for improvement. All interviews were conducted via Zoom with informed consent and audio-recorded for analysis.

% 访谈开始于研究目标、系统结构与使用方式的介绍。随后参与者进行首次自由探索（通常约十分钟），以熟悉界面与核心功能。在此之后，
% 为了确保参与者能够在接近真实课堂的环境下体验系统，我们从真实 user study 的课堂数据中提取了三个“高密度模式调整阶段”。这些阶段是教师在某些时间段内针对多名学生连续进行模式切换的区间，反映了教师对学生学习状态变化的即时判断。
% 每个片段都包含该阶段学生与 TA Agent 的完整交互记录，与user study中教员看到的内容一致。
% 参与者依次查看上述三个课堂片段。对于每个片段，参与者会像在真实教学中一样观察学生的学习状态，分析学生的困难点，并以 think-aloud 的方式操作 \textit{ClassAid} 来调整 TA Agent 模式，同时说明其教学决策的依据和预期效果。该过程通常会引发多轮交互、反复试用与讨论，使参与者能够在具体情境下深入理解系统的功能与模式设计。
% 在每轮基于片段的操作任务之后，我们向参与者展示 user study 中教师在相同课堂阶段的真实操作视频，帮助他们将自己的判断与真实课堂中的教学实践进行对照。最后，我们基于这些实践体验，引导参与者讨论系统设计目标（DG1–DG4）、有用性、可用性、局限性以及改进建议。所有访谈均在 Zoom 上进行，并在取得知情同意后录音以供分析。

\RR{
Each interview (90 minutes) began with an introduction to the study's goals, a system overview, and guidance on using \textit{ClassAid}. Participants then conducted an brief free exploration to familiarize themselves with both the student interface and the instructor dashboard.
% To help participants experience the system in a context closer to real classroom use, we selected three high-density mode adjustment phases from the actual user study classroom data.
To help participants experience the system in a context closer to real classroom use, we selected three segments from the user study data that ended at the timestamps where the instructor initiated high-frequency mode adjustments. 
These segments captured the student–AI interaction patterns that immediately preceded those adjustments, and each segment included the complete interaction records visible to instructors during the user study.
% These phases represent intervals in which the instructor rapidly switched modes for multiple students, reflecting immediate instructional decisions based on students' learning states. Each phase contained the complete student–AI interaction records visible to instructors during the user study.
% Participants reviewed the three classroom segments in sequence. For each segment, they observed students' learning states, analyzed their difficulties, and used a think-aloud protocol to adjust TA Agent modes in \textit{ClassAid}, explaining the reasoning behind their decisions. This process involved multiple rounds of interaction, repeated exploration, and discussion, allowing participants to develop a deeper understanding of the system's functionalities, mode differences, and usage scenarios.
Participants examined the three segments in sequence. For each segment, they reviewed the interaction logs, observed students' learning states, analyzed their difficulties, and used a think-aloud protocol to adjust TA Agent modes, explaining the reasoning behind their decisions. 
% After each phase-based task, we showed participants the corresponding classroom footage from the user study, allowing them to compare their reasoning with authentic instructional practices. Finally, informed by these experiences, we invited participants to discuss the system's design goals (DG1–DG4), usefulness, usability, limitations, and suggestions for improvement. All interviews were conducted via Zoom and audio-recorded with informed consent.
% After each segment-based task, we showed participants the corresponding classroom video footage, enabling them to compare their reasoning with authentic instructional decisions. Finally, informed by these experiences, participants discussed the system's design goals (DG1–DG4), usefulness, usability, limitations, and suggestions for improvement. All interviews were conducted via Zoom and audio recorded with informed consent.
After each segment, we showed the corresponding classroom video for comparison with real instructional decisions. The interview concluded with a discussion of design goals, usability, limitations, and suggestions. All interviews were conducted via Zoom and audio recorded with consent.

}

\section{Results}

\RR{\subsection{TA Agents' Cognitive-Level Assessment and Feedback Quality}}

\RR{This section summarizes the quantitative evaluation results of the TA Agent based on instructors' ratings across several dimensions.}

\RR{\subsubsection{Accuracy of of Cognitive-Level}
% To evaluate the accuracy of the TA Agent's cognitive-level classification, we randomly sampled 274 student questions and asked two instructors (I1 and I2) to independently label each question according to Bloom's revised taxonomy. 
We evaluated the accuracy of the TA Agent's cognitive-level predictions by randomly sampling 274 student questions. For each question, the Agent produced a Bloom-level classification, and two instructors (I1, I2) labeled the correctness of that classification using a three-level rubric.
As shown in Table~\ref{table:cognitive-level}, the results demonstrated 95.99\% agreement which indicates high a high rater consistency and supports the reliability of the TA Agent's cognitive-level estimates.

\begin{table*}[htbp]
\centering
\caption{\RR{Inter-rater agreement on the TA Agent's cognitive-level classification, including detailed rating distributions and overall reliability.}}
\label{table:cognitive-level}
\begin{tabular}{l l c c c c}
\toprule
\textbf{Cognitive-level} 
& \textbf{Rating Level} 
& \textbf{I1} 
& \textbf{I2} 
& \textbf{Agreement Count} 
% & \textbf{Agreement/$\kappa$}\\
& \textbf{Agreement Ratio}\\
\midrule

\multirow{3}{*}{Overall (n = 274)} 
% & Correct (1)             & 91 & 88 & 87 & \multirow{3}{*}{94.29\%/0.8161} \\
& Correct (1)             & 255 & 249 & 247 & \multirow{3}{*}{95.99\%} \\
& Partially Correct (0.5) & 12  & 16  & 10  &  \\
& Incorrect (0)           & 7  & 9  & 6  &  \\

\bottomrule
\end{tabular}
\end{table*}

}
\subsubsection{Correctness of Feedback}
\RR{
% We randomly sampled 274 student questions, each accompanied by one heuristic and one technical feedback response generated by the TA Agent. 
We randomly sampled 274 student questions, each with one heuristic and one technical feedback response.
% Two instructors (I1 and I2) independently rated each response using a three-level rubric (Correct, Partially Correct, Incorrect), as shown in Table~\ref{table:feedback}. 
Two instructors (I1, I2) independently rated all responses using a three-level rubric, as shown in Table~\ref{table:feedback}.
% The average scores for heuristic feedback were 0.95 (I1) and 0.95 (I2), and for technical feedback, 0.97 (I1) and 0.96 (I2).
% The average rubric scores were 0.95 (I1) and 0.95 (I2) for heuristic feedback and 0.97 (I1) and 0.96 (I2) for technical feedback. 
Overall agreement reached 94.71\%. These results indicate strong inter-rater consistency and suggest that the TA Agent provides reliable feedback across both modes.}
% Overall, the results demonstrated 94.71\% agreement.
% with a Cohen's $\kappa$ coefficient of 0.6774. 
% These findings indicate a high level of consistency among instructors and suggest that the TA Agent provides reliable feedback across both heuristic and ?technical dimensions.

% Inter-rater agreement was highest for Incorrect labels (95.5\%), followed by Correct (70.85\%) and Partially Correct (41.5\%). These results suggest that the TA Agent generally produces accurate and reliable feedback.

\begin{table*}[htbp]
\centering
\caption{\RR{Inter-rater agreement between I1 and I2 across feedback types, including fine-grained ratings and overall reliability.}}
\label{table:feedback}
\begin{tabular}{l l c c c c}
\toprule
\textbf{Feedback Type} 
& \textbf{Rating Level} 
& \textbf{I1} 
& \textbf{I2} 
& \textbf{Agreement Count} 
% & \textbf{Agreement/$\kappa$}\\
& \textbf{Agreement Ratio}\\
\midrule

\multirow{3}{*}{Heuristic (n = 274)} 
% & Correct (1)             & 95 & 99 & 94 & \multirow{3}{*}{95.24\%/0.6886} \\
& Correct (1)             & 253 & 249 & 245 & \multirow{3}{*}{95.26\%} \\
& Partially Correct (0.5) & 16  & 21  & 12  &  \\
& Incorrect (0)           & 5  & 4  & 4  &  \\
\midrule

\multirow{3}{*}{Technical (n = 274)} 
% & Correct (1)             & 99 & 98 & 97 & \multirow{3}{*}{96.19\%/0.6774} \\
& Correct (1)             & 263 & 258 & 250 & \multirow{3}{*}{94.16\%} \\
& Partially Correct (0.5) & 6  & 12  & 4  &  \\
& Incorrect (0)           & 5  & 4  & 4  &  \\
\midrule

\textbf{Total} 
% & -- & -- & -- & \textbf{201} & \textbf{95.71\%/0.6830} \\
& -- & -- & -- & \textbf{519} & \textbf{94.71\%} \\
\bottomrule
\end{tabular}
\end{table*}

% \begin{table*}[htbp]
% \centering
% \caption{Agreement between I1 and I2 on across different mode feedback}
% \begin{tabular}{l l c c c c}
% \toprule
% \textbf{Feedback Type} & \textbf{Rating Level} & \textbf{I1 Count} & \textbf{I2 Count} & \textbf{Agreement Count} & \textbf{Agreement} \\
% \midrule
% Heuristic (n = 105)     & Correct (1)     & 95 & 99 & 94 & 94.0\% \\
%               & Partially Correct (0.5) & 6  & 3  & 3  & 50.0\% \\
%               & Incorrect (0)   & 4  & 3  & 3  & 75.0\% \\
% \midrule
% Technical (n = 105)     & Correct (1)     & 99 & 98 & 97 & 97.0\% \\
%               & Partially Correct (0.5) & 3  & 5  & 2  & 33.0\% \\
%               & Incorrect (0)   & 3  & 2  & 2  & 66.7\% \\
% \bottomrule
% \end{tabular}
% \label{table:feedback}
% \end{table*}
% \begin{figure}
%     \centering
%     \includegraphics[width=1\linewidth]{figs/table3.pdf}
%     % \caption{Enter Caption}
%     % \label{fig:placeholder}
% \end{figure}

\RR{\subsubsection{Appropriateness of Auto Mode Selections}
To assess the TA Agent's feedback-selection in Auto mode, we analyzed 50\% of automatically generated feedback instances (n = 106), including 66 technical and 40 heuristic responses (in Table~\ref{table:auto}).
I1 and I2 independently rated each response using a predefined three-level rubric.
% For heuristic responses, the average scores were 0.93 (I1) and 0.91 (I2); for technical responses, 0.91 (I1) and 0.86 (I2). 
The overall percent agreement was 88.68\% 
% and the weighted Cohen's $\kappa$ was 0.68
, indicating high rater consistency and reliable evaluation of the TA Agent's feedback quality.}

% 
% Agreement on Correct responses was high (89.2\%), while Partially Correct and Incorrect labels had agreement rates of 37.5\% and 65.0\%, respectively.

% \begin{table*}[htbp]
% \centering
% \caption{Agreement between I1 and I2 on TA Agent's Auto-mode feedback}
% \begin{tabular}{l l c c c c}
% \toprule
% \textbf{Feedback Type} & \textbf{Rating Level} & \textbf{I1 Count} & \textbf{I2 Count} & \textbf{Agreement Count} & \textbf{Agreement} \\
% \midrule
% Heuristic (n=25)     & Correct (1)              & 20 & 22 & 20 & 90.9\% \\
%               & Partially Correct (0.5)  & 4  & 1  & 1  & 25.0\% \\
%               & Incorrect (0)            & 1  & 2  & 1  & 50.0\% \\
% \midrule
% Technical (n=42)     & Correct (1)              & 31 & 29 & 28 & 87.5\% \\
%               & Partially Correct (0.5)  & 6  & 9  & 5  & 50.0\% \\
%               & Incorrect (0)            & 5  & 4  & 4  & 80.0\% \\
% \bottomrule
% \end{tabular}
% \label{table:auto}
% \end{table*}

\begin{table*}[htbp]
\centering
\caption{\RR{Inter-rater agreement between I1 and I2 for Auto-mode feedback responses, including detailed ratings and overall reliability.}}
\label{table:auto}
\begin{tabular}{l l c c c c}
\toprule
\textbf{Feedback Type} 
& \textbf{Rating Level} 
& \textbf{I1} 
& \textbf{I2} 
& \textbf{Agreement Count} 
% & \textbf{Agreement/$\kappa$}\\
& \textbf{Agreement Ratio}\\
\midrule

\multirow{3}{*}{Heuristic (n = 40)} 
& Correct (1)             & 36 & 35 & 34 & \multirow{3}{*}{90.00\%} \\
% & Correct (1)             & 20 & 22 & 20 & \multirow{3}{*}{88.00\% / 0.5810} \\
& Partially Correct (0.5) & 2  & 3  & 1  &  \\
& Incorrect (0)           & 2  & 2  & 1  &  \\
\midrule

\multirow{3}{*}{Technical (n = 66)} 
% & Correct (1)             & 31 & 29 & 28 & \multirow{3}{*}{88.10\% / 0.7345} \\
& Correct (1)             & 57 & 52 & 50 & \multirow{3}{*}{87.88\%} \\
& Partially Correct (0.5) & 6  & 10  & 5  &  \\
& Incorrect (0)           & 3  & 4  & 3  &  \\
\midrule

\textbf{Total} 
% & -- & -- & -- & \textbf{59} & \textbf{88.06\% / 0.6772} \\
& -- & -- & -- & \textbf{94} & \textbf{88.68\%} \\
\bottomrule
\end{tabular}
\end{table*}

% \begin{table*}[htbp]
% \centering
% \caption{Inter-rater agreement between I1 and I2 on TA Agent feedback, including fine-grained ratings and overall reliability}
% \begin{tabular}{l l c c c c c}
% \toprule
% \textbf{Feedback Type} 
% & \textbf{Rating Level} 
% & \textbf{I1 Count} 
% & \textbf{I2 Count} 
% & \textbf{Agreement Count} 
% & \textbf{Agreement (\%)} 
% & \textbf{Cohen's $\kappa$} \\
% \midrule

% \multirow{3}{*}{Heuristic (n = 25)} 
% & Correct (1)             & 20 & 22 & 20 & \multirow{3}{*}{88.00} & \multirow{3}{*}{0.5810} \\
% & Partially Correct (0.5) & 4  & 1  & 1  &                     &                           \\
% & Incorrect (0)           & 1  & 2  & 1  &                     &                           \\
% \midrule

% \multirow{3}{*}{Technical (n = 42)} 
% & Correct (1)             & 31 & 29 & 28 & \multirow{3}{*}{88.10} & \multirow{3}{*}{0.7345} \\
% & Partially Correct (0.5) & 6  & 9  & 5  &                     &                           \\
% & Incorrect (0)           & 5  & 4  & 4  &                     &                           \\
% \midrule

% \textbf{Total} 
% & -- & -- & -- & 59 & 88.06 & 0.6772 \\
% \bottomrule
% \end{tabular}
% \label{table:combined-agreement}
% \end{table*}

% \begin{figure}
%     \centering
%     \includegraphics[width=1\linewidth]{figs/table4.pdf}
%     % \caption{Enter Caption}
%     % \label{fig:placeholder}
% \end{figure}

\subsubsection{Accuracy of Question Type}

% Among the 274 student questions, the system classified 175 as critical thinking and 99 as direct answer-seeking. I1 and I2 evaluated the responses using the same three-level rubric \RR{(shown in Table~\ref{table:thinking_agreement})}. 

\RR{Among the 274 student questions, the system classified 175 as critical thinking and 99 as direct answer-seeking. I1 and I2 evaluated the accuracy of these classifications using the same three-level rubric (Table~\ref{table:thinking_agreement}). The overall agreement was 95.99\%, indicating that the system can reliably distinguish between critical-thinking and direct answer-seeking questions.}

\begin{table*}[htbp]
\centering
\caption{\RR{Inter-rater agreement between I1 and I2 on thinking types, including fine-grained ratings and overall reliability.}}
\label{table:thinking_agreement}
\begin{tabular}{l l c c c c}
\toprule
% \textbf{Thinking Type} & \textbf{Rating Level} & \textbf{I1} & \textbf{I2} & \textbf{Agree} & \textbf{Agreement / $\kappa$} \\
\textbf{Thinking Type} & \textbf{Rating Level} & \textbf{I1} & \textbf{I2} & \textbf{Agreement Count} 
% & \textbf{Agreement/$\kappa$}\\
& \textbf{Agreement Ratio}\\
\midrule
 
\multirow{3}{*}{Critical Thinking (n=175)}
& Correct (1)             & 166 & 164 & 163 & \multirow{3}{*}{98.29\%} \\
% & Correct (1)             & 66 & 64 & 64 & \multirow{3}{*}{96.97\% / 0.0000} \\
& Partially Correct (0.5) & 9  & 10  & 8  & \\
& Incorrect (0)           & 0  & 1  & 1  & \\
\midrule

\multirow{3}{*}{Direct Answer Seeking (n=99)}
& Correct (1)             & 91 & 87 & 85 & \multirow{3}{*}{91.92\%} \\
% & Correct (1)             & 24 & 28 & 24 & \multirow{3}{*}{84.62\% / 0.7677} \\
& Partially Correct (0.5) & 3  & 5  & 3  & \\
& Incorrect (0)           & 5  & 7  & 3  & \\
\midrule

\textbf{Total}
& -- & -- & -- & \textbf{267} & \textbf{95.99\%} \\
% & -- & -- & -- & \textbf{97} & \textbf{92.38\% / 0.2851} \\
\bottomrule
\end{tabular}
\end{table*}

Across all evaluated components, the TA Agent demonstrated consistently strong performance. 
It provided accurate assessments of students' learning states and generated high-quality responses. 
Although some subjectivity remained in Auto-mode feedback selection, particularly in borderline cases, this does not imply insufficient performance. In many situations, human instructors also differ in determining whether a student would benefit more from technical or heuristic guidance, because such pedagogical decisions do not have a single correct answer.

\subsection{System Evaluation}

% In this study, we conducted a systematic analysis of both the student-side and instructor-side interactions with the system.

In this study, we conducted a systematic analysis of system interactions from both the student and instructor perspectives.

\subsubsection{Student-Side Performance and Feedback Analysis}

Table~\ref{tab:student_behavior} summarizes student performance, TA Agent triggers, and feedback across the two programming tasks. 
Students achieved accuracy scores of 3.87 and 3.97, with average completion times of 21 and 17 minutes. 
In total, they made 6,841 code edits, 1,102 code executions, and 82 pauses.
The TA Agent generated 
% 209 edit-monitor events, 526 run-time checks, 138 predictive triggers, 
394 proactive triggers, and 28 passive triggers (except student request).
Students submitted 547 questions, resulting in 1,107 feedback messages, including 731 heuristic responses, 165 technical responses, and 211 Auto-mode responses (132 technical and 79 heuristic). 
Students provided 29 likes and 24 dislikes, corresponding to rating frequencies of 5.5\% and 3.3\%.
Despite receiving over 1,100 feedback messages, students only rated a small fraction, consistent with the ``extreme feedback bias''~\cite{harber1998feedback}, where moderate experiences are less likely to elicit reactions.

% \begin{figure}
%     \centering
%     \includegraphics[width=0.8\linewidth]{figs/table6.pdf}
%     % \caption{Enter Caption}
%     % \label{fig:placeholder}
% \end{figure}

\begin{table*}[htbp]
\centering
\caption{Summary of student performance, TA Agent triggers, and feedback across Task1 and Task2.}
\label{tab:student_behavior}
\begin{tabular}{l c c c}
\toprule
\textbf{Measure} & \textbf{Task1} & \textbf{Task2} & \textbf{Total} \\
\midrule
Accuracy (1--5) & 3.87 & 3.97 & -- \\
Completion Time (min) & 21 & 17 & -- \\
\midrule
Code Edits & 3,538 & 3,003 & 6,841 \\
Code Executions & 674 & 428 & 1,102 \\
Pauses & 25 & 57 & 82 \\
\midrule
% Edit Monitors & 123 & 86 & 209 \\
% Run-time Monitors & 328 & 198 & 526 \\
Predictive Triggers & 75 & 63 & 138 \\
Proactive Triggers & 267 & 127 & 394 \\
Passive Triggers & 5 & 23 & 28 \\
\midrule
Questions Submitted & 393 & 154 & 547 \\
Feedback Received & 740 & 367 & 1,107 \\
Heuristic Feedback & 517 & 214 & 731 \\
Technical Feedback & 101 & 64 & 165 \\
Auto-mode Feedback & 122 & 89 & 211 (132 Tech., 79 Heur.) \\
\midrule
Feedback Likes & 21 & 8 & 29 \\
Feedback Dislikes & 20 & 4 & 24 \\
Rating Rate & 5.5\% & 3.3\% & -- \\
\bottomrule
\end{tabular}
\end{table*}

\subsubsection{Instructor-Side Interventions and Feedback Mode Adjustment Strategies}

% At the beginning of the in-class programming activity, the course instructor and two teaching assistants (TAs) collaboratively used the \textit{ClassAid} instructor interface to monitor student progress and manage behavior. During the activity, the system triggered a total of 15 alerts, including 4 agent alerts, 3 process alerts, and 8 outcomes alerts. Of these, the instructional team addressed 4 agent alerts, 3 process alerts, and 5 code alerts, ensuring timely intervention in response to key issues throughout the session.
At the start of the in-class programming activity, the instructor and two TAs collaboratively used the \textit{ClassAid} instructor dashboard to monitor student progress and manage TA Agents. 
As previously introduced (in Sec.~\ref{alert}), the \textit{ClassAid} supports three types of alerts.
The system triggered 15 alerts (four agent, three process, and eight outcome alerts), of which the team addressed four agent, three process, and five code alerts, enabling timely interventions.

% To manage feedback mode for a large cohort of students, the instructional team implemented a total of 8 classroom-wide feedback mode adjustments, based on students' progress and system observations. Initially, to encourage independent thinking and reduce the reliance on AI-generated solutions, all students were placed in heuristic feedback mode. However, ten minutes into the activity, no students had completed Task1. After discussion, the instructors decided to switch the entire class to the auto mode, which provides a dynamic mix of feedback styles.
% Following this adjustment, some students began successfully completing Task1. To support those who were still struggling, the instructors transitioned the remaining students to the technical feedback mode, aiming to accelerate progress. Once students moved on to Task2, their feedback mode was again switched back to the heuristic mode to continue promoting independent problem-solving. This staged and responsive adjustment strategy was applied throughout the session and demonstrated strong adaptability in a live teaching context.

To manage feedback mode at scale, the instructional team made eight classroom-wide feedback mode adjustments based on student progress and system cues. Initially, all students were set to heuristic mode to encourage independent and critical thinking. However, after 10 minutes with no task completions, the team switched to auto mode to offer more dynamic support. When half of the students completed Task 1, those still struggling were transitioned to technical mode. Later, when students moved on to Task 2, the class reverted to heuristic mode to reinforce independent problem-solving. This adaptive strategy supported diverse learning needs in real time.

% Beyond these class-level changes, the instructional team also made individual feedback mode adjustments based on specific student behavior. During the activity, the instructor viewed detailed interaction logs for 46 students. The most common reason for reviewing a student's record was repeated ``answer-seeking'' behavior in their questions, followed by system-generated alerts, code-related issues, and low performance. Based on these insights, 22 students had their feedback mode individually adjusted. For example, one student was identified through question issues as having repeatedly asked for direct answers. Upon reviewing the interaction history, the instructor confirmed the pattern and, after discussion with a TA, switched the student's feedback mode to silent in order to reduce overreliance on AI support.

In addition to class-level adjustments, the instructor reviewed logs for 46 students, most commonly due to repeated answer-seeking behavior, system alerts, code issues, or low performance. Based on these reviews, 22 students had their feedback modes individually adjusted. For example, one student exhibiting persistent answer-seeking was switched to silent mode to reduce AI overreliance.
% In addition, instructors noticed that some students made little progress even after being placed in the technical mode. In such cases, the instructor or a TA provided direct human support—engaging with students to identify their difficulties, guiding them on how to better interact with the TA Agent, or offering direct assistance. In total, 5 students received individualized support from a human instructor or TA during the session.
Finally, for students making little progress even under technical mode, instructors or TAs provided additional human support. This included identifying difficulties, advising on TA Agent use, or offering direct assistance. In total, five students received personalized in-person help during the session.

\subsection{Evaluation on Student Interface and Instructor Dashboard}
% 需要修改，增加和其他AI工具的比较
\subsubsection{Feedback from Students' Perspective}
% As illustrated in Fig.~\ref{fig:result}, we analyzed student ratings of the feedback, proactive feedback, feedback mode, and overall system using a 7-point Likert scale (1 = strongly disagree, 7 = strongly agree), and reported score distributions across each dimension.
As shown in Fig.~\ref{fig:result}, student ratings of feedback, proactive feedback, feedback modes, and overall system experience were collected on a 7-point Likert scale. 

\textit{General feedback} received positive evaluations, with average scores above 5.0 for quality, intelligence, coherence, and accuracy, except for response speed. These results indicate that students generally found the TA Agent's feedback effective, although some reported dissatisfaction with latency when timely support was needed.

% \textit{General feedback} received consistently positive evaluations, with average scores above 5.0 for quality, intelligence, coherence, and accuracy. The only exception was response speed.
% This indicates that the TA Agent's feedback was generally perceived as effective and helpful. 
% However, some students expressed dissatisfaction about latency, particularly when timely support was critical.
% \textit{ One student commented: ``The TA Agent was a bit slow to respond, possibly due to the number of students in the class. Real-time services like this require strong back-end processing capabilities; I suggest expanding the system's deployment in the future.''}
% \textit{One student noted: ``The TA Agent responded slowly, likely due to class size. Real-time services need stronger back-end support; I suggest scaling up deployment.''}

% \textit{Proactive feedback} was also rated favorably. 
% Quality received the highest score (M = 5.20), while frequency was rated lowest (M = 4.98).
% Some students felt that the proactive feedback appeared too quickly, which left little time for independent reflection before receiving hints. 
% Others, particularly beginners, appreciated the timely guidance and corrections. 
% For example, one student recalled that during Task 2, when asked to ``color a bar chart by category,'' she initially tried to assign colors manually. The Agent proactively suggested a more concise specification:
% \texttt{color: {"field": "category", "type": "nominal"}}.

\textit{Proactive feedback} was rated positively overall, with quality receiving the highest rating (M = 5.20) and frequency the lowest (M = 4.98). 
Some students felt the feedback arrived too quickly and limited independent reflection, whereas beginners appreciated the timely guidance. 
For instance, one student initially tried to assign colors manually until the Agent suggested a more concise specification: \texttt{color: {``field'': ``category'', ``type'': ``nominal''}}.
She reported being pleasantly surprised because the Agent detected her difficulty and offered timely support without being prompted.

% \textit{On the topic of feedback mode adjustment,} most students could clearly distinguish among the different modes, noticed when instructors switched the Agent's mode, and believed that these adjustments significantly influenced their task performance. 
% In a preference survey, auto mode and technical feedback were each preferred by 42.6\% of students, while 14.8\% favored heuristic feedback. 

\textit{Regarding feedback mode adjustment}, most students could distinguish among the modes, noticed instructor-initiated changes, and believed these shifts affected their performance. In the preference survey, Auto mode and technical feedback were each preferred by 42.6\% of students, whereas 14.8\% preferred heuristic feedback.
% Interviews revealed that students who preferred heuristic feedback valued the space it left for independent problem solving and appreciated its concise and readable format. They felt it offered just enough guidance without giving away the answer too early. In contrast, technical feedback was seen as more direct and efficient, providing specific code examples and concrete suggestions—though some students felt it made them overly reliant on the AI. For those who wanted to complete tasks more quickly, technical feedback was often the preferred option. Students who favored the auto mode appreciated its ability to dynamically adapt feedback based on context, offering a smoother overall experience even when they were unsure which style of feedback best suited their needs.
% Interviews revealed that students who preferred heuristic feedback appreciated its concise, readable format and the space it left for independent problem solving. They felt it offered just enough guidance without revealing answers prematurely. 
% By contrast, technical feedback was viewed as more direct and efficient, offering specific code suggestions and concrete fixes. However, some students expressed concern that it encouraged overreliance on the AI. 
% For those prioritizing task completion speed, technical feedback was often the preferred choice. 
% Students who selected auto mode valued its ability to adapt dynamically to context, offering a smoother experience, especially when they were uncertain which feedback style suited them best.
Interviews showed that students who preferred heuristic feedback appreciated its concise and readable style, as well as the space it provided for independent problem solving. 
In contrast, technical feedback was viewed as direct and efficient, offering specific code suggestions and concrete fixes, although some students worried it might encourage overreliance on the AI. 
Students prioritizing task completion often preferred technical feedback, whereas those choosing Auto mode valued its dynamic adaptation, especially when unsure which style suited them.
% We also examined students' willingness to manually adjust the Agent's feedback mode. The average rating was 5.44, reflecting a strong desire for greater control. Many students expressed interest in selecting modes themselves, rather than being passively subject to instructor-driven adjustments, particularly when the system's responses did not meet their expectations. One student noted that in large classes, it is difficult for instructors to track individual Agent settings and suggested adding a request feature to allow students to formally apply for mode changes.
We also examined students' willingness to adjust the feedback mode manually. The average rating was 5.44, indicating strong interest in having more control. Many students expressed a desire to select modes themselves when the system's responses did not meet their expectations. One student noted that instructors may struggle to track individual settings in large classes and suggested adding a request feature to allow students to formally ask for mode changes.

\textit{Regarding the overall system evaluation}, all aspects received high scores except for personality, with friendliness rated highest (M = 5.70). Students found the interface clean, intuitive, and beginner-friendly. 
% \textit{One student commented: ``The interface is clean and easy to use, especially for someone like me who is new to data visualization. The layout of the AI assistant and its prompts were clear and helpful, which allowed me to quickly understand how to complete the tasks.''}
The lower rating for personality likely reflects the Agent's uniform and repetitive response style. 
Although students appreciated the content, the rigid formatting created a more robotic impression, which may have negatively influenced their perception of its personality.
% The lower personality rating likely reflects the Agent's uniform and repetitive responses, which created a robotic impression despite generally helpful content.

% 此外在后续的访谈中，我们也询问了学生关于我们系统和其他AI工具的使用体验比较（例如ChatGPT）。
% 大部分学生认为在课堂学习过程中使用ClassAid更好，而在急需快速得到答案时其他AI工具效率更高。
% 他们表示，在学习过程中需要启发式的、主动的反馈以便更好的理解并掌握知识，而不是直接获取答案，因此他们普遍认为ClassAid比较于其他AI工具更有教育意义。
% One student commented: ``学习过程中我更希望得到hint而不是answer。''
% 此外，他们表示当他们遇到困难的时候由于害羞等社会原因会使用AI工具，这些工具可以解决他们部分的问题，但是这些工具不受教师的监督，因此常常让教员无法及时了解他们的困惑和学习情况。
% 而ClassAid可以很好的平衡这一点，在提供AI反馈的同时，也受到教员的监督。
% 一个学生表示：当我遇到学习问题时，我并不是不想要让教员了解我的问题，而是向教员提出问题这个过程让我感觉害怕。
% 这表明ClassAid相比较于其他AI工具，在xxx上都xxxx

\RX{

\textit{In follow-up interviews,} we asked students to compare \textit{ClassAid} with other AI tools, such as ChatGPT. Eighty percent of the students reported that \textit{ClassAid} was better suited for classroom learning contexts, whereas general-purpose chatbots were more efficient when quick answers were needed. Students' comparisons primarily focused on several key dimensions.

\textit{Response time and usage efficiency.}
Most students noted that general chatbots had a clear advantage in providing immediate answers, making them suitable for situations with time pressure. However, students perceived this efficiency as often coming at the cost of reduced instructional guidance during the learning process.

\textit{Feedback form and analytical quality.}
Approximately 70\% of the students emphasized that during learning, they preferred heuristic or proactive feedback that supported understanding rather than direct solutions. Students generally perceived the feedback provided by \textit{ClassAid} as placing greater emphasis on reasoning processes and problem decomposition, which better supported conceptual understanding. As S1 noted, \textit{``During learning, I prefer hints rather than answers.''}

\textit{Learning support and instructor visibility.}
When encountering learning difficulties, more than half of the students (approximately 60\%) reported that they were more likely to turn to AI tools rather than directly approach instructors, primarily due to feelings of shyness or discomfort. Although general AI tools could address some immediate issues, multiple students pointed out that the lack of instructor oversight made it difficult for instructors to identify students' misconceptions or monitor their learning progress. In contrast, approximately 70\% of the students viewed \textit{ClassAid} as providing a better balance between learning support and accountability by offering AI assistance while remaining visible to instructors. S4 explained, \textit{``When I encounter learning problems, it is not that I do not want my instructor to know, but asking directly feels intimidating.''}

These findings suggest that compared with other AI tools, \textit{ClassAid} is uniquely positioned to support educational goals by offering pedagogically aligned feedback while maintaining transparency for instructors.

}

% The system's code analysis function received moderately positive feedback (M = 5.31). Some students noted its limited ability to recognize alternative but correct coding approaches.
% \textit{One student shared: ``The system often fails to recognize different correct ways of writing code. For example, when I used \texttt{'aggregate': 'mean'} instead of \texttt{'aggregate': 'average'}, it marked the code as incorrect, even though it executed properly and produced the correct chart.''}
% This suggests the current evaluation logic may be overly rigid, highlighting the need for improved flexibility and tolerance in future iterations.

\subsubsection{Feedback from Instructor Participants in the User Study}

The instructional team expressed highly positive feedback on \textit{ClassAid}, especially praising its real-time visualization of individual progress and classroom dynamics.
% The instructor noted that, unlike previous classes where student engagement was largely invisible, \textit{ClassAid} immediately revealed who had begun working. This real-time responsiveness was described as ``a surprising and impressive moment'' during initial use.
% The instructor noted that, unlike in previous classes where student engagement was difficult to observe, \textit{ClassAid} immediately revealed who had begun working. This real-time responsiveness was described as ``a surprising and impressive moment'' during initial use.
The instructor noted that, unlike in earlier classes where engagement was hard to monitor, \textit{ClassAid} instantly showed who had started working, which she described as ``a surprising and impressive moment.''

% By clearly displaying each student's status, task progress, and difficulties, the system significantly reduced the instructor's cognitive load, allowing her to focus on students who really needed support and respond to class-wide issues. 
% For example, in one session, an omitted explanation of the ``scheme'' concept led to widespread errors. The system quickly surfaced this trend, prompting the instructor to clarify it in a timely manner.

By clearly displaying each student's progress and difficulties, the system significantly reduced the instructor's cognitive load, enabling her to focus on students who required support and address class-wide issues more effectively. 
In user study, for instance, an omitted explanation of the ``scheme'' concept led to widespread errors. The system quickly identified this pattern, prompting the instructor to provide timely clarification.
% The system also helped the instructor form a clearer impression of students, especially those who consistently excelled or struggled. One particularly notable case involved a student unable to complete the task even in technical mode. Further investigation revealed the student had no prior experience with AI tools and lacked effective interaction strategies. The instructor personally guided the student on how to pose questions to the agent, and with support from both the instructor and the Agent, the student was eventually able to complete the task successfully.
The system also helped the instructor gain a clearer understanding of students, particularly those who consistently struggled. One notable case involved a student who was unable to complete the task even with technical mode enabled. Further investigation revealed that the student had no prior experience with AI tools and lacked effective strategies for interacting with the system. The instructor provided guidance on how to formulate questions for the Agent, and with support from both the instructor and the Agent, the student ultimately completed the task.

% Regarding feedback mode adjustment, the instructor emphasized that the system strengthened her pedagogical agency. In one instance, a student repeatedly asked the Agent for direct answers, prompting the instructor to switch their mode to silent. When the student inquired about the change, it opened space for a conversation on the value of independent thinking. This moment not only encouraged student reflection but also reinforced the instructor's role in shaping learning behaviors in AI-supported classrooms.

Regarding feedback mode adjustment, the instructor emphasized that the system enhanced her instructional agency. In one instance, a student repeatedly asked the Agent for direct answers, prompting the instructor to switch the student's mode to silent. When the student inquired about the change, it created an opportunity to discuss the value of independent thinking. This interaction not only encouraged student reflection but also reinforced the instructor's role in shaping learning behaviors in AI-supported classrooms.

% \vspace{-2mm}

\subsubsection{Feedback from Educators}

\RR{
% Based on a thematic analysis of interviews with eight educators~\cite{guest2011applied}, we found that they expressed a consistently positive attitude toward ClassAid, highlighting its controllability, flexibility, and the benefits of dynamic Agent orchestration. They believed the system helps align AI support with pedagogical goals, manage class progress more effectively, and reduce the burden of providing personalized feedback in programming-intensive settings. 
% E1 noted that by allowing real-time adjustments to the AI Agent, ClassAid reinforces the instructor's central role and remains highly controllable and adaptable.

Based on a thematic analysis of interviews with eight educators~\cite{guest2011applied}, we found that they expressed a broadly positive attitude toward \textit{ClassAid}, emphasizing its controllability, flexibility, and the benefits of dynamic orchestration. They believed the system helps align AI support with pedagogical goals, manage class progress more effectively, and reduce the burden of providing personalized feedback in programming-intensive settings. 
E1 noted that the ability to adjust the AI Agent in real time reinforces the instructor's central role and ensures that the system remains highly controllable and adaptable.

\textit{Feedback Quality.}
% Although both assistant and full professors expected high-quality feedback, their tolerance for the TA Agent's limitations differed. Assistant professors were generally more accepting of imperfect responses, whereas full professors showed lower tolerance. E1 noted that most students are familiar with tools such as ChatGPT and therefore understand that AI-generated feedback may contain inaccuracies. E6 added that instructors still need to remind students to treat the AI's responses as references rather than authoritative answers.
Although both assistant and full professors expected high-quality feedback, their tolerance for the TA Agent's limitations differed. Assistant professors were generally more tolerant of imperfect responses, whereas full professors showed lower tolerance. E1 noted that most students are familiar with tools such as ChatGPT and therefore understand that AI-generated feedback may contain inaccuracies. E6 added that instructors still need to remind students to view AI responses as references rather than authoritative answers.
However, E3 argued that relying solely on prompt engineering is insufficient for reliably distinguishing heuristic from technical feedback in Auto mode. He suggested using large-scale fine-tuning or retrieval-augmented generation to improve decision accuracy and ensure more appropriate guidance.
E8 further observed that Auto-mode decisions depend on the system's accumulated understanding of students' learning states. 
Early in a course, this information is limited, which may lead to unstable mode selection. He therefore recommended enabling Auto mode only after sufficient student data has been collected. 
Building on this perspective, E7 noted that expectations for feedback quality depend on the type of question. 
Errors are unacceptable for problems with standardized answers, whereas flexible feedback is appropriate for open-ended~problems.

\textit{System Usage.} In contrast, assistant professors focused more on \textit{ClassAid}'s visual monitoring capabilities.
E7 praised the layout and information presentation of the instructor dashboard, noting that thoughtful visualization design reduced the burden of classroom monitoring.
E6 observed that although the dashboard presents task scores clearly, richer behavioral information is scattered across multiple detailed views. This fragmentation makes it difficult for instructors to form a coherent understanding of students' learning processes and increases the effort needed to make fine-grained adjustments during class.
% E4 added that although the dashboard already provides a summary layer through bar charts of question types and error types, connecting these summaries more tightly with the other views is still challenging. For instance, while the system can show which students frequently ask for answers directly, identifying the specific content areas where this behavior occurs still requires instructors to explore manually. These observations suggest a need for better integration between the existing summary layer and the detailed behavioral views, helping instructors interpret student progress more efficiently and adjust the Agent accordingly.
E4 added that although the dashboard includes a summary layer through bar charts of question and error types, integrating these summaries with the other views remains challenging. For example, while the system can highlight students who frequently request direct answers, identifying and summarizing the specific questions they posed still requires manual exploration.
These observations suggest a need for tighter integration between summary-level analytics and detailed behavioral views to help instructors interpret student progress more efficiently and adjust the Agent accordingly.

\textit{Agent Control.} 
% Regarding the low frequency of student feedback on the TA Agent, instructors were not surprised. They noted that students naturally focus on completing tasks rather than evaluating AI responses during class. E2 suggested shifting attention toward the Agent's ability to evaluate its own effectiveness by analyzing student code behavior, which may offer a more reliable supervisory mechanism. E5 further proposed examining the relationship between student code and AI feedback as an indirect way to assess the TA Agent's performance.
% Instructors were unsurprised by the low frequency of student feedback on the TA Agent, noting that students tend to prioritize task completion. E2 suggested shifting attention toward the Agent's ability to evaluate its own effectiveness through student code behavior, which may offer a more reliable supervisory mechanism. E5 recommended using the relationship between code and feedback as an indirect performance indicator.
Instructors were unsurprised by the low frequency of student feedback on the TA Agent, noting that students tend to prioritize task completion. E2 suggested focusing instead on the Agent's ability to assess its own effectiveness through students' behavior, which may provide a more reliable supervisory mechanism. E5 further recommended using the relationship between code and feedback as an indirect indicator of the Agent's performance.
% Based on these concerns, we proposed allowing students limited control over their own Agent mode. E3 responded cautiously, expressing concern that students might misuse this functionality, such as by repeatedly selecting the technical mode to obtain final answers. He suggested that usage restrictions would be necessary if such a feature were to be implemented.
When we proposed giving students limited control over their Agent mode, E3 raised concerns about potential misuse, such as repeatedly choosing technical mode to obtain answers. He suggested that usage restrictions would be necessary if the feature were implemented.
}

\section{Discussion}

\subsection{Design Implications}
\textbf{Structured Design of Multi-stage Pedagogical Agents.}
\textit{ClassAid} introduces a six-stage intelligent pedagogical agent grounded in formative and dynamic assessment theories~\cite{black2009developing,minick1987implications}, as shown in Fig.~\ref{fig:s-agent}.
By abstracting how human instructors provide classroom feedback, the design aligns with the generative process of LLMs. 
This approach improves feedback quality and enhances the interpretability of what is typically a 'black box' in LLM-based responses~\cite{khosravi2022explainable}. 
Our findings highlight the value of refining and modularizing the individual levels of such agents. In the current design, only the \textit{Consider} level allows instructor intervention, while the remaining levels follow fixed rules. 
Expanding configurability across multiple levels could give instructors finer control over how feedback is generated, including how student issues are identified, reviewed, and selected, thereby supporting more precise management of pacing and granularity.
Such flexibility enables more personalized and context-sensitive learning experiences rather than relying on a uniform agent design.

\textbf{Real-Time Class-Level Oversight through Feedback Monitoring.}
% While recent research has explored student behavior tracking and feedback generation~\cite{tang2024sphere}, there remains a lack of mechanisms for aggregating these behaviors into actionable class-level insights, particularly in live, in-person settings~\cite{meyer2024using}. 
While recent research has explored student behavior tracking and feedback generation~\cite{tang2024sphere}, mechanisms for aggregating these behaviors into actionable class-level insights remain limited, especially in in-person classroom settings~\cite{meyer2024using}. 
Our observations suggest that summarizing irregular student behaviors into interpretable visual feedback can help educators quickly identify issues and intervene in a timely manner. 
In our implementation, these summaries supported monitoring at both the individual and class levels. 
% Future systems may extend this approach by incorporating rich, multi-layered visualizations that capture behavioral patterns across different levels of granularity and temporal scope, supporting more informed and responsive classroom decision-making.
Future systems may extend this approach by incorporating multi-layered visualizations that capture behavioral patterns across varying degrees of granularity and time, enabling more informed and responsive classroom decision-making.

\textbf{Instructor-AI-Student Triangular Supervision of AI Agents.}
Although the potential of LLMs in education is widely acknowledged~\cite{park2024promise}, their real-world deployment requires external oversight to ensure appropriate and accountable use~\cite{williamson2024time}. 
One promising direction is a triangular supervision framework in which instructors and students jointly monitor and regulate the AI agent's feedback.
Our implementation shows that such collaborative oversight can enhance transparency and foster trust in AI-supported instruction. Looking ahead, systems might incorporate self-supervised capabilities that allow AI agents to assess the effectiveness of their feedback based on student responses and adapt in real time to improve in-class performance.

\subsection{Future Work}
\textbf{Fine-Grained Feedback for Personalized Learning Support.}
By integrating heuristic and technical feedback modes, \textit{ClassAid} has shown promise in reducing students' overreliance on AI and limiting direct answer-seeking behavior.
However, as students' needs become more nuanced, current feedback strategies remain too coarse-grained to fully support scaffolded learning pathways grounded in theories such as the Zone of Proximal Development~\cite{shabani2010vygotsky} and instructional scaffolding~\cite{belland2017instructional}. 
Future work should explore fine-grained learner modeling to support tiered TA Agent responses. For example, within the technical feedback mode, systems should distinguish among sample code, pseudocode, and full solutions, allowing the depth of feedback to adapt dynamically to students' proficiency levels~\cite{kazemitabaar2024codeaid}. 
% Moreover, different instructional tasks entail varying pedagogical objectives, underscoring the need for aligning feedback content with specific learning goals.
Additionally, instructional tasks vary in their pedagogical objectives, underscoring the need to align feedback content with specific learning goals.

% \textbf{Enhancing Instructor Control in Proactive Feedback.}
% % Although the current system can automatically generate proactive feedback through observation-based triggers, this process remains heavily dependent on LLM inference and rule-based logic, resulting in limited stability and adaptability. Future iterations should incorporate instructor intervention mechanisms, allowing teachers to flexibly adjust or override proactive feedback while preserving the benefits of automated triggering. This human-in-the-loop design not only enables greater personalization but also reinforces the instructor's agency and control within the feedback process.
% Although the current system can automatically generate proactive feedback through observation-based triggers, the process remains heavily dependent on LLM inference and rule-based logic, which limits stability and adaptability\RR{~\cite{}}. Future iterations should incorporate instructor intervention mechanisms, allowing instructors to adjust, approve, or override AI-generated feedback while retaining the benefits of automated detection. For example, instructors could control the timing of feedback delivery, such as delaying or pausing feedback during lectures, or selectively disable certain types of automated triggers that may not align with specific instructional goals. Such a human-in-the-loop design would not only enhance personalization but also reinforce instructors' agency and control within the feedback process.

% 可以在这里补充一点--1204

\textbf{Exploring Shared Control Mechanisms Between Instructors and Students.}
% 在现有的规模中，教员并没有提到工作负担的问题。但是很显然在更大规模的 classroom settings, 
% 由于学生的增加，教师的workload会骤然增加，因为他需要去manage all TA Agent modes for all students in a timely and accurate manner，不及时的调整potentially compromising learning experiences and 个性化 feedback quality. （人多了会增加调整负担）
% At the same time, students who are active agents in their own learning, may benefit from limited control over the agent's behavior~\cite{yang2023pair,borchers2025learner}. 
% However, as Holstein and Olsen note~\cite{holstein2023human}, 如果granting students full control could 更会 increase instructors' cognitive load. （其实就是会增加监督负担）
% Students also vary widely in their capacity to manage such control effectively, depending on their educational background and prior experience with AI-supported environments~\cite{lawrence2024teachers}.
% A more balanced approach would allow students to request control permissions or make a limited number of feedback adjustments, supporting autonomy while maintaining necessary instructor oversight，或者增强Agent的自监督机制。
\RR{Managing TA Agent modes becomes increasingly burdensome in larger classroom settings, where untimely adjustments may compromise both learning experiences and the quality of personalized feedback.
Students who actively regulate their learning may benefit from limited control over the agent~\cite{yang2023pair,borchers2025learner}, but granting full control could raise instructors' supervisory load~\cite{holstein2023human}.}
Students also vary widely in their capacity to manage such control effectively, depending on their educational background and prior experience with AI-supported environments~\cite{lawrence2024teachers}. A more balanced approach would allow students to request permissions or make a small number of adjustments, or rely on stronger self-supervision mechanisms within the agent, thereby supporting autonomy without increasing instructor workload.

\textbf{Integrating Multi-Source Learning Data to Broaden System Applicability.}
The future development of TA Agents should extend beyond prompt engineering by incorporating diverse learning data such as student history, behavioral traces, and performance patterns~\cite{ye2022using}.
Techniques such as fine-tuning and retrieval-augmented generation (RAG) can further enhance the agent's ability to interpret educational contexts~\cite{lewis2020retrieval}.
To support a broader range of courses, a modular configuration system is also needed, one that enables automatic updates to knowledge and prompt libraries based on course content. 
Finally, allowing instructors to fine-tune the agent's feedback according to specific learning objectives will enhance the relevance and flexibility of AI support across diverse classroom environments.

\subsection{Limitation}
% \textbf{Limited Supervision from Students and Instructors Constrains the Evaluation of AI Feedback.}
% Since students primarily focus on task completion rather than evaluating system feedback, resulting in limited subjective assessments of the TA Agent's responses. 
% The scarcity of feedback hinders a comprehensive evaluation of the Agent's effectiveness. 
% While instructors could access student–agent interaction logs, limited time and cognitive resources made it difficult to review all exchanges in detail. These constraints collectively limit current supervision mechanisms. Future work could explore auxiliary supervisory agents to help instructors identify and filter abnormal feedback or student behavior, thereby improving oversight efficiency and scalability.

\RR{

\textit{Scalability}
The current deployment involved 54 students in a single session, but larger-scale or extended deployments may introduce additional challenges. As classroom size increases, system response time, server load, and the instructor's capacity to monitor more student–AI interactions may become bottlenecks. Managing TA Agent modes for many students could also increase instructor workload and reduce the timeliness of pedagogically meaningful adjustments. Multi-session and higher-enrollment studies are needed to evaluate scalability and identify potential constraints in long-term or institution-wide use.

\textit{Generalizability}
Our study examined how the TA Agent supported students in completing relatively simple Vega-Lite tasks. Thus, it remains unclear how well the system would generalize to more complex programming environments, diverse toolchains, or other pedagogical contexts. The study was also limited to a single in-person session with one student group and no longitudinal observation, restricting insights into long-term stability and broader applicability. Although the deployment focused on an in-person setting, future work should explore online or hybrid classrooms, where real-time orchestration may be especially valuable. Broader evaluations across subject areas, task complexity, teaching modalities, and learner profiles are needed to assess the system's generalizability.
}
\section{Conclusion}
This paper introduces \textit{ClassAid}, a real-time classroom orchestration system that integrates an intelligent TA Agent and an instructor dashboard to facilitate responsive feedback and instructor-AI collaboration. At its core is a six-stage framework that enables the TA Agent to monitor student behavior, infer cognitive states, select feedback strategies, and deliver targeted interventions to support metacognitive development.
We deployed \textit{ClassAid} in a graduate-level programming course (n = 54) and found that the TA Agent provided accurate and personalized feedback, while the instructor dashboard enabled real-time oversight and dynamic control of feedback modes. Together, these features supported instructors in maintaining pedagogical authority, addressing emerging learning needs, and reducing the cognitive load of managing large classrooms.
Quantitative results demonstrate the TA Agent's strong performance in feedback generation and decision-making, and qualitative feedback from students and educators highlights the system's usability, instructional value, and potential for responsible AI integration.
Overall, this study proposes a novel instructor-AI collaboration model for programming classrooms, offering a scalable, adaptive solution to the challenges of real-time feedback and classroom orchestration in the era of generative AI.

\begin{acks}
Guodao Sun and Meng Xia are the corresponding authors.
The work was supported by the National Natural Science Foundation of China, (62422607, 62372411, 62432014) and the Zhejiang Provincial Natural Science Foundation of China (QKWL25F0301).

\end{acks}

\bibliographystyle{ACM-Reference-Format}
\bibliography{CAref}

@article{minick1987implications,
  title={Implications of Vygotsky's theories for dynamic assessment.},
  author={Minick, Norris},
  year={1987},
  publisher={The Guilford Press}
}

@article{sperling2004metacognition,
  title={Metacognition and self-regulated learning constructs},
  author={Sperling, Rayne A and Howard, Bruce C and Staley, Richard and DuBois, Nelson},
  journal={Educational research and evaluation},
  volume={10},
  number={2},
  pages={117--139},
  year={2004},
  publisher={Taylor \& Francis}
}

@article{noroozi2018promoting,
  title={Promoting argumentation competence: Extending from first-to second-order scaffolding through adaptive fading},
  author={Noroozi, Omid and Kirschner, Paul A and Biemans, Harm JA and Mulder, Martin},
  journal={Educational psychology review},
  volume={30},
  number={1},
  pages={153--176},
  year={2018},
  publisher={Springer}
}

@article{benson2011socratic,
  title={Socratic method},
  author={Benson, Hugh H},
  journal={The Cambridge companion to socrates},
  pages={179--200},
  year={2011},
  publisher={Cambridge University Press Cambridge}
}

@book{carnine1997direct,
  title={Direct instruction reading},
  author={Carnine, Douglas and Silbert, Jerry and Kameenui, Edward J and Tarver, Sara G},
  year={1997},
  publisher={Merrill Columbus, OH}
}

@article{hein1991constructivist,
  title={Constructivist learning theory},
  author={Hein, George E},
  journal={Institute for Inquiry},
  volume={14},
  year={1991}
}

@inproceedings{zhang2023vizprog,
  title={Vizprog: Identifying misunderstandings by visualizing students’ coding progress},
  author={Zhang, Ashley Ge and Chen, Yan and Oney, Steve},
  booktitle={Proceedings of the 2023 CHI Conference on Human Factors in Computing Systems},
  pages={1--16},
  year={2023}
}

@article{hmelo2004problem,
  title={Problem-based learning: What and how do students learn?},
  author={Hmelo-Silver, Cindy E},
  journal={Educational psychology review},
  volume={16},
  number={3},
  pages={235--266},
  year={2004},
  publisher={Springer}
}

@article{wang2021puzzleme,
  title={Puzzleme: Leveraging peer assessment for in-class programming exercises},
  author={Wang, April Yi and Chen, Yan and Chung, John Joon Young and Brooks, Christopher and Oney, Steve},
  journal={Proceedings of the ACM on Human-Computer Interaction},
  volume={5},
  number={CSCW2},
  pages={1--24},
  year={2021},
  publisher={ACM New York, NY, USA}
}

@inproceedings{kim2015rimes,
  title={RIMES: Embedding interactive multimedia exercises in lecture videos},
  author={Kim, Juho and Glassman, Elena L and Monroy-Hern{\'a}ndez, Andr{\'e}s and Morris, Meredith Ringel},
  booktitle={Proceedings of the 33rd annual ACM conference on human factors in computing systems},
  pages={1535--1544},
  year={2015}
}

@article{zheng2022exploratory,
  title={An exploratory study on fade-in versus fade-out scaffolding for novice programmers in online collaborative programming settings},
  author={Zheng, Lanqin and Zhen, Yuanyi and Niu, Jiayu and Zhong, Lu},
  journal={Journal of Computing in Higher Education},
  volume={34},
  number={2},
  pages={489--516},
  year={2022},
  publisher={Springer}
}

@inproceedings{liffiton2023codehelp,
  title={Codehelp: Using large language models with guardrails for scalable support in programming classes},
  author={Liffiton, Mark and Sheese, Brad E and Savelka, Jaromir and Denny, Paul},
  booktitle={Proceedings of the 23rd Koli Calling International Conference on Computing Education Research},
  pages={1--11},
  year={2023}
}

@article{black2009developing,
  title={Developing the theory of formative assessment},
  author={Black, Paul and Wiliam, Dylan},
  journal={Educational Assessment, Evaluation and Accountability (formerly: Journal of personnel evaluation in education)},
  volume={21},
  pages={5--31},
  year={2009},
  publisher={Springer}
}

@inproceedings{tan2024more,
  title={More than model documentation: uncovering teachers' bespoke information needs for informed classroom integration of ChatGPT},
  author={Tan, Mei and Subramonyam, Hari},
  booktitle={Proceedings of the 2024 CHI Conference on Human Factors in Computing Systems},
  pages={1--19},
  year={2024}
}

@incollection{holstein2023human,
  title={Human-AI co-orchestration: the role of artificial intelligence in orchestration},
  author={Holstein, Ken and Olsen, Jennifer K},
  booktitle={Handbook of artificial intelligence in education},
  pages={309--321},
  year={2023},
  publisher={Edward Elgar Publishing}
}

@article{harber1998feedback,
  title={Feedback to minorities: Evidence of a positive bias.},
  author={Harber, Kent D},
  journal={Journal of personality and social psychology},
  volume={74},
  number={3},
  pages={622},
  year={1998},
  publisher={American Psychological Association}
}

@book{moore2014effective,
  title={Effective instructional strategies: From theory to practice},
  author={Moore, Kenneth D},
  year={2014},
  publisher={Sage Publications}
}

@inproceedings{park2024promise,
  title={The promise and peril of ChatGPT in higher education: opportunities, challenges, and design implications},
  author={Park, Hyanghee and Ahn, Daehwan},
  booktitle={Proceedings of the 2024 CHI Conference on Human Factors in Computing Systems},
  pages={1--21},
  year={2024}
}

@article{meyer2024using,
  title={Using LLMs to bring evidence-based feedback into the classroom: AI-generated feedback increases secondary students’ text revision, motivation, and positive emotions},
  author={Meyer, Jennifer and Jansen, Thorben and Schiller, Ronja and Liebenow, Lucas W and Steinbach, Marlene and Horbach, Andrea and Fleckenstein, Johanna},
  journal={Computers and Education: Artificial Intelligence},
  volume={6},
  pages={100199},
  year={2024},
  publisher={Elsevier}
}

@inproceedings{borchers2025learner,
  title={How Learner Control and Explainable Learning Analytics About Skill Mastery Shape Student Desires to Finish and Avoid Loss in Tutored Practice},
  author={Borchers, Conrad and Ooge, Jeroen and Peng, Cindy and Aleven, Vincent},
  booktitle={Proceedings of the 15th International Learning Analytics and Knowledge Conference},
  pages={810--816},
  year={2025}
}

@article{lewis2020retrieval,
  title={Retrieval-augmented generation for knowledge-intensive nlp tasks},
  author={Lewis, Patrick and Perez, Ethan and Piktus, Aleksandra and Petroni, Fabio and Karpukhin, Vladimir and Goyal, Naman and K{\"u}ttler, Heinrich and Lewis, Mike and Yih, Wen-tau and Rockt{\"a}schel, Tim and others},
  journal={Advances in neural information processing systems},
  volume={33},
  pages={9459--9474},
  year={2020}
}

@book{solso2005cognitive,
  title={Cognitive psychology},
  author={Solso, Robert L and MacLin, M Kimberly and MacLin, Otto H},
  year={2005},
  publisher={Pearson Education New Zealand}
}

@article{ye2022using,
  title={Using trace data to enhance Students' self-regulation: A learning analytics perspective},
  author={Ye, Dan and Pennisi, Svoboda},
  journal={The Internet and Higher Education},
  volume={54},
  pages={100855},
  year={2022},
  publisher={Elsevier}
}

@article{khosravi2022explainable,
  title={Explainable artificial intelligence in education},
  author={Khosravi, Hassan and Shum, Simon Buckingham and Chen, Guanliang and Conati, Cristina and Tsai, Yi-Shan and Kay, Judy and Knight, Simon and Martinez-Maldonado, Roberto and Sadiq, Shazia and Gasevic, Dragan},
  journal={Computers and education: artificial intelligence},
  volume={3},
  pages={100074},
  year={2022},
  publisher={Elsevier}
}

@article{mogavi2024chatgpt,
  title={ChatGPT in education: A blessing or a curse? A qualitative study exploring early adopters’ utilization and perceptions},
  author={Mogavi, Reza Hadi and Deng, Chao and Kim, Justin Juho and Zhou, Pengyuan and Kwon, Young D and Metwally, Ahmed Hosny Saleh and Tlili, Ahmed and Bassanelli, Simone and Bucchiarone, Antonio and Gujar, Sujit and others},
  journal={Computers in Human Behavior: Artificial Humans},
  volume={2},
  number={1},
  pages={100027},
  year={2024},
  publisher={Elsevier}
}

@article{beck2023backtalk,
  title={Backtalk: ChatGPT: A powerful technology tool for writing instruction},
  author={Beck, Sarah W and Levine, Sarah R},
  journal={Phi Delta Kappan},
  volume={105},
  number={1},
  pages={66--67},
  year={2023},
  publisher={SAGE Publications Sage CA: Los Angeles, CA}
}

@inproceedings{yang2021surveying,
  title={Surveying teachers’ preferences and boundaries regarding human-AI control in dynamic pairing of students for collaborative learning},
  author={Yang, Kexin Bella and Lawrence, LuEttaMae and Echeverria, Vanessa and Guo, Boyuan and Rummel, Nikol and Aleven, Vincent},
  booktitle={Technology-Enhanced Learning for a Free, Safe, and Sustainable World: 16th European Conference on Technology Enhanced Learning, EC-TEL 2021, Bolzano, Italy, September 20-24, 2021, Proceedings 16},
  pages={260--274},
  year={2021},
  organization={Springer}
}

@article{lawrence2024teachers,
  title={How teachers conceptualise shared control with an AI co-orchestration tool: A multiyear teacher-centred design process},
  author={Lawrence, LuEttaMae and Echeverria, Vanessa and Yang, Kexin and Aleven, Vincent and Rummel, Nikol},
  journal={British Journal of Educational Technology},
  volume={55},
  number={3},
  pages={823--844},
  year={2024},
  publisher={Wiley Online Library}
}

@book{belland2017instructional,
  title={Instructional scaffolding in STEM education: Strategies and efficacy evidence},
  author={Belland, Brian R},
  year={2017},
  publisher={Springer Nature}
}

@article{shabani2010vygotsky,
  title={Vygotsky's zone of proximal development: Instructional implications and teachers' professional development.},
  author={Shabani, Karim and Khatib, Mohamad and Ebadi, Saman},
  journal={English language teaching},
  volume={3},
  number={4},
  pages={237--248},
  year={2010},
  publisher={ERIC}
}

@article{yan2024practical,
  title={Practical and ethical challenges of large language models in education: A systematic scoping review},
  author={Yan, Lixiang and Sha, Lele and Zhao, Linxuan and Li, Yuheng and Martinez-Maldonado, Roberto and Chen, Guanliang and Li, Xinyu and Jin, Yueqiao and Ga{\v{s}}evi{\'c}, Dragan},
  journal={British Journal of Educational Technology},
  volume={55},
  number={1},
  pages={90--112},
  year={2024},
  publisher={Wiley Online Library}
}

@article{williamson2024time,
  title={Time for a Pause: Without Effective Public Oversight, AI in Schools Will Do More Harm Than Good.},
  author={Williamson, Ben and Molnar, Alex and Boninger, Faith},
  journal={Commercialism in Education Research Unit},
  year={2024},
  publisher={ERIC}
}

@book{given2008sage,
  title={The Sage encyclopedia of qualitative research methods},
  author={Given, Lisa M},
  year={2008},
  publisher={Sage publications}
}

@article{lan2024teachers,
  title={Teachers’ agency in the era of LLM and generative AI},
  author={Lan, Yu-Ju and Chen, Nian-Shing},
  journal={Educational Technology \& Society},
  volume={27},
  number={1},
  pages={I--XVIII},
  year={2024},
  publisher={JSTOR}
}

@article{pu2025assistance,
  title={Assistance or Disruption? Exploring and Evaluating the Design and Trade-offs of Proactive AI Programming Support},
  author={Pu, Kevin and Lazaro, Daniel and Arawjo, Ian and Xia, Haijun and Xiao, Ziang and Grossman, Tovi and Chen, Yan},
  journal={arXiv preprint arXiv:2502.18658},
  year={2025}
}

@misc{vega-lite,
  author = {Satyanarayan, Arvind and Moritz, Dominik and Wongsuphasawat, Kanit and Heer, Jeffrey},
  title = {{Vega-Lite}: A Grammar of Interactive Graphics},
  howpublished = {\url{https://vega.github.io/vega-lite/}},
  year = {2017},
  note = {Accessed: [Insert date you accessed the site]}
}

@article{kasneci2023chatgpt,
  title={ChatGPT for good? On opportunities and challenges of large language models for education},
  author={Kasneci, Enkelejda and Se{\ss}ler, Kathrin and K{\"u}chemann, Stefan and Bannert, Maria and Dementieva, Daryna and Fischer, Frank and Gasser, Urs and Groh, Georg and G{\"u}nnemann, Stephan and H{\"u}llermeier, Eyke and others},
  journal={Learning and individual differences},
  volume={103},
  pages={102274},
  year={2023},
  publisher={Elsevier}
}

@article{forehand2010bloom,
  title={Bloom’s taxonomy},
  author={Forehand, Mary},
  journal={Emerging perspectives on learning, teaching, and technology},
  volume={41},
  number={4},
  pages={47--56},
  year={2010}
}

@inproceedings{head2017writing,
  title={Writing reusable code feedback at scale with mixed-initiative program synthesis},
  author={Head, Andrew and Glassman, Elena and Soares, Gustavo and Suzuki, Ryo and Figueredo, Lucas and D'Antoni, Loris and Hartmann, Bj{\"o}rn},
  booktitle={Proceedings of the Fourth (2017) ACM Conference on Learning@ Scale},
  pages={89--98},
  year={2017}
}

@inproceedings{guo2015codeopticon,
  title={Codeopticon: Real-time, one-to-many human tutoring for computer programming},
  author={Guo, Philip J},
  booktitle={Proceedings of the 28th Annual ACM Symposium on User Interface Software \& Technology},
  pages={599--608},
  year={2015}
}

@article{glassman2015overcode,
  title={OverCode: Visualizing variation in student solutions to programming problems at scale},
  author={Glassman, Elena L and Scott, Jeremy and Singh, Rishabh and Guo, Philip J and Miller, Robert C},
  journal={ACM Transactions on Computer-Human Interaction (TOCHI)},
  volume={22},
  number={2},
  pages={1--35},
  year={2015},
  publisher={ACM New York, NY, USA}
}

@article{harvey2025don,
  title={" Don't Forget the Teachers": Towards an Educator-Centered Understanding of Harms from Large Language Models in Education},
  author={Harvey, Emma and Koenecke, Allison and Kizilcec, Rene F},
  journal={arXiv preprint arXiv:2502.14592},
  year={2025}
}

@article{chen2024need,
  title={Need Help? Designing Proactive AI Assistants for Programming},
  author={Chen, Valerie and Zhu, Alan and Zhao, Sebastian and Mozannar, Hussein and Sontag, David and Talwalkar, Ameet},
  journal={arXiv preprint arXiv:2410.04596},
  year={2024}
}

@article{dakhel2023github,
  title={Github copilot ai pair programmer: Asset or liability?},
  author={Dakhel, Arghavan Moradi and Majdinasab, Vahid and Nikanjam, Amin and Khomh, Foutse and Desmarais, Michel C and Jiang, Zhen Ming Jack},
  journal={Journal of Systems and Software},
  volume={203},
  pages={111734},
  year={2023},
  publisher={Elsevier}
}

@inproceedings{moore2024teaching,
  title={Teaching artificial intelligence in extracurricular contexts through narrative-based learnersourcing},
  author={Moore, Dylan Edward and Moore, Sophia RR and Ireen, Bansharee and Iskandar, Winston P and Artazyan, Grigory and Murnane, Elizabeth L},
  booktitle={Proceedings of the 2024 CHI Conference on Human Factors in Computing Systems},
  pages={1--28},
  year={2024}
}

@article{ji2023survey,
  title={Survey of hallucination in natural language generation},
  author={Ji, Ziwei and Lee, Nayeon and Frieske, Rita and Yu, Tiezheng and Su, Dan and Xu, Yan and Ishii, Etsuko and Bang, Ye Jin and Madotto, Andrea and Fung, Pascale},
  journal={ACM computing surveys},
  volume={55},
  number={12},
  pages={1--38},
  year={2023},
  publisher={ACM New York, NY}
}

@article{denny2023promptly,
  title={Promptly: Using prompt problems to teach learners how to effectively utilize ai code generators},
  author={Denny, Paul and Leinonen, Juho and Prather, James and Luxton-Reilly, Andrew and Amarouche, Thezyrie and Becker, Brett A and Reeves, Brent N},
  journal={arXiv preprint arXiv:2307.16364},
  year={2023}
}

@inproceedings{woodrow2024ai,
  title={Ai teaches the art of elegant coding: Timely, fair, and helpful style feedback in a global course},
  author={Woodrow, Juliette and Malik, Ali and Piech, Chris},
  booktitle={Proceedings of the 55th ACM Technical Symposium on Computer Science Education V. 1},
  pages={1442--1448},
  year={2024}
}

@article{neyem2024towards,
  title={Towards an AI knowledge assistant for context-aware learning experiences in software capstone project development},
  author={Neyem, Andr{\'e}s and Gonz{\'a}lez, Luis A and Mendoza, Marcelo and Alcocer, Juan Pablo Sandoval and Centellas, Leonardo and Paredes, Carlos},
  journal={IEEE Transactions on Learning Technologies},
  year={2024},
  publisher={IEEE}
}

@inproceedings{chen2024learning,
  title={Learning agent-based modeling with LLM companions: Experiences of novices and experts using ChatGPT \& NetLogo chat},
  author={Chen, John and Lu, Xi and Du, Yuzhou and Rejtig, Michael and Bagley, Ruth and Horn, Mike and Wilensky, Uri},
  booktitle={Proceedings of the 2024 CHI Conference on Human Factors in Computing Systems},
  pages={1--18},
  year={2024}
}

@inproceedings{lyu2024evaluating,
  title={Evaluating the effectiveness of llms in introductory computer science education: A semester-long field study},
  author={Lyu, Wenhan and Wang, Yimeng and Chung, Tingting and Sun, Yifan and Zhang, Yixuan},
  booktitle={Proceedings of the Eleventh ACM Conference on Learning@ Scale},
  pages={63--74},
  year={2024}
}

@article{liao2024scaffolding,
  title={Scaffolding computational thinking with ChatGPT},
  author={Liao, Jian and Zhong, Linrong and Zhe, Longting and Xu, Handan and Liu, Ming and Xie, Tao},
  journal={IEEE Transactions on Learning Technologies},
  year={2024},
  publisher={IEEE}
}

@inproceedings{kuramitsu2023kogi,
  title={Kogi: A seamless integration of ChatGPT into Jupyter environments for programming education},
  author={Kuramitsu, Kimio and Obara, Yui and Sato, Miyu and Obara, Momoka},
  booktitle={Proceedings of the 2023 ACM SIGPLAN International Symposium on SPLASH-E},
  pages={50--59},
  year={2023}
}

@inproceedings{jacobs2024evaluating,
  title={Evaluating the application of large language models to generate feedback in programming education},
  author={Jacobs, Sven and Jaschke, Steffen},
  booktitle={2024 IEEE Global Engineering Education Conference (EDUCON)},
  pages={1--5},
  year={2024},
  organization={IEEE}
}

@inproceedings{hou2024codetailor,
  title={Codetailor: Llm-powered personalized parsons puzzles for engaging support while learning programming},
  author={Hou, Xinying and Wu, Zihan and Wang, Xu and Ericson, Barbara J},
  booktitle={Proceedings of the Eleventh ACM Conference on Learning@ Scale},
  pages={51--62},
  year={2024}
}

@inproceedings{jury2024evaluating,
  title={Evaluating llm-generated worked examples in an introductory programming course},
  author={Jury, Breanna and Lorusso, Angela and Leinonen, Juho and Denny, Paul and Luxton-Reilly, Andrew},
  booktitle={Proceedings of the 26th Australasian computing education conference},
  pages={77--86},
  year={2024}
}

@inproceedings{liu2024teaching,
  title={Teaching CS50 with AI: leveraging generative artificial intelligence in computer science education},
  author={Liu, Rongxin and Zenke, Carter and Liu, Charlie and Holmes, Andrew and Thornton, Patrick and Malan, David J},
  booktitle={Proceedings of the 55th ACM technical symposium on computer science education V. 1},
  pages={750--756},
  year={2024}
}

@article{zhang2025cpvis,
  title={CPVis: Evidence-based Multimodal Learning Analytics for Evaluation in Collaborative Programming},
  author={Zhang, Gefei and Ji, Shenming and Li, Yicao and Tang, Jingwei and Ding, Jihong and Xia, Meng and Sun, Guodao and Liang, Ronghua},
  journal={arXiv preprint arXiv:2502.17835},
  year={2025}
}

@inproceedings{kazemitabaar2024codeaid,
  title={Codeaid: Evaluating a classroom deployment of an llm-based programming assistant that balances student and educator needs},
  author={Kazemitabaar, Majeed and Ye, Runlong and Wang, Xiaoning and Henley, Austin Zachary and Denny, Paul and Craig, Michelle and Grossman, Tovi},
  booktitle={Proceedings of the 2024 chi conference on human factors in computing systems},
  pages={1--20},
  year={2024}
}

@article{liu2024proactive,
  title={Proactive Conversational Agents with Inner Thoughts},
  author={Liu, Xingyu Bruce and Fang, Shitao and Shi, Weiyan and Wu, Chien-Sheng and Igarashi, Takeo and Chen, Xiang Anthony},
  journal={arXiv preprint arXiv:2501.00383},
  year={2024}
}

@article{chen2024stugptviz,
  title={StuGPTViz: A Visual Analytics Approach to Understand Student-ChatGPT Interactions},
  author={Chen, Zixin and Wang, Jiachen and Xia, Meng and Shigyo, Kento and Liu, Dingdong and Zhang, Rong and Qu, Huamin},
  journal={IEEE Transactions on Visualization and Computer Graphics},
  year={2024},
  publisher={IEEE}
}

@article{tang2024sphere,
  title={SPHERE: Scaling Personalized Feedback in Programming Classrooms with Structured Review of LLM Outputs},
  author={Tang, Xiaohang and Wong, Sam and Huynh, Marcus and He, Zicheng and Yang, Yalong and Chen, Yan},
  journal={arXiv preprint arXiv:2410.16513},
  year={2024}
}

@incollection{vanlehn2018can,
  title={How can FACT encourage collaboration and self-correction?},
  author={VanLehn, Kurt and Burkhardt, Hugh and Cheema, Salman and Pead, Daniel and Schoenfeld, Alan and Wetzel, Jon},
  booktitle={Deep Comprehension},
  pages={114--127},
  year={2018},
  publisher={Routledge}
}

@inproceedings{yang2023pair,
  title={Pair-up: prototyping human-AI co-orchestration of dynamic transitions between individual and collaborative learning in the classroom},
  author={Yang, Kexin Bella and Echeverria, Vanessa and Lu, Zijing and Mao, Hongyu and Holstein, Kenneth and Rummel, Nikol and Aleven, Vincent},
  booktitle={Proceedings of the 2023 CHI conference on human factors in computing systems},
  pages={1--17},
  year={2023}
}

@inproceedings{kumar2023bridging,
  title={Bridging the gap in ai-driven workflows: The case for domain-specific generative bots},
  author={Kumar, Akit and Devi, MS Lakshmi and Saltz, Jeffrey S},
  booktitle={2023 IEEE International Conference on Big Data (BigData)},
  pages={2421--2430},
  year={2023},
  organization={IEEE}
}

@inproceedings{holstein2019designing,
  title={Designing for complementarity: Teacher and student needs for orchestration support in AI-enhanced classrooms},
  author={Holstein, Kenneth and McLaren, Bruce M and Aleven, Vincent},
  booktitle={Artificial Intelligence in Education: 20th International Conference, AIED 2019, Chicago, IL, USA, June 25-29, 2019, Proceedings, Part I 20},
  pages={157--171},
  year={2019},
  organization={Springer}
}

@article{sweller1988cognitive,
  title={Cognitive load during problem solving: Effects on learning},
  author={Sweller, John},
  journal={Cognitive science},
  volume={12},
  number={2},
  pages={257--285},
  year={1988},
  publisher={Elsevier}
}

@book{guest2011applied,
  title={Applied thematic analysis},
  author={Guest, Greg and MacQueen, Kathleen M and Namey, Emily E},
  year={2011},
  publisher={sage publications}
}

@article{hattie2007power,
  title={The power of feedback},
  author={Hattie, John and Timperley, Helen},
  journal={Review of educational research},
  volume={77},
  number={1},
  pages={81--112},
  year={2007},
  publisher={Sage Publications Sage CA: Thousand Oaks, CA}
}

@article{vanlehn2011relative,
  title={The relative effectiveness of human tutoring, intelligent tutoring systems, and other tutoring systems},
  author={VanLehn, Kurt},
  journal={Educational psychologist},
  volume={46},
  number={4},
  pages={197--221},
  year={2011},
  publisher={Taylor \& Francis}
}

@article{shute2008focus,
  title={Focus on formative feedback},
  author={Shute, Valerie J},
  journal={Review of educational research},
  volume={78},
  number={1},
  pages={153--189},
  year={2008},
  publisher={Sage Publications}
}

@article{van2019orchestration,
  title={Orchestration tools to support the teacher during student collaboration: a review},
  author={van Leeuwen, Anouschka and Rummel, Nikol and others},
  journal={Unterrichtswissenschaft},
  volume={47},
  number={2},
  pages={143--158},
  year={2019},
  publisher={Juventa Verlag GmbH}
}

@article{yangspark,
  title={SPARK: Real-Time Monitoring of Multi-Faceted Programming Exercises},
  author={Yang, Yinuo and Zhang, Ashley Ge and Oney, Steve and Wang, April Yi},
  year={2025},
}

@article{holstein2019co,
  title={Co-designing a real-time classroom orchestration tool to support teacher-AI complementarity.},
  author={Holstein, Kenneth and McLaren, Bruce M and Aleven, Vincent},
  journal={Grantee Submission},
  year={2019},
  publisher={ERIC}
}

@article{martinez2014mtfeedback,
  title={MTFeedback: providing notifications to enhance teacher awareness of small group work in the classroom},
  author={Martinez-Maldonado, Roberto and Clayphan, Andrew and Yacef, Kalina and Kay, Judy},
  journal={IEEE Transactions on Learning Technologies},
  volume={8},
  number={2},
  pages={187--200},
  year={2014},
  publisher={IEEE}
}

@article{lui2023facilitated,
  title={Facilitated model-based reasoning in immersive virtual reality: Meaning-making and embodied interactions with dynamic processes},
  author={Lui, Michelle and Chong, Kit-Ying Angela and Mullally, Martha and McEwen, Rhonda},
  journal={International Journal of Computer-Supported Collaborative Learning},
  volume={18},
  number={2},
  pages={203--230},
  year={2023},
  publisher={Springer}
}

@article{wei2022emergent,
  title={Emergent abilities of large language models},
  author={Wei, Jason and Tay, Yi and Bommasani, Rishi and Raffel, Colin and Zoph, Barret and Borgeaud, Sebastian and Yogatama, Dani and Bosma, Maarten and Zhou, Denny and Metzler, Donald and others},
  journal={arXiv preprint arXiv:2206.07682},
  year={2022}
}

@article{wei2022chain,
  title={Chain-of-thought prompting elicits reasoning in large language models},
  author={Wei, Jason and Wang, Xuezhi and Schuurmans, Dale and Bosma, Maarten and Xia, Fei and Chi, Ed and Le, Quoc V and Zhou, Denny and others},
  journal={Advances in neural information processing systems},
  volume={35},
  pages={24824--24837},
  year={2022}
}

@inproceedings{wu2022ai,
  title={Ai chains: Transparent and controllable human-ai interaction by chaining large language model prompts},
  author={Wu, Tongshuang and Terry, Michael and Cai, Carrie Jun},
  booktitle={Proceedings of the 2022 CHI conference on human factors in computing systems},
  pages={1--22},
  year={2022}
}

@article{park2024generative,
  title={Generative agent simulations of 1,000 people},
  author={Park, Joon Sung and Zou, Carolyn Q and Shaw, Aaron and Hill, Benjamin Mako and Cai, Carrie and Morris, Meredith Ringel and Willer, Robb and Liang, Percy and Bernstein, Michael S},
  journal={arXiv preprint arXiv:2411.10109},
  year={2024}
}

@article{estevez2024evaluation,
  title={Evaluation of LLM Tools for Feedback Generation in a Course on Concurrent Programming},
  author={Est{\'e}vez-Ayres, Iria and Callejo, Patricia and Hombrados-Herrera, Miguel {\'A}ngel and Alario-Hoyos, Carlos and Delgado Kloos, Carlos},
  journal={International Journal of Artificial Intelligence in Education},
  pages={1--17},
  year={2024},
  publisher={Springer}
}

@inproceedings{nguyen2024beginning,
  title={How Beginning Programmers and Code LLMs (Mis) read Each Other},
  author={Nguyen, Sydney and Babe, Hannah McLean and Zi, Yangtian and Guha, Arjun and Anderson, Carolyn Jane and Feldman, Molly Q},
  booktitle={Proceedings of the CHI Conference on Human Factors in Computing Systems},
  pages={1--26},
  year={2024}
}

@article{tang2024vizgroup,
  title={VizGroup: An AI-Assisted Event-Driven System for Real-Time Collaborative Programming Learning Analytics},
  author={Tang, Xiaohang and Wong, Sam and Pu, Kevin and Chen, Xi and Yang, Yalong and Chen, Yan},
  journal={arXiv preprint arXiv:2404.08743},
  year={2024}
}

\appendix
\newpage
\section{Formative Study}
\subsection{Participant Demographics}

\begin{table}[h]
\centering
\caption{Instructor Demographics and Classroom Information}
\Description{Instructor Demographics and Classroom Information}
\resizebox{\linewidth}{!}{ % 自动适应列宽
\begin{tabular}{cccccc}
\toprule
\textbf{ID} & \textbf{Gender} & \textbf{Age} & \textbf{Experience (Years)} & \textbf{Class Size} & \textbf{No. of Activities} \\
\midrule
T1 & Male   & 36 & 8  & 40  & 5  \\
T2 & Male   & 29 & 5  & 50  & 4  \\
T3 & Female & 49 & 23 & 100 & 8  \\
T4 & Female & 33 & 6  & 45  & 5  \\
T5 & Male   & 36 & 6  & 100 & 10 \\
T6 & Female & 34 & 5  & 100 & 5  \\
T7 & Female & 31 & 4  & 70  & 4  \\
\bottomrule
\end{tabular}
}
\label{tab:instructor-info}
\end{table}

\RR{

\subsection{Interview Question for Instructor}
\label{Instructor}
\textbf{1. Basic Information}

\begin{enumerate}
    \item What is your age and gender?
    \item How many years of programming teaching experience do you have?
    \item How many students are typically in your class?
    \item How many programming practice sessions do you usually organize during a course?
\end{enumerate}

\textbf{2. Understanding Your Teaching Experience}

\begin{enumerate}
    \setcounter{enumi}{4}
    \item In programming courses, do you prefer students to complete lab assignments individually or in groups?
    \item Why do you prefer individual lab work or group lab work?\\
    \textit{Follow-up:} What do you think are the advantages and disadvantages of individual lab work?\\
    Advantages for students and instructors:\\
    Disadvantages for students and instructors:
\end{enumerate}

\textbf{3. Barriers to Teacher Student Interaction}

\begin{enumerate}
    \setcounter{enumi}{6}
    \item Do you observe any barriers that make students reluctant to ask questions, such as social distance or fear of asking simple questions?\\
    \textit{Follow-up:} Have you found ways to address this, or is it difficult to change?
\end{enumerate}

\textbf{4. Instructional Guidance}

\begin{enumerate}
    \setcounter{enumi}{7}
    \item How do you usually guide students during programming activities? For example, prompting them to reason or giving direct solutions? Could you provide an example?
    \item How do you decide the level of guidance to provide when a student asks for help? \\
    \textit{Follow-up:} Does your decision depend on student performance, time constraints, or other factors?
    % \item In what situations do you prefer to give direct solutions, and when do you withhold answers to encourage exploration?
    % \item In what situations do you think it is appropriate to provide a specific solution? In what situations is it better to encourage students to explore on their own?
    \item Besides giving concrete solutions and offering heuristic encouragement, do you have other ways of guiding students?
    % \item Ideally, what should guidance for students look like in your teaching?
    % \item In your classes, what type of guidance do you think students need more? Or in what scenarios do they need each type of guidance?
    \item Do you proactively offer help during labs? What factors influence your decision to intervene?
    % \item What factors do you consider when you proactively offer help, such as course progress or common difficulties among students?
    
\end{enumerate}

\textbf{5. During Student Agent Interactions}

\begin{enumerate}
    \setcounter{enumi}{11}
    \item If an AI agent provided feedback to students, what capabilities or types of feedback would you want it to have?
    \item What information about student–agent interactions would you want access to (e.g., conversation logs, task progress, question types)?
    \item Do you think it is necessary to monitor or evaluate the agent's performance? How would you assess whether it is helping students effectively?
    \item Do you need information about the students' task completion status?
    \item In general, do you believe an LLM-based agent can support student learning effectively?
\end{enumerate}

\textbf{6. Perspectives on Agents}

\begin{enumerate}
    \setcounter{enumi}{16}
    \item If an AI agent could monitor or participate in student discussions, what functions or characteristics would you want it to have?
    \item What aspects of the agent's behavior should be adjustable, and who should have control (instructors, students, or both)?
    \item Would you prefer the agent to actively participate in student collaboration or to provide feedback only when needed?
    \item What aspects of the agent's behavior should be adjustable, and who should have control (instructors, students, or both)?
    \item Do you trust the outputs of LLM-based agents? Under what conditions would you need explanations or insights from them (e.g., participation analysis, concept understanding, off-topic detection)?
\end{enumerate}

}
\section{Task Design}
\label{task}

We deployed ClassAid in a graduate-level Data Visualization course offered by the Department of Computer Science at the local research university. The course combines theoretical instruction with in-class programming activities to introduce key principles of data visualization and their practical implementation.
To prepare for deployment, we met with the course instructor over three one-hour online meetings. After discussing the feasibility of integration, the instructor agreed to adopt ClassAid for in-class use, and we collaboratively defined the classroom programming tasks.

\subsection{Task Design}
\label{task}
Both Task 1 and Task 2 used the same synthetic dataset consisting of 100 two-tuples, where one element is a numeric score and the other is a categorical label. The dataset exhibited clusters of values within specific ranges, making it particularly suitable for binned statistics and aggregation-based visual analysis. The tasks were designed to help students uncover score distributions and explore relationships between categories.

\subsubsection{Task 1: Score Distribution Chart.}

In Task 1, students were asked to use Vega-Lite to create a bar chart showing the distribution of scores across predefined ranges. The x-axis was required to display binned score intervals, while the y-axis represented the count of scores within each bin. Additionally, appropriate axis labels and a chart title were expected. This task assessed students' ability to create basic bar charts using Vega-Lite, focusing on binning continuous variables, aggregating values per bin, and presenting the results in a readable format.

\subsubsection{Task 2: Class Average Scores}

Task 2 required students to create a bar chart depicting the average score for each category (A, B, C, D, E). The x-axis represented categories, the y-axis displayed average scores, and bars were colored based on category. Building on the first task, Task 2 evaluated students' ability to perform data aggregation, assign categorical values to the x-axis, apply color encoding, and present average values in a clear and interpretable~chart.

\section{Interview Questions for Students}

\textbf{1. Prior Experience with AI Tools}
\begin{enumerate}
    \item Have you used AI tools to assist with programming before? How about during classroom programming activities?
    \item What motivated you to use AI? Was it mainly to get direct answers?
    \item Did you find the AI-generated responses helpful?
    \item Do you feel that relying on AI allowed you to complete the task without truly learning?
    \item Have you used AI tools in other classes? Were those uses allowed by your instructors?
\end{enumerate}

\textbf{2. Comparing Our System with Other AI Tools}
\begin{enumerate}
    \setcounter{enumi}{5}
    \item When using our system, did you interact with it the same way you would with other AI tools (e.g., directly asking for answers or requesting code fixes)?
    \item Did you notice any differences between our system's responses and those from other AI tools? What were they?
    \item Which type of response did you prefer, and why?
\end{enumerate}

\textbf{3. Social and Emotional Reactions}
\begin{enumerate}
    \setcounter{enumi}{8}
    \item Did the fact that teachers could view your AI interactions make you feel uncomfortable or less willing to ask direct questions?
\end{enumerate}

\textbf{4. System Usage Patterns and Feedback Perception}
\begin{enumerate}
    \setcounter{enumi}{9}
    \item Did you frequently check the provided tutorials? Did you rely more on the AI or the tutorials to complete tasks?
    \item How would you evaluate the AI responses in our system? What were the strengths and weaknesses?
    \item What had the greatest impact on your experience? Do you have any examples to share?
    \item When you received proactive feedback, did it arrive in a timely and useful manner? Why or why not?
\end{enumerate}

\textbf{5. Feedback Evaluation and AI Mode Awareness}
\begin{enumerate}
    \setcounter{enumi}{13}
    \item When did you feel inclined to rate the AI's feedback? Or why did you choose not to rate it?
    \item In which cases were you most likely to give a rating (e.g., when the feedback was very wrong or particularly helpful)?
    \item Did you notice when the teacher adjusted your AI mode? Could you guess why? Was the adjustment timely?
    \item Did you perceive differences between the AI modes? If so, what were they?
    \item Which mode did you prefer, and why?
    \item If you could choose your own AI mode, what factors would guide your decision?
\end{enumerate}

\textbf{6. Learning Outcomes and Teacher Interaction}
\begin{enumerate}
    \setcounter{enumi}{19}
    \item Compared to before, do you feel you gained more knowledge or had a better grasp of the material?
    \item Did the system enhance your interaction with the teacher? Besides adjusting the AI, did the teacher provide any extra support?
\end{enumerate}

\textbf{7. Reflections and Future Expectations}
\begin{enumerate}
    \setcounter{enumi}{21}
    \item Do you have any suggestions for improving the system?
    \item Would you prefer the system to help you expand your knowledge or simply assist in completing the task?
    \item If the system were applied to other subjects such as Python learning, would you be willing to use it?
\end{enumerate}

\section{Interview Questions for Instructors}

\textbf{1. Teaching Practices and Existing Challenges}
\begin{enumerate}
    \item How did you typically organize programming activities in your previous classes? What challenges did you face?
    \item Were you able to monitor students' learning behaviors and progress in real time? Did they actively ask for help?
    \item Were you previously aware of students' weaknesses in specific areas? How did you evaluate their programming skills then, and how does that compare to now?
\end{enumerate}

\textbf{2. System Impact and Perceived Changes}
\begin{enumerate}
    \setcounter{enumi}{3}
    \item Do you think our system has helped address some of the issues you faced before? Could you elaborate?
    \item Has the system helped you better understand students' programming abilities and individual differences?
    \item Were you able to identify students who needed additional support beyond the AI? What actions did you take?
    \item Besides adjusting the AI feedback mode, did you have any other interactions with students?
    \item Which part of the system did you find most useful? Why?
    \item Which part did you find less useful? Why?
\end{enumerate}

\textbf{3. Information Presentation and System Performance}
\begin{enumerate}
    \setcounter{enumi}{9}
    \item Do you find the system's information sufficiently detailed? Were there any insights you wished to see but couldn't?
    \item Did this information help create more opportunities for interacting with students?
    \item Was the data visualization intuitive and easy to interpret?
    \item How would you evaluate the system's responsiveness and real-time performance?
    \item Did you find the system workflow smooth? Was it overly complex at any point?
\end{enumerate}

\textbf{4. AI Feedback Quality and Control Strategy}
\begin{enumerate}
    \setcounter{enumi}{14}
    \item From your observation, was the AI-generated feedback helpful to students? How would you rate its quality? Did it help students complete tasks or improve their coding skills?
    \item Was the system's \textit{proactive feedback} effective? Did you observe students' attitudes toward it? Were there students who disliked it, or none who explicitly liked it?
    \item What was your strategy for adjusting the AI feedback modes? Were there general rules or specific student behaviors that guided your decisions?
\end{enumerate}

\textbf{5. Notable Events and Student Reactions}
\begin{enumerate}
    \setcounter{enumi}{17}
    \item Were there any interesting or memorable events during your use of the system? (e.g., the student in Silent Mode who still struggled even after switching to Technical Mode)
    \item Did you receive any direct feedback from students? What did they say?
\end{enumerate}

\textbf{6. Teaching Load and Trust in AI}
\begin{enumerate}
    \setcounter{enumi}{19}
    \item Did the system help reduce your teaching burden? For example, by simplifying feedback delivery or improving awareness of student performance?
    \item After being given control over the AI, did you feel more confident or trusting in its role?
\end{enumerate}

\textbf{7. Insights and Reflections}
\begin{enumerate}
    \setcounter{enumi}{21}
    \item Did you discover anything new through using the system? For example, students not knowing how to ask questions or struggling to articulate their problems to the AI?
    \item Would you be willing to continue using the system in the future? Why or why not?
\end{enumerate}

\section{Prompt for Reviewing and Assessing Student History}

\begin{lstlisting}
prompt = """
You are a ReviewlAgent responsible for reviewing and summarizing a student's learning trajectory in response to system triggers. Your goal is to assess the student's current and past progress to support the delivery of targeted, pedagogically aligned feedback.

You will receive:
- Recent activity context: includes latest interactions, code submissions, and AI feedback traces.
- Historical learning profile: includes completed tasks, concept mastery, performance patterns, and learner preferences.

Your job is to produce a structured summary across five dimensions:
1. **Cognitive Analysis**: Determine the current Bloom level and confidence trend based on recent interactions. Identify signs of stagnation or regression. Use Bloom's taxonomy to classify student questions or actions into one of six levels: Remember, Understand, Apply, Analyze, Evaluate, Create. Include reasoning for the classification.
2. **Error Analysis**: Extract error types, frequency, and distribution. Identify recurring mistakes and infer potential conceptual misunderstandings.
3. **Learning History**: Report preferred feedback mode, completed tasks, success rate, and learning style.
4. **Current State**: Assess current task status, recent triggers, activity level, and code quality.
5. **Knowledge State**: Distinguish mastered vs. struggling concepts, and highlight areas needing further support.

--- Input Schema ---
{
  "recent_activity": { ... },
  "historical_profile": { ... }
}

--- Output Format ---
{
  "cognitive_analysis": {
    "level": "apply",
    "confidence": 0.8,
    "reasoning": "The student is asking how to apply a specific encoding to a Vega-Lite chart, which aligns with the Apply level."
  },
  "error_analysis": { ... },
  "learning_history": { ... },
  "current_state": { ... },
  "knowledge_state": { ... },
  "metadata": { "is_auto_generated": true }
}

Behavioral Rules:
- Use Bloom's taxonomy to infer and track cognitive progression.
- Identify stagnation if levels drop or remain unchanged.
- Analyze code execution success and progress to infer task phase.
- Detect high-frequency error patterns linked to conceptual gaps.
- Maintain structured, valid output suitable for downstream reasoning agents.
"""

\end{lstlisting}
\section{Prompt for Considering Appropriate Forms of Learning Support}
\begin{lstlisting}
    """
You are a ThoughtFormation agent that transforms user input, student state, and code context
into structured 'thoughts' for downstream tutoring agents. You generate multiple categorized
responses to support reflective feedback, scaffolded guidance, and proactive corrections.

Your job is to segment the incoming data into actionable thinking paths for three feedback agents:
1. TechnicalAgent (for code-based explanation and fixes)
2. HeuristicAgent (for reflective, question-based prompts)
3. MetaAgent (for cognitive-level insights and metadata packaging)

**Input Schema**:
{
    "user_message": "Why is my chart not showing any bars?",
    "current_code": "{ \"mark\": \"bar\", \"data\": {}, \"encoding\": {} }",
    "response_mode": "auto",
    "retrieval_result": {
        "cognitive_analysis": {
            "level": "understand",
            "confidence": 0.8
        },
        "error_analysis": {
            "patterns": [
                { "type": "data", "message": "Missing values field" }
            ],
            "most_common": "data"
        },
        "learning_history": {
            "preferred_mode": "technical",
            "completed_tasks_count": 4
        },
        "current_state": {
            "recent_triggers": [{"type": "run", "is_auto_generated": true}]
        },
        "metadata": {
            "is_auto_generated": true
        }
    },
    "task_id": "task1"
}

**Output Schema** (Auto Mode Example):
{
    "is_automatic": true,
    "mode_used": "auto",
    "timestamp": 1712854012.123,
    "thoughts": {
        "technical": [
            "Your chart doesn't render because the 'data' field is empty. Try adding a 'values' object with your data.",
            "You may be missing 'encoding' definitions. Vega-Lite needs both 'x' and 'y' channels to position bars.",
            "Make sure you include a mark type and fields for encoding. A minimal bar chart requires: 'mark', 'data', 'encoding'."
        ],
        "heuristic": [
            "What fields are you trying to display on the x and y axes?",
            "How does your data structure match the encoding definition?",
            "Have you defined both the mark and the encoding channels required for this chart?"
        ]
    },
    "metadata": {
        "cognitive_level": "understand",
        "error_patterns": [
            { "type": "data", "message": "Missing values field" }
        ],
        "learning_history": {
            "preferred_mode": "technical",
            "completed_tasks_count": 4
        },
        "current_analysis": {
            "code": {
                "has_mark": true,
                "has_data": false,
                "has_encoding": false
            },
            "question": {
                "types": ["debug", "visualization"],
                "thinking_type": "A",
                "has_hypothesis": false
            },
            "is_automatic": true,
            "task_id": "task1"
        }
    }
}

**Output Example (Technical Mode Only)**:
{
    "is_automatic": false,
    "mode_used": "technical",
    "timestamp": 1712854012.345,
    "thoughts": {
        "technical": [
            "Try specifying your data like this:\n    \"data\": { \"values\": [ { \"x\": 1, \"y\": 2 }, { \"x\": 2, \"y\": 3 } ] }",
            "Your encoding might be missing. Add:\n    \"encoding\": { \"x\": { \"field\": \"x\", \"type\": \"quantitative\" }, \"y\": { \"field\": \"y\", \"type\": \"quantitative\" } }",
            "Also make sure your mark is set to 'bar'. Try this:\n    \"mark\": \"bar\""
        ]
    },
    "metadata": {
        "cognitive_level": "understand",
        "error_patterns": [
            { "type": "data", "message": "Missing values field" }
        ],
        "learning_history": {
            "preferred_mode": "technical",
            "completed_tasks_count": 4
        },
        "current_analysis": {
            "code": {
                "has_mark": true,
                "has_data": false,
                "has_encoding": false
            },
            "question": {
                "types": ["debug", "visualization"],
                "thinking_type": "A",
                "has_hypothesis": false
            },
            "is_automatic": false,
            "task_id": "task1"
        }
    }
}

**Behavior Rules**:

1. **Silent Mode**:
   - Do not generate any textual response.
   - Only return updated metadata (e.g., error pattern, cognitive level, etc.)

2. **Auto Mode**:
   - Generate both technical and heuristic responses.
   - Each response category must include 3 distinct, task-aligned responses.
   - Use historical error patterns and Bloom-level classification to customize difficulty.

3. **Technical Mode**:
   - Provide 3 fix-oriented, code-grounded responses.
   - Include complete but minimal code examples per response.
   - Ground each suggestion in task goals and data validity checks.

4. **Heuristic Mode**:
   - Provide 3 concise, question-led responses to stimulate student reflection.
   - Encourage self-correction and exploration without directly giving answers.
   - Align questions with the user's current cognitive level and task intent.

5. **Metadata Packaging**:
   - Always include structured metadata for downstream agent reasoning:
     * Bloom cognitive level
     * Error types and frequency
     * Learning history and style
     * Automatic vs. manual origin of message
     * Code validity and structure

**Technical Prompt Construction Example (in ThoughtFormationSystem)**

Depending on whether the message is automatically triggered or manually asked by the user,
the following templates are used to construct prompt for OpenAI API.

Case 1: Automatically Detected (is_automatic = true)

Prompt Template:
"""
As a proactive technical tutor, generate 3 different responses to automatically detected issues.
Each response MUST be completely separated from others using the '---RESPONSE---' marker.

Current Context:
- System Message: {user_message}
- Code: {current_code}
- Code Analysis: {code_analysis_dict}
- Question Analysis: {question_analysis_dict}
- Data Status: {"Valid" or "Invalid"}
{task_specific_context}

Historical Context:
- Cognitive Level: {cognitive_level}
- Error Patterns: {error_list}
- Learning process: {student_process}

For EACH response, please:
1. Proactively identify potential issues or improvements specific to the current task
2. Provide clear, actionable suggestions that align with the task's visualization goals
3. Include specific code examples that match the task requirements
4. Focus on best practices for the specific type of visualization needed
5. Keep explanations concise but comprehensive

IMPORTANT FORMAT REQUIREMENTS:
- DO NOT use any markdown formatting (no **, *, _, etc.)
- Use plain text only
- For emphasis, use UPPERCASE words instead of markdown
- For code examples, simply indent them with 4 spaces
- Keep line breaks for readability

Format your response exactly like this:

---RESPONSE---
[First proactive technical response in plain text]

---RESPONSE---
[Second proactive technical response in plain text]

---RESPONSE---
[Third proactive technical response in plain text]
"""

Case 2: User-Initiated (is_automatic = false)

Prompt Template:
"""
As a technical tutor, please generate 3 different technical responses to the user's question.
Each response MUST be completely separated from others using the '---RESPONSE---' marker.

Current Context:
- Question: {user_message}
- Code: {current_code}
- Code Analysis: {code_analysis_dict}
- Question Analysis: {question_analysis_dict}
- Data Status: {"Valid" or "Invalid"}
{task_specific_context}

Historical Context:
- Cognitive Level: {cognitive_level}
- Error Patterns: {error_list}
- Learning process: {student_process}

For EACH response, please:
1. Directly answer the user's question with explanation
2. Provide working code examples that align with the task's specific requirements
3. Build upon the student's historical learning
4. Explain why the suggested code works and is appropriate for this specific visualization task
5. Include complete, working code examples with necessary explanations, also need concise

Format your response exactly like this:

---RESPONSE---
[First technical response here with explanation and reasoning]

---RESPONSE---
[Second technical response here with explanation and reasoning]

---RESPONSE---
[Third technical response here with explanation and reasoning]
"""

"""

\end{lstlisting}
\section{Prompt for Selecting Adaptive Feedback Modes}
\begin{lstlisting}
    prompt = """
You are a teaching assistant AI responsible for selecting the most appropriate feedback style (technical or heuristic) and choosing the best response for the student.

You will receive:
1. Student's current question and code.
2. A set of generated technical and heuristic responses.
3. Analysis results about the student's cognitive level, past learning behavior, error patterns, and code structure.

Your task is to:
A. Determine whether the technical or heuristic feedback mode is more suitable for the current student situation.
B. Select the single best response from the chosen mode that maximally aligns with the student's needs.

Use the following decision framework:
- Mode selection is based on cognitive psychology and instructional strategy principles.
- Apply a weighted scheme to assess:
  * Current cognitive level (50%)
  * Error types (20%)
  * Learning history (30%)

You can refer to the principles:
- If the student is at the Apply level or above, with mostly logic/design errors and steady progress, prefer heuristic feedback.
- If the student is at lower cognitive levels, has frequent syntax errors, or shows inconsistent progress, prefer technical feedback.

Once a mode is selected, rank the candidate responses from that mode using the following five criteria:
1. Relevance to the student's question and code (40%)
2. Complexity appropriate to the cognitive level (20%)
3. Consistency with prior behavior and learning history (20%)
4. Clarity of explanation (15%)
5. Urgency based on current errors or stagnation (5%)

Choose the top-ranked response and justify your choice.

--- Input Example ---
student_question: Why is my bar chart blank?
student_code: { "mark": "bar", "encoding": {}, "data": {} }
thoughts: {
  "technical": ["Try adding a 'values' array to your data field.", "Ensure 'encoding' has 'x' and 'y' fields defined.", "The chart is empty because Vega-Lite can't draw bars without data values."],
  "heuristic": ["What is missing from the data definition?", "How does your encoding connect to the dataset?", "Can you see if you've declared both axes?"]
}
cognitive_info: {"level": "understand", "confidence": 0.7, "has_hypothesis": false}
error_info: {"patterns": [{"type": "data", "message": "Missing values field"}]}
learning_history: {"preferred_mode": "heuristic", "completed_tasks_count": 3, "success_rate": 0.6}
code_analysis: {"has_mark": true, "has_data": false, "has_encoding": false}

--- Output Format ---
{
  "selected_mode": "heuristic" or "technical",
  "selected_response": "...chosen response string from that mode...",
  "justification": "...explanation of why this response and mode were selected based on cognitive and code context."
}

--- Output Example ---
{
  "selected_mode": "heuristic",
  "selected_response": "What is missing from the data definition?",
  "justification": "The student's question is vague and the code lacks both data and encoding, which suggests early confusion. Their preferred mode is also heuristic, and their Bloom level is 'understand'. A reflective prompt would encourage self-discovery more effectively than a direct fix."
}

Please analyze carefully. Consider if the student is looking for a direct fix or needs guidance. If the question is vague or the code is incomplete, heuristic guidance may be better. If the error is obvious and the student has shown technical preference, technical feedback might be better.
"""
\end{lstlisting}
\section{Prompt for Intervening to Support Learning Progress}
\begin{lstlisting}
     prompt = """
You are a classroom intervention assistant AI responsible for deciding whether an instructor should intervene to support a student during programming activities.

You will receive:
1. The student's current feedback mode (e.g., auto, technical, heuristic)
2. Cognitive analysis (level, confidence, understanding)
3. Error analysis (types, frequency, severity)
4. Learning history (success rate, help frequency, completion rate)
5. Trigger event information (type: active, passive, predictive, and whether auto-generated)

Your task is to:
A. Decide whether intervention is necessary at this time.
B. Provide a justification for your decision based on cognitive need, error risk, history, and trigger information.
C. Assign an intervention score (0.0 to 1.0) and suggest an appropriate intervention mode ("proactive", "passive").

Use the following decision framework:
- Immediately intervene if explicit help is requested (passive, not auto-generated), or if the student is stagnant (active trigger with duration > 60s).
- Otherwise, compute a motivation-to-intervene score using:
  * Error severity and frequency (40%)
  * Cognitive level, confidence, and understanding (30%)
  * Historical performance (30%)
- If the combined score exceeds 0.5, initiate proactive intervention.
- If the score is below threshold, refrain from intervening to encourage autonomy and metacognitive development.

--- Input Example ---
current_mode: "auto"
cognitive_analysis: {
  "level": 2,
  "confidence": 0.4,
  "understanding": 0.3
}
error_analysis: {
  "patterns": [{"type": "syntax"}, {"type": "runtime"}],
  "frequency": {"syntax": 3, "runtime": 2}
}
learning_history: {
  "success_rate": 0.4,
  "completion_rate": 0.5,
  "help_frequency": 0.6
}
trigger_info: {
  "type": "active",
  "details": {"is_stagnant": true, "duration": 140}
}

--- Output Format ---
{
  "should_intervene": true or false,
  "intervention_score": float (0.0 - 1.0),
  "mode": "technical" | "heuristic" | "auto",
  "reason": "...rationale for the decision...",
  "timestamp": "YYYY-MM-DDTHH:MM:SS"
}

--- Output Example ---
{
  "should_intervene": true,
  "intervention_score": 0.85,
  "mode": "proactive",
  "reason": "Student shows low confidence and understanding, has high help frequency, and experienced stagnation for 140 seconds. Proactive intervention recommended to re-engage the student and provide timely support.",
  "timestamp": "2025-04-11T22:45:00"
}

Please weigh all inputs thoughtfully and act in the student's best interest.
"""

\end{lstlisting}

\RX{
\section{Performance Comparison of Different LLMs}
\label{comparison}
To clarify the engineering considerations underlying our choice of the large language model (LLM), we conducted a controlled comparison of multiple GPT models using five representative query types derived from classroom use. These query types include basic visualization construction, error diagnosis and repair, feature enhancement, conceptual explanation, and task oriented specification generation.
It is important to emphasize that this comparison focuses on system level metrics rather than semantic correctness or task optimality. Specifically, we evaluated average response latency, token usage, and output format compliance. Format compliance measures whether model outputs strictly adhere to the response schema required by \textit{ClassAid}, such as correct delimiters, complete response blocks, and parsable code. This property is critical for enabling instructor mediated orchestration and downstream processing in real time classroom settings.
}
\begin{lstlisting}
[
  {
    "query_type": "Q1: Basic Visualization Construction",
    "user_message": "How do I create a bar chart in Vega-Lite?",
    "current_code": "",
    "description": "Constructing a basic bar chart specification from scratch"
  },
  {
    "query_type": "Q2: Error Diagnosis and Repair",
    "user_message": "My code has an error: Missing encoding specification",
    "current_code": {
      "$schema": "https://vega.github.io/schema/vega-lite/v5.json",
      "data": {"values": [{"x": 1, "y": 2}]},
      "mark": "bar"
    },
    "description": "Identifying and repairing missing encoding fields"
  },
  {
    "query_type": "Q3: Feature Enhancement",
    "user_message": "How can I add colors to distinguish different categories?",
    "current_code": {
      "$schema": "https://vega.github.io/schema/vega-lite/v5.json",
      "data": {"values": [{"category": "A", "value": 10}, {"category": "B", "value": 20}]},
      "mark": "bar",
      "encoding": {
        "x": {"field": "category", "type": "nominal"},
        "y": {"field": "value", "type": "quantitative"}
      }
    },
    "description": "Extending an existing visualization with color encoding"
  },
  {
    "query_type": "Q4: Conceptual Explanation",
    "user_message": "What's the difference between bar and column charts?",
    "current_code": "",
    "description": "Explaining visualization concepts without code generation"
  },
  {
    "query_type": "Q5: Task-Oriented Specification Generation",
    "user_message": "I want to create a histogram showing score distribution",
    "current_code": {
      "$schema": "https://vega.github.io/schema/vega-lite/v5.json",
      "data": {"values": [{"score": 85}, {"score": 90}, {"score": 75}]}
    },
    "description": "Generating a complete visualization specification based on task intent"
  }
]
\end{lstlisting}

\RX{As summarized in Table~\ref{tab:llm_performance_cost}, all evaluated models were compared based on the average results from 30 experimental runs. While all the models were generally capable of generating valid responses, their performance varied significantly across different query types.
Other models, such as GPT 4 and GPT 4 Turbo, tended to generate longer and more detailed explanations but exhibited significantly higher response latency, particularly for complex task oriented queries. This characteristic makes them less suitable for real time classroom deployment. In contrast, GPT 4o consistently demonstrated a more balanced trade off among response latency, instructional richness, format compliance, and per query cost.

Based on these engineering considerations, we adopted GPT 4o for the classroom deployment of \textit{ClassAid}. We emphasize that our contribution does not depend on any specific LLM, but rather on the system design that enables real time instructor control over student AI interaction. The appendix level model comparison is intended to improve transparency and support future extensions of this work under alternative model choices.}

\begin{table}[t]
\centering
\caption{Performance and Cost Comparison of LLMs Across Query Types (Q1--Q5)}
\label{tab:llm_performance_cost}
\small % Change font size here
\resizebox{\linewidth}{!}{
\begin{tabular}{llccccc}
\toprule
\textbf{Model} & \textbf{Metric} & \textbf{Q1} & \textbf{Q2} & \textbf{Q3} & \textbf{Q4} & \textbf{Q5} \\
\midrule
\multirow{3}{*}{GPT-4o} 
& Avg. Response Time (s) & 7.33 & 10.51 & 9.00 & 5.51 & 10.72 \\
& Avg. Tokens / Request  & 563  & 966   & 1096 & 512  & 1101 \\
& Format Compliance Rate (\%) & 100 & 100 & 90 & 100 & 100 \\
& Input Cost (\$/1M tokens)  & \multicolumn{5}{c}{5.00} \\
& Output Cost (\$/1M tokens) & \multicolumn{5}{c}{15.00} \\
\midrule
\multirow{3}{*}{GPT-4-turbo} 
& Avg. Response Time (s) & 12.50 & 14.40 & 17.81 & 12.95 & 19.01 \\
& Avg. Tokens / Request  & 698   & 989   & 1129  & 651   & 1142 \\
& Format Compliance Rate (\%) & 100 & 100 & 100 & 100 & 53 \\
& Input Cost (\$/1M tokens)  & \multicolumn{5}{c}{10.00} \\
& Output Cost (\$/1M tokens) & \multicolumn{5}{c}{30.00} \\
\midrule
\multirow{3}{*}{GPT-4-0125-preview} 
& Avg. Response Time (s) & 10.71 & 19.94 & 41.48 & 11.65 & 19.86 \\
& Avg. Tokens / Request  & 602   & 1119  & 1214  & 662   & 1170 \\
& Format Compliance Rate (\%) & 100 & 93 & 100 & 100 & 83 \\
& Input Cost (\$/1M tokens)  & \multicolumn{5}{c}{10.00} \\
& Output Cost (\$/1M tokens) & \multicolumn{5}{c}{30.00} \\
\bottomrule
\end{tabular}
}
\end{table}

\end{document}